\documentclass[11pt,a4paper]{article} 
\pdfoutput=1
\usepackage{jheppub}
\usepackage[T1]{fontenc}
\usepackage{mathbbol}

\usepackage{bm}
\usepackage{booktabs}
\usepackage{colortbl}
\usepackage{enumerate}
\usepackage{latexsym}
\usepackage{lscape}
\usepackage{multirow}
\usepackage{psfrag}

\def\beq{\begin{equation}}
\def\eeq{\end{equation}}
\def\beqa{\begin{eqnarray}}
\def\eeqa{\end{eqnarray}}

\def\be{\begin{equation}}
\def\ee{\end{equation}}
\def\bea{\begin{eqnarray}}
\def\eea{\end{eqnarray}}

\def\nn{\nonumber}

\newcommand\Eqn[1]     {Eq.\,(\ref{#1})}

\newcommand\eqn[1]     {eq.\,(\ref{#1})}
\newcommand\eqns[2]    {eqs.\,(\ref{#1}) and~(\ref{#2})}

\newcommand{\bei}{\begin{itemize}}
\newcommand{\eei}{\end{itemize}}

\newcommand{\RE}{{\rm Re}}
\newcommand{\IM}{{\rm Im}}
\newcommand{\TR}{{\rm Tr}}

\def\T{{\bf T}}
\def\Tt{{\bf T}_t^2} 
\def\Ts{{\bf T}_s^2} 
\def\Tu{{\bf T}_u^2} 
\def\Tsu{{\bf T}_{s{-}u}^2}

\def\dbar{\,{\mathchar'26\mkern-12mu d}}  
\def\rGamma{{r_{\Gamma}}}

\newcommand{\CA}{C_{A}}

\newcommand{\nf}{n_f}

\def \MS {\overline{\rm MS}}

\newcommand{\as}{\alpha_s}

\newcommand{\eps}{\epsilon}

\newcommand{\ord}{{\cal O}}

\def\bra#1{\langle#1|}
\def\ket#1{|#1\rangle}
\def\bracket#1#2{\langle#1|#2\rangle}

\newcommand{\LO}{{\rm(LO)}}
\newcommand{\NLO}{{\rm(NLO)}}
\newcommand{\NNLO}{{\rm(NNLO)}}

\def\II{\mathcal{I}}
\def\MM{\mathcal{M}}
\def\Mreduced{\hat{\mathcal{M}}}

\def\Hhard{\mathcal{H}} 

\def\Hhat{\hat H} 

\def\Log{L}

\allowdisplaybreaks

\title{Two-parton scattering in the high-energy limit}

\author[a]{Simon Caron-Huot,}
\author[b]{Einan Gardi,}
\author[b]{Leonardo Vernazza}

\affiliation[a]{Department of Physics, McGill University, 3600 rue University, Montr\'eal, QC Canada H3A 2T8}
\affiliation[b]{Higgs Centre for Theoretical Physics, 
School of Physics and Astronomy, 
The University of Edinburgh, Edinburgh EH9 3FD, Scotland, UK}

\emailAdd{schuot@physics.mcgill.ca}
\emailAdd{Einan.Gardi@ed.ac.uk}
\emailAdd{lvernazz@staffmail.ed.ac.uk}

\abstract{Considering $2\to 2$ gauge-theory scattering with general colour in the high-energy limit, we 
compute the Regge-cut contribution to three loops through 
next-to-next-to-leading high-energy logarithms (NNLL) in the signature-odd sector. Our 
formalism is based on using the non-linear Balitsky-JIMWLK 
rapidity evolution equation to derive an effective Hamiltonian 
acting on states with a fixed number of Reggeized gluons. A 
new effect occurring first at NNLL is mixing between states 
with $k$ and $k+2$ Reggeized gluons due non-diagonal 
terms in this Hamiltonian.
 Our results are consistent with a recent determination of the 
 infrared structure of scattering amplitudes at three loops, as 
 well as a computation of $2\to 2$ gluon scattering in ${\cal N}=4$ 
 super Yang-Mills theory. Combining the latter with our Regge-cut 
 calculation we extract the three-loop Regge trajectory in this 
 theory. Our results open the way to predict high-energy 
 logarithms through NNLL at higher-loop orders.}


\begin{document} 

\begin{flushright}
Edinburgh 2017/01
\vspace*{-25pt}
\end{flushright}

\maketitle
\flushbottom


\section{Introduction}

The high-energy limit of gauge-theory scattering amplitudes 
has long been understood to offer a unique insight into gauge 
dynamics. In this kinematic limit, amplitudes drastically simplify 
and factorise in rapidity, giving rise to new degrees of freedom 
in two dimensions.
Within perturbative QCD, BFKL \cite{Kuraev:1977fs,Balitsky:1978ic} 
and related rapidity evolution equations allow us to translate 
concepts from Regge theory \cite{Collins:1977jy} into calculation 
tools, leading to concrete predictions.  The simplest example is 
that of the Reggeized gluon, the effective interaction which 
governs the behaviour of $2\to 2$ scattering amplitudes in 
QCD in the limit where the energy $s$ is much larger than 
the momentum transfer $-t$. In the leading logarithmic 
approximation the exchange of a single Reggeized gluon 
leads to a trivial evolution equation in rapidity, which amounts 
to straightforward exponentiation of logarithms of $|s/t|$ to all 
orders in the coupling.
At higher logarithmic accuracy more complex analytic structure 
emerges, which can be understood in QCD as compound states of two 
or more Reggeized gluons~\cite{Low:1975sv,Nussinov:1975mw,Gunion:1976iy}. 
In contrast to the single Reggeon case,
these are difficult to solve in general \cite{Lipatov:1993yb,Faddeev:1994zg}.
Nevertheless, they  can be integrated iteratively, thus generating perturbative 
high-energy amplitudes order-by-order in the coupling.

Taking the high-energy limit, $s\gg -t$, a fast moving 
projectile can be seen as a cloud of partons, each of 
which is dressed by a Wilson line, sourcing additional 
radiation. The high-energy limit corresponds to forward 
scattering, where recoil is neglected, hence the effective 
description is in terms of straight infinite lightlike Wilson 
lines~\cite{Korchemskaya:1994qp,Korchemskaya:1996je}. 
The number and transverse positions of these Wilson lines 
are not fixed, since the projectile can contain an arbitrary 
number of quantum fluctuations. The evolution of the 
system in rapidity is controlled by the Balitsky-JIMWLK 
equation~\cite{Balitsky:1995ub,Kovchegov:1999yj,JalilianMarian:1996xn,JalilianMarian:1997gr,Iancu:2001ad}. 
In Ref.~\cite{Caron-Huot:2013fea} it was shown how 
to translate the latter into evolution equations controlling 
a given number of Reggeized gluons. These equations 
are in general coupled, and in particular, the evolution 
of three Reggeized gluons involves mixing with a single 
Reggeized gluon. In the present paper we explore this 
mixing for the first time. We use the leading-order
Balitsky-JIMWLK equation to derive the effective 
Hamiltonians governing the diagonal and next-to-diagonal 
evolution terms describing $k$ Reggeized gluon evolution 
into $k$ and $k+2$ ones, respectively, and use symmetry considerations
to obtain the mixing into $k-2$ ones. We then use these 
evolution equations to explicitly compute three-loop 
corrections to the signature odd $2\to 2$ amplitude 
in the high-energy limit, and compare them to other 
recent results.

It is well known that gauge-theory amplitudes have 
long-distance singularities, which cancel in physical 
observables such as sufficiently inclusive cross 
sections. Owing to the factorization properties of 
fixed-angle scattering amplitudes~\cite{Collins:1989gx,Botts:1989kf} 
these singularities are largely process-independent. 
Furthermore, they admit evolution equations leading 
to exponentiation. Of special interest are soft 
singularities, which in contrast to collinear ones, 
are sensitive to the colour flow of the underlying 
hard process. Soft singularities can be computed 
by considering correlators of semi-infinite Wilson 
lines~\cite{Korchemsky:1985xj,Korchemsky:1985xu,Korchemsky:1987wg,Kidonakis:1997gm,Kidonakis:1998nf,Gardi:2013ita,Gardi:2011yz,Mitov:2009sv,Becher:2009kw}.
The corresponding \emph{soft anomalous dimension} 
encodes the structure of these singularities to all orders 
in perturbation theory. 
In recent years there has been significant 
progress~\cite{Catani:1998bh,Aybat:2006mz,Gardi:2009qi,Gardi:2009zv,Becher:2009qa,Dixon:2009ur,Ahrens:2012qz,Almelid:2015jia,Gardi:2016ttq} 
in determining the precise structure of long-distance 
singularities to massless gauge theories. Through a 
recent explicit computation of the soft anomalous 
dimension, these are now known in full for amplitudes 
with any number of legs in general kinematics through 
three loops~\cite{Almelid:2015jia,Gardi:2016ttq}.  

While infrared factorization of fixed-angle 
scattering and high-energy factorization start from different 
kinematic set ups, and are based on different evolution 
equations, they lead to partially overlapping predictions 
for the structure of scattering amplitudes. In recent years 
the complementary nature of these two factorization 
pictures has been put to 
use~\cite{Bret:2011xm,DelDuca:2011ae,Caron-Huot:2013fea,DelDuca:2013ara,DelDuca:2014cya}. 
For example, Refs.~\cite{Bret:2011xm,DelDuca:2011ae} 
showed that infrared factorization
excludes the simplest form of Regge factorization where 
the amplitude in the high-energy limit is governed by a so-called
Regge pole, and predicts that contributions associated 
with a Regge cut appear starting from the next-to-leading 
logarithmic (NLL) accuracy for the imaginary part of the 
amplitude and starting from the next-to-next-to-leading 
logarithmic (NNLL) accuracy for its real part. Conversely, 
it was shown how the Regge limit can constrain the 
(then unknown) three-loop soft anomalous dimension. 
Ref.~\cite{Caron-Huot:2013fea} used the Balitsky-JIMWLK 
equation to computed the first few orders in the Regge 
cut of the signature even part of the amplitude at NLL 
accuracy, and predicted a corresponding correction to 
the soft anomalous dimension in the high-energy limit 
at four loops. In this paper we use a similar technique 
to predict the signature odd amplitude at NNLL accuracy. 
This requires us to address for the first time the effect of 
non-diagonal terms in the effective Hamiltonian. We are 
then able to compute three-loop corrections generated 
by the evolution of three Reggeized gluons and their 
mixing with a single Reggeized gluon.  
Finally, we contrast our result with other recent calculations 
at three loops. First, the infrared singularities are compared with 
predictions based on the soft anomalous 
dimension~\cite{Almelid:2015jia,Gardi:2016ttq}, finding 
full consistency. Second, considering the case of gluon 
scattering in ${\cal N}=4$ Supersymmetric Yang-Mills 
theory (SYM), we find full agreement with the results 
of Ref.~\cite{Henn:2016jdu}, expanded in the high-energy 
limit. The latter, in combination with the Regge cut we 
computed, allows us to fix the three-loop gluon Regge 
trajectory in this theory.  

The outline of the paper is as follows. Section 
\ref{Sec:Aspects} introduces the relevant aspects 
of Regge and BFKL theory. This includes, in section 
\ref{signature}, a review and analysis of the relation 
between reality properties and signature within Regge 
theory. Section \ref{Regge_limit_perturbative} then 
focuses on reviewing the perturbative description of 
gluon Reggeisation and the structure of $2 \to 2$ 
scattering amplitudes in the high-energy limit. We 
conclude the introduction in section \ref{BFKL_abridged} 
where we explain how we use the Balitsky-JIMWLK 
equation to obtain information on the (non-diagonal) 
evolution of states with a fixed number of Reggeized 
gluons. The computation itself is described in 
section~\ref{Regge}, which starts with a derivation 
of the explicit form of the Hamiltonian for $k$ goes to $k$, $k+2$ and $k-2$
Reggeized gluons, and concludes 
with a calculation of all the relevant signature-odd 
matrix elements contributing through three loops. 
Finally, section~\ref{dip_comparison} is dedicated 
to a detailed comparison between the results of 
section~\ref{Regge} with the theory of infrared 
factorization. We begin by reviewing the latter, 
specializing the results of~\cite{Almelid:2015jia,Gardi:2016ttq} 
to the high-energy limit. We then systematically 
determine the ``infrared renormalized'' hard function 
based on our results of section~\ref{Regge} for the 
amplitude in the high-energy limit, and verify that the 
result is indeed finite. Explicit expressions for the 
anomalous dimensions are quoted in 
Appendix~\ref{GammaAndZ}, while 
Appendices~\ref{3LoopPredictOrtho} 
and~\ref{3LoopPredictTrace} collect the hard function in QCD 
gluon-gluon scattering in the $t$-channel 
colour flow basis and the ``trace'' basis, respectively. Finally 
Appendix~\ref{TrajectoryImpact} collects 
the results for high-energy factorization 
in ${\cal N}=4$ SYM. Our conclusions and 
some open questions are discussed in 
section~\ref{conclusion}.

\section{Aspects of \boldmath $2 \to 2$ scattering 
amplitudes in the high-energy limit\label{Sec:Aspects}}

In this paper we explore properties of $2\to 2$ 
QCD scattering amplitudes in the high-energy 
limit. This kinematical configuration is interesting 
because of the appearance of large logarithms 
of the centre of mass energy $s$ over the momentum 
transfer $t$, $\log |s/t|$. It is a well-known fact that 
these logarithms exponentiate at leading logarithmic 
(LL) accuracy, and also at the next-to-leading logarithmic 
(NLL) order, for some parts of the amplitude. A deeper 
understanding of their factorisation and exponentiation 
relies however on non-trivial properties of scattering 
amplitudes, that we discuss in this section. 

Our starting point is the study of analytic properties 
of scattering amplitudes. This is historically one 
of the first approaches to the study of amplitudes, 
which leads to the concepts of signature and of 
Regge poles, Regge cuts and Regge trajectories, 
that we briefly review below. Next, we explain how 
these concepts relate to the standard calculation 
of QCD scattering amplitudes as a perturbative 
expansion in the strong coupling constant. We 
introduce then the modern framework in which 
the factorisation of amplitudes in the the high-energy 
limit needs to be discussed, namely, the treatment 
of QCD radiation as originating from Wilson lines 
associated to the direction of the incoming and 
outgoing quarks and gluons. This framework allows 
one to link the origin of high-energy logarithms 
to the renormalisation-group evolution of amplitudes 
with respect to the rapidity, which is governed by 
BFKL theory, more specifically by the Balitsky-JIMWLK 
equation. 

\subsection{\boldmath
Signature and the high-energy limit of $2 \to 2$ amplitudes}
\label{signature}

\begin{figure}[t]
\begin{center}
  \includegraphics[width=0.34\textwidth]{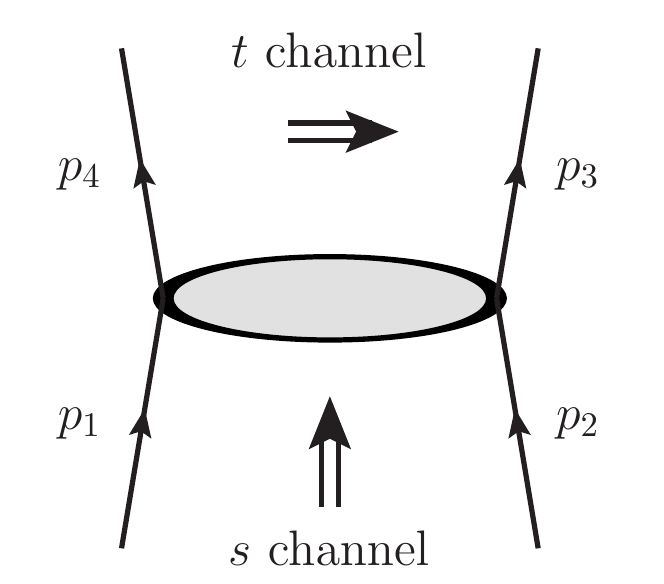}
  \end{center}
  \caption{The $t$-channel exchange dominating the high-energy limit, 
  $s\gg -t>0$. The figure also defined our conventions for momenta assignment 
  and Mandelstam invariants. We shall assume that particles $2$ and $3$ are 
  of the same type, and similarly for particles $1$ and $4$.}
\label{setup_fig}
\end{figure}

We consider $2\to 2$ scattering amplitudes, 
${\cal M}_{ij\to ij}$, where $i$, $j$ can be a quark 
or a gluon. In the following we will suppress these 
indices $i$, $j$, unless explicitly needed. In the 
high-energy limit the Mandelstam variables satisfy 
$s\gg -t>0$. The various terms of the amplitude 
will have definite reality properties, which are related 
to the properties of the amplitude under crossing. 
This is a consequence of the analytic structure, which 
is conveniently summarised via dispersion relations:
\begin{equation}
\label{Adisp}
{\cal M}(s,t)=\frac{1}{\pi}\int_0^{\infty} \frac{d\hat{s}}{\hat{s} - s -i0} \, D_{s}(\hat{s},t)
+ \frac{1}{\pi}\int_0^\infty \frac{d\hat{u}}{\hat{u}+s+t-i0} \, D_{u}(\hat{u},t)
\end{equation}
where $D_{s}$ and $D_{u}$ are the discontinuities of ${\cal M}(s,t)$ in the 
$s$- and $u$-channels, respectively. In general the lower limit 
of integration should of course be a positive threshold, 
and there could be subtraction terms, but this would 
not matter for our discussion. The important fact is that 
the discontinuities $D_{s}$ and $D_u$ are real, 
having a physical interpretation as spectral density of positive 
energy states propagating in the $s$ and $u$ channel 
respectively. To see the consequence on the amplitude, 
let us parametrize the discontinuities as a sum of power 
laws by means of a Mellin transformation:
\begin{subequations}
\begin{align}
a^{s}_j(t)&=\frac{1}{\pi} \,\int_0^{\infty} \frac{d \hat{s}}{\hat{s}} \, 
D_{s}(\hat{s},t) \, \left(\frac{\hat{s}}{-t}\right)^{-j}\label{Mellin_trans}\,,  \\
D_{s}(s,t)&=\frac{1}{2i} \int_{\gamma- i\infty}^{\gamma+i\infty} dj 
\, a^{s}_j(t) \, \left(\frac{s}{-t}\right)^{j}\,, \label{invMellin_trans}
\end{align}
\end{subequations}
and similarly for $a^u$ and $D_u$. Note that the reality condition 
of $D_{s}(s,t)$ implies that the Fourier coefficients admit 
\be\label{reality-condition}
\left(a_{j^*}^s(t) \right)^*=a_j^s(t),
\ee
and similarly for $a_j^u(t)$. Substituting the 
inverse transform \eqn{invMellin_trans} into the 
dispersive representation \eqn{Adisp}, swapping the order of 
integration and performing the $\hat{s}$ and $\hat{u}$ integrals, 
one obtains a Mellin representation of the amplitude:
\begin{equation}
\label{Ast_mellin}
{\cal M}(s,t)=\frac{-1}{2i}\int_{\gamma- i \infty}^{\gamma + i \infty} \frac{dj}{\sin (\pi j)} \,
\left(a_j^{s}(t)  \,\left(\frac{-s-i0}{-t}\right)^{j}+ a_j^{u}(t)  \,\left(\frac{s+t-i0}{-t}\right)^{j}\right)\,.
\end{equation}
Since the coefficients $a_j^{s,u}$ are real (for real $j$), 
and $(-s-i0)^j=e^{-i\pi j}|s|^j$ for $s>0$, we see that the 
phase of each power law contribution is related to its exponent.
The statement simplifies when one projects the amplitude onto 
eigenstates of \emph{signature}, that is crossing symmetry
$s\leftrightarrow u$:
\begin{equation}\label{Odd-Even-Amp-Def}
 {\cal M}^{(\pm)}(s,t) = \tfrac12\Big( {\cal M}(s,t) \pm {\cal M}(-s-t,t) \Big),\\
\end{equation}
where ${\cal M}^{(+)}$, ${\cal M}^{(-)}$
are refered respectively to as the \emph{even}
and \emph{odd} amplitudes. Restricting to the 
region $s>0$ and working to leading 
power as $s\gg |t|$, the formula then evaluates to
\begin{subequations}
\label{even_odd_Inv_Mellin}
\begin{align}
\label{even_Inv_Mellin}
{\cal M}^{(+)}(s,t) &= i \,\int_{\gamma- i\infty}^{\gamma+i\infty} 
 \frac{dj}{\sin (\pi j)} \cos\Big(\tfrac{\pi j}{2}\Big)\, a^{(+)}_j(t) \, e^{jL}\,, \\
\label{odd_Inv_Mellin}
{\cal M}^{(-)}(s,t) &=  \int_{\gamma- i\infty}^{\gamma+i\infty} 
 \frac{dj}{\sin (\pi j)}\sin\Big(\tfrac{\pi j}{2}\Big)\, a^{(-)}_j(t) \, e^{jL}\,,
\end{align}
\end{subequations}
where we have defined $a^{(\pm)}_j(t) \equiv \tfrac12(a^s_j(t) \pm a^u_j(t))$ 
and $L$ is the natural signature-even combination of logarithms:
\begin{equation}\begin{aligned}
\label{L-def}
L &\equiv  \log\left|\frac{s}{t}\right| -i\frac{\pi}{2}
\\ &= \frac12\left(\log\frac{-s-i0}{-t}+\log\frac{-u-i0}{-t}\right)\,.
\end{aligned}
\end{equation}
Let us interpret \eqn{even_odd_Inv_Mellin}. First of all, 
we notice that the reality properties of $a^s_j(t)$, $a^u_j(t)$
stated in \eqn{reality-condition} implies that the coefficients 
of powers of $L$ in ${\cal M}^{(+)}$ and ${\cal M}^{(-)}$ are imaginary and 
real, respectively. Note, however, that it is important for 
these reality properties to express results in terms of $L$ 
defined in \eqn{L-def}, which has an extra imaginary part, 
rather than in terms of the large logarithm $\log|s/t|$ itself. 
This simple observation will remove many explicit $i\pi$'s 
from expressions in this paper, and facilitate non-trivial 
checks of the results. Moreover, for gluon scattering, 
invoking Bose symmetry we deduce that ${\cal M}^{(+)}$, 
which is symmetric under permutation of the kinematic 
variables $s$ and $u$, picks out the colour component 
which are symmetric under permutation of the indices 
of particles 2 and 3, and~${\cal M}^{(-)}$, which is 
antisymmetric upon swapping $s$ and $u$, picks 
out the colour-antisymmetric part.

In this paper we focus on the leading power in $t/s$, 
and in this limit the Mellin variable $j$ used above is 
identical to the spin $j$ which enters conventional partial 
wave functions\footnote{
The standard partial wave decomposition of a 
$2\to 2$ scattering amplitude is given by (see e.g. 
\cite{Collins:1977jy})
\begin{subequations}
\begin{align}
\label{Alt}
{\cal M}_j(t)&=\frac{1}{16\pi}\frac12 \int_{-1}^{1} 
dz_t P_j(z_t) {\cal M}(s(z_t,t),t), \qquad j=0,1,2,\ldots \\
{\cal M}(s,t)&=16\pi \sum_{j=0}^{\infty}(2j+1){\cal M}_j(t) P_{j}(z_t),
\end{align}
\end{subequations}
where $P_j(z_t)$ are Lagendre polynomials obeying 
$P_j(-z)=(-1)^jP_j(z)$, and $z_t=\cos(\theta_t)$ where 
$\theta_t$ is the $t$-channel scattering angle (namely, 
using the conventions of Fig.~\ref{setup_fig}, it is the 
angle between the $p_1$ and $p_2$ in the centre-of-mass 
frame of $p_1$ and $p_4$). For massless scattering 
considered here, where $s+t+u=0$, 
\begin{equation}
z_t=1+\frac{2s}{t}=-1-\frac{2u}{t}\,.
\end{equation}
The symmetry $z_t\to -z_t$ relates scattering with
angle $\theta_t$ to scattering with angle $\pi-\theta_t$; 
in terms of the Mandelstam invariants, it corresponds to 
$s\leftrightarrow u$. We see that under an 
$s\leftrightarrow u$ interchange ${\cal M}_j(t)$ of 
\eqn{Alt} is even for even $j$ and odd for odd $j$. }.
This explains our notation.  One could easily extend the 
above discussion to subleading powers, but one would 
have to replace the Mellin transform by the partial wave 
expansion. For example, $(s/t)^{-j-1}$ and $(s/t)^j$ in 
\eqns{Mellin_trans}{Ast_mellin} would 
be replaced respectively by the associated Legendre 
function $Q_j(1+2s/t)$ and Legendre polynomials 
$P_j(1+2s/t)$, see \cite{Collins:1977jy}.

The simplest conceivable asymptotic behaviour would 
be a pure power law, whose Mellin transform is a simple 
Regge pole, namely
\be\label{ReggePole}
a_j^{(-)}(t) \simeq \frac{1}{j-1-\alpha(t)}.
\ee
The leading perturbative behaviour is obtained upon taking
the residue of \eqn{odd_Inv_Mellin} about the Regge pole,
getting
\be\label{ReggePoleAmp}
{\cal M}^{(-)}(s,t)|_{\rm Regge \,pole} \simeq 
\frac{\pi }{\sin \frac{\pi\, \alpha(t)}{2}}  \frac{s}{t}  
\, e^{L\, \alpha(t)} + \ldots,
\ee
where the ellipsis indicated subleading contributions. 
Regge poles give the correct behaviour of the $2\to 2$ 
amplitude at leading logarithm accuracy in perturbation theory, 
where $\alpha(t)$ is interpreted as the gluon Regge trajectory, 
$\alpha(t) \equiv \alpha_g(t) \sim {\cal O}(\alpha_s(t))$. In order
to get the precise behavior at higher orders in perturbation theory
one needs to take into account the contribution of Regge cuts, 
which arises from $a_j^{(-)}(t)$ of the form 
\be\label{ReggeCut}
a_j^{(-)}(t) \simeq \frac{1}{(j-1-\alpha(t))^{1+\beta(t)}},
\ee
which has a branch point from $1+\alpha(t)$ to $-\infty$, 
or a multiple pole if $\beta(t)$ is a positive integer.  
Integrating along the discontinuity one gets 
\be\label{ReggeCutAmp}
{\cal M}^{(-)}(s,t)|_{\rm Regge\, cut} \simeq 
\frac{\pi }{\sin \frac{\pi\, \alpha(t)}{2}}  \frac{s}{t} 
\, \frac{1}{\Gamma\left(1+\beta(t)\right)} L^{\beta(t)}  
\, e^{L\, \alpha(t)} + \mbox{subleading logs}.
\ee
While Regge poles contribute to LL accuracy,
therefore to the odd amplitude, Regge cuts start 
contributing at the NLL order, to the even amplitude. 
A complete treatment of scattering amplitudes up to 
NNLL accuracy requires to take into account the 
contribution of Regge cuts both to the odd and the 
even amplitude. In order to clarify this structure, we 
are now going to explore the implications of Regge 
poles and cuts in perturbation theory.

\subsection{The Regge limit in perturbation theory}
\label{Regge_limit_perturbative}

\begin{figure}[t]
\begin{center}
  \includegraphics[width=0.38\textwidth]{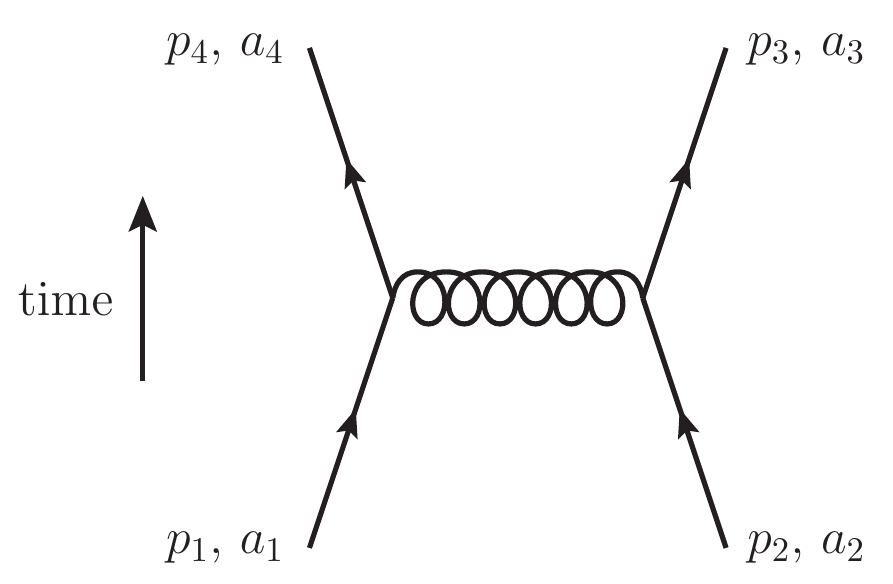}
  \end{center}
  \caption{Tree-level t-channel exchange contributing in the 
  high-energy limit to quark-quark, quark-gluon or gluon-gluon 
  scattering. The solid external lines represent either quarks or 
  gluons, depending on the process considered.}
\label{tree-t}
\end{figure}

We write the perturbative expansion of a $2 \to 2$ scattering 
amplitude in the high-energy limit as
\be\label{Mpower_expansion}
{\cal M}(s,t) = 4\pi\alpha_s\, \sum_{n = 0}^{\infty} \left( \frac{\as}{\pi} \right)^n 
{\cal M}^{(n)}(s,t),
\ee
where we systematically neglect any powers suppressed 
terms in $t/s$. This perturbative expansion correspond to the 
ultraviolet-renormalised scattering amplitude, with the strong 
coupling $\alpha_s$ renormalized for convenience at the 
momentum-transfer scale, $\mu^2=-t$. Infrared divergences 
are regulated in $d = 4-2\eps$ dimensions.
 
In the previous section we have shown that an amplitude 
can always be written as the sum of its signature 
odd and even component, 
\be\label{MOddPlusEven}
{\cal M}(s,t)  = {\cal M}^{(-)}(s,t)  + {\cal M}^{(+)}(s,t), 
\ee
as defined in \eqn{Odd-Even-Amp-Def}. 
Moreover, the reality condition in \eqn{reality-condition}
guarantees that, upon expressing the amplitude in 
terms of the variable $L$ defined in \eqn{L-def}, 
its real and imaginary parts are separately fixed by 
its odd and even components, respectively, see 
\eqn{even_odd_Inv_Mellin}. As a consequence, 
the perturbative expansion of ${\cal M}^{(-)}$ and 
${\cal M}^{(+)}$ is of the form 
\be\label{MOddEvenExpansion}
{\cal M}^{(\pm)}(s,t)  = 4\pi \alpha_s \,\sum_{l,m} \left( \frac{\as}{\pi} \right)^l
\,L^m\, {\cal M}^{(\pm,l,m)}\,,
\ee
where the coefficients ${\cal M}^{(-,l,m)}$
and ${\cal M}^{(+,l,m)}$ are purely real and 
imaginary, respectively. 

\begin{figure}[t]
\begin{center}
  \includegraphics[width=0.80\textwidth]{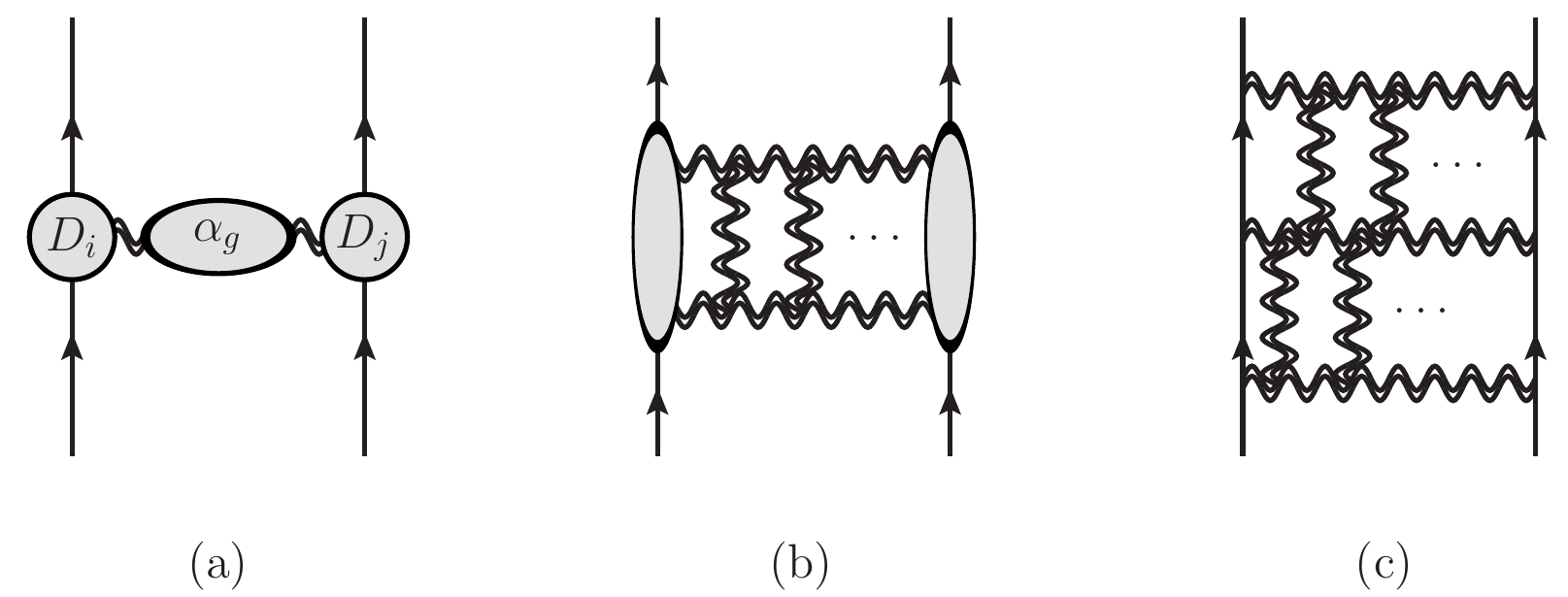}
  \end{center}
  \caption{From left to right, exchange of one, two and 
  three Reggeized gluons, respectively. We draw the 
  Reggeized gluons as double wavy lines, in order to 
  distinguish them from standard gluon exchange in 
  perturbation theory. Single Reggeon exchange 
  in the first diagram contribute at LL accuracy, while 
  two-Reggeon exchange in the second diagram contribute 
  at NLL accuracy. Last, three Reggeons exchange start 
  contributing at NNLL accuracy. The shaded blobs in the first 
  and second diagram account for single- and two-Reggeon
  impact factors, which give additional contributions at 
  subleading logaritmic accuracy to these diagrams.}
\label{123-Reggeons}
\end{figure}

At tree level, in the high-energy limit, the amplitude 
reduces to the $t$-channel exchange represented 
in figure \ref{tree-t}. Moreover, only helicity conserving 
scattering processes are leading in the high energy 
limit. This gives 
\be\label{treeConvention}
{\cal M}_{ij \to ij}^{(0)} =
{\cal M}_{ij \to ij}^{(-,0)} = \frac{2s}{t} 
\,  (T_i^b)_{a_1a_4} (T_j^b)_{a_2a_3} \,
\delta_{\lambda_1 \lambda_4}
\delta_{\lambda_2 \lambda_3},
\qquad \quad  {\cal M}_{ij \to ij}^{(+,0)} = 0, 
\ee
where $T_i$, $T_i$ are color generators 
in the representation of the corresponding
particle: $(T_i^b)_{a_1a_4} = t^b_{a_1 a_4}$
for quarks, $(T_i^b)_{a_1a_4} = - t^b_{a_4 a_1}$
for antiquarks, and $(T_i^b)_{a_1a_4} = i f^{a_1 b a_4}$
for gluons, and the factor $\delta_{\lambda_1 \lambda_4}
\delta_{\lambda_2 \lambda_3}$ represents helicity 
conservation. It is a well-known fact that, at higher orders, 
the leading logaritmic (LL) contribution is due to a Regge
pole term of the type in \eqn{ReggePole}. Such term 
contributes to the odd part of the amplitude, and 
one has 
\beq\label{LL-odd}
\MM_{ij\to ij}|_{\rm LL} = \MM^{(-)}_{ij\to ij}|_{\rm LL} 
= \left(\frac{s}{-t}\right)^{\frac{\as}{\pi}\, 
C_A\, \alpha^{(1)}_g(t)}\,  4\pi\alpha_s\, {\cal M}_{ij \to ij}^{(0)},
\eeq
which is interpreted as the exchange of a 
Reggeized gluon, or ``Reggeon'', as represented 
by the double wavy line in diagram (a) of 
figure~\ref{123-Reggeons}. The function 
$\alpha^{(1)}_g(t)$ in \eqn{LL-odd} represents 
the leading order contribution to the gluon 
Regge trajectory\footnote{Compared to the 
standard definition in literature, we single out 
a factor $C_A$ from the definition of the 
Regge trajectory, see \eqn{LL-odd}, in order
to simplify comparison with the infrared 
factorisation formula in section \ref{dip_comparison}.}
\be\label{GluonRegge}
\alpha_g(t)  \,= \,
\sum_{n = 1}^{\infty} \left( \frac{\as}{\pi} \right)^n 
\alpha^{(n)}_g(t), \qquad 
\alpha^{(1)}_g(t) \,= \,\frac{\rGamma}{2\eps} 
\left(\frac{-t}{\mu^2}\right)^{-\eps} \,\,
\stackrel{\mu^2 \to -t}{=}\,\,\frac{\rGamma}{2\eps},
\ee
where $\rGamma$ is a ubiquitous loop factor
\beq \label{rGamma}
\rGamma= e^{\eps\gamma_{\rm E}} \,
\frac{\Gamma(1-\eps)^2\Gamma(1+\eps)}{\Gamma(1-2\eps)} 
\approx 1-\frac12\zeta_2\, \eps^2-\frac73\zeta_3 \,\eps^3 + \ldots
\eeq

\begin{figure}[t]
\begin{center}
  \includegraphics[width=0.74\textwidth]{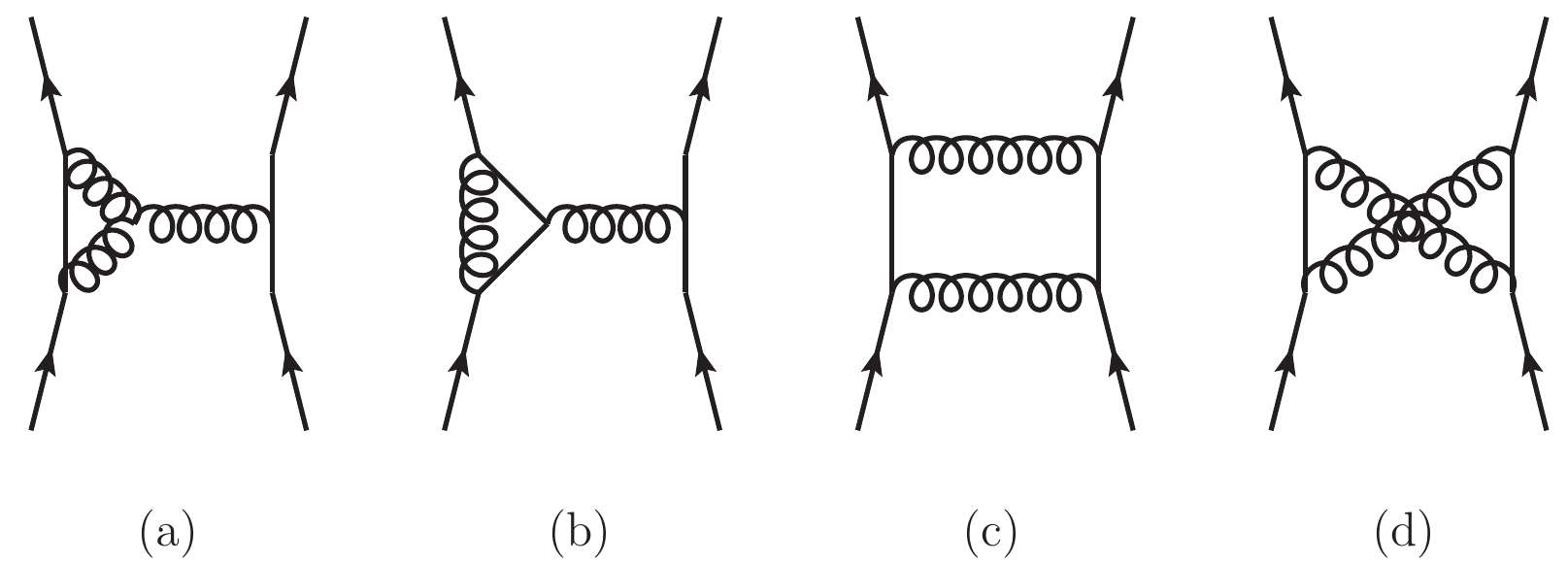}
  \end{center}
  \caption{A few sample of one-loop diagram contributing to
  quark and gluon scattering at next-to-leading order, in the 
  high-energy limit. Diagrams such as (a) and (b) have the 
  same color structure of the tree-level diagram, and contribute 
  to the one-Reggeon impact factor.
  Diagrams such as (c) and (d) introduce color structures 
  different from the color structure of the tree-level amplitude, 
  and contribute to the two-Reggeon exchange.}
\label{1loop-t}
\end{figure}

At next to leading logarithmic (NLL) accuracy 
the single Reggeon exchange described by 
\eqn{LL-odd} receives corrections, which, 
based on our discussion in section \ref{signature},
are expected to be of the form 
\beq\label{Regge-Pole-General}
\MM^{(-)}_{ij\to ij}  
\sim e^{C_A\, \alpha_g(t)\, L} \, 
Z_i(t) D_i(t) \, Z_j(t) D_j(t)\, 
4\pi\alpha_s\,  {\cal M}_{ij \to ij}^{(0)} , 
\eeq
where $\alpha_g(t)$ is the Regge trajectory 
defined in \eqn{GluonRegge}, and the factors 
$Z_{i/j}(t) D_{i/j}(t)$ represent corrections to 
the scattering amplitude independent of the 
centre of mass energy~$s$. These corrections 
contain in general collinear divergences, 
which factorise according to the 
infrared factorisation formula, 
\cite{Gardi:2009qi,Becher:2009qa,Gardi:2009zv},
to be introduced in section \ref{IRfact}, 
see in particular \eqn{Zidef}.
Anticipating our analysis below, 
it proves useful to make the form
of this factorisation manifest, such 
that the factors $Z_{i/j}(t)$ contain 
the collinear singularities, while the 
terms $D_{i/j}(t)$, to which we will refer 
in the following as ``impact factors'', 
represent the finite correction. 
In perturbation theory these objects 
are calculated as an expansion in the 
strong coupling constant, according to
\be\label{impact_factors_def}
Z_i(t) = \sum_{n = 0}^{\infty} 
\left( \frac{\as}{\pi} \right)^n Z^{(n)}_i(t), \qquad
D_i(t) = \sum_{n = 0}^{\infty} 
\left( \frac{\as}{\pi} \right)^n D^{(n)}_i(t).
\ee
A graphical representation of these 
corrections is given in diagram (a) of 
figure~\ref{123-Reggeons}. More in details, 
\Eqn{Regge-Pole-General} involves three types 
of subleading corrections to \eqn{LL-odd}: first of 
all, there is a NLL contribution which arises because 
of the exponentiation pattern, which involve 
$L = \log |s/t| - i \pi/2$ instead of just $\log |s/t|$, 
as a consequence of symmetry with respect 
to the signature, discussed in section 
\ref{signature}. Next, there are contributions 
arising from higher-order corrections to the \emph{gluon
Regge trajectory}, indicated as a shaded blob denoted 
by $\alpha_g$ in diagram (a) of figure~\ref{123-Reggeons}. 
At NLL, such a correction arises from the next-to-leading 
order (NLO) contribution ${\cal O}(\as^2)$ to the Regge 
trajectory, i.e. $\alpha_g^{(2)}(t)$ in \eqn{GluonRegge}. 
As we will discuss below, beyond NLO the Regge 
trajectory corresponding to a single Reggeon 
exchange is not uniquely defined; clarifying 
this issue is one of the goals of this papers. 
For now, it suffices to say that \eqn{Regge-Pole-General}
can be interpreted consistently only up to NLL accuracy. 
The third type of subleading corrections is due to the 
impact factor $D_i(t)$, which can be seen as an
``effective vertex'' associated to the emission 
(or absorption) of a single Reggeon, indicated 
by the shaded blobs in diagram (a) of figure~\ref{123-Reggeons}.
These type of corrections, which depend only on 
the momentum transfer $t$ (and not on the energy $s$), 
arise in perturbation theory for instance from diagrams 
like (a) and (b) in figure~\ref{1loop-t}. 

Starting at NLL accuracy there are new 
corrections, which cannot be interpreted as the exchange 
of a single-Reggeized gluon, and originate instead from 
Regge cuts as in~\eqn{ReggeCutAmp} corresponding to 
the exchange of two or more Reggeized gluons, as 
indicated by diagrams (b) and (c) in figure~\ref{123-Reggeons}. 
This paper focuses on the determination of these corrections. 

Restricting for now to NLL accuracy, 
the Regge cut contribution involves the exchange 
of two Reggeized gluons, and the symmetry 
properties of this state dictate that it contributes to the 
even amplitude, i.e. to $\MM^{(+)}_{ij\to ij}$. 
From the point of view of perturbation theory 
this can be understood by inspecting diagrams (c) 
and (d) in figure~\ref{1loop-t}. These diagrams 
introduce new color structures compared to 
the tree-level color factor $(T_i^b)_{a_1a_4} 
(T_j^b)_{a_2a_3}$ in \eqn{treeConvention}. 
To proceed and characterise these corrections, 
let us briefly review some aspects of color 
decomposition of scattering amplitudes.
 
Scattering amplitudes can be seen as vectors 
in color-flow space, 
\be\label{AmpColor}
{\cal M}(s,t) =\sum_i \, c^{[i]} \, {\cal M}^{[i]}(s,t),
\ee 
where $c^{[i]}$ represent the elements 
of a color basis, and ${\cal M}^{[i]}(s,t)$
are the corresponding amplitude coefficients.
Examples of color bases are the $t$-channel 
exchange orthonormal basis provided in 
appendix \ref{3LoopPredictOrtho}, 
or the ``trace'' basis provided in appendix 
\ref{3LoopPredictTrace}. From the point of 
view of Regge theory it is convenient to 
focus on the former, in which the color operator 
(defined in (\ref{TtTsTu})) in the $t$ channel, 
$\T_t^2$, is diagonal (see in (\ref{Ttgggg})), 
hence providing insight into the factorisation 
structure of the amplitude in the high-energy limit. 

An orthonormal color basis in the $t$-channel
can be obtained by decomposing the direct 
product of the color representations 
associated to the incoming and outgoing particle 
1 and 4 (see Figure~\ref{tree-t}) into a direct 
sum. For instance, in case of gluon-gluon 
scattering the amplitude lives in the space 
of the $8 \otimes 8$ color representation. 
An orthonormal color basis is obtained 
decomposing it into a direct sum, i.e. 
$8 \otimes 8 = 1 \oplus 8_{s} \oplus 8_{a} 
\oplus 10 \oplus \overline{10} \oplus 27 \oplus 0$.
At this point it is useful to make contact 
with the discussion following
\eqn{L-def}: because of Bose symmetry, 
the symmetry of the color structure mirrors the 
signature of the corresponding amplitude coefficients,
which can thus be separated into signature odd and even:
\begin{equation} \label{odd_even_Ms}
\mbox{odd: } {\cal M}^{[8_a]}, {\cal M}^{[10+\overline{10}]},
\qquad
\mbox{even: } 
{\cal M}^{[1]},
{\cal M}^{[8s]}, 
{\cal M}^{[27]},
{\cal M}^{[0]} \qquad \mbox{($gg$ scattering)\,.}
\end{equation}
Here $8_{s}$ and $8_{a}$ represent respectively 
a symmetric and antisymmetric octet representation, 
and $0$ is a ``null'' representation, which is present 
in general for SU(N), and vanishes for N = 3.
A more exhaustive discussion on how to decompose 
the amplitude into orthonormal color basis, together 
with explicit expressions for the orthonormal color 
basis of quark-quark, quark-gluon and gluon-gluon
scattering have been given in 
\cite{Beneke:2009rj,Keppeler:2012ih,DelDuca:2014cya}, 
to which we refer for further details, as well as appendix 
\ref{3LoopPredictOrtho}.

For our discussion, it suffices to note that 
the exchange of one Reggeized gluon contributes only
to the antisymmetric octet, so that at leading-log order
only this structure is nonzero:
\be \label{8a_LL}
{\cal M}_{gg\to gg}(s,t)\big|_{LL} = 
c_{[8_a]} \,\,  {\cal M}^{[8_a]}(s,t)\big|_{LL}
= \frac{2s}{t} \,  (T_i^b)_{a_1a_4} (T_j^b)_{a_2a_3} \,
\delta_{\lambda_1 \lambda_4}
\delta_{\lambda_2 \lambda_3}\,.
\ee
At NLL order, certain diagrams like (a) and (b) in 
figure~\ref{1loop-t} contribute only to the $8_a$ 
color structure also, but others like (c) and (d) 
contribute in addition to the even structures listed 
in eq.~(\ref{odd_even_Ms}). These signature-even 
contributions represent the exchange of a pair of 
Reggeized gluons and do not exponentiate in a 
simple way. Rather they contribute a Regge cut 
which can be calculated order by order in 
perturbation theory within a framework developed 
in \cite{Caron-Huot:2013fea}, based on BFKL theory, 
and reviewed shortly.

This paper will focus on the three-Reggeon exchange 
at the NNLL order, which contributes to both the $8_a$ 
and $10+\overline{10}$ color structures. At NNLL order, 
presently unknown corrections to single-Reggeon exchange
also enter but they only contribute to the $8_a$ color structure.
We will therefore unambiguously predict the $10+\overline{10}$ 
amplitude. Furthermore, the relationship between the $8_a$ 
contributions to gluon-gluon, quark-gluon and quark-quark
amplitudes will be unambiguously predicted.

In order to display the Regge-cut contributions in the most 
transparent way, it proves useful to define a ``reduced'' 
amplitude by removing from it the Reggeized gluon and 
collinear divergences as follows:
\be\label{Mreduced}
\Mreduced_{ij\to ij} \equiv 
\left(Z_i  Z_j \right)^{-1} 
\,e^{-\,\T_t^2\, \alpha_g(t) \, L} \, \MM_{ij\to ij}\,,
\ee
where $\T_t^2$ represents the colour charge of a Reggeized 
gluon exchanged in the $t$ channel (see eq.~(\ref{TtTsTu}) below) 
and $Z_i$ and $Z_j$ stand for collinear divergences, defined in 
(\ref{Zidef}) below. At tree-level one obviously has
$\Mreduced^{(0)} = \MM^{(0)}$, and based on our 
discussion so far the odd component of the reduced 
amplitude up to NLL reads 
\cite{DelDuca:2001gu,Caron-Huot:2013fea}
\be\label{Odd_NLL}
\Mreduced^{(-)}_{ij\to ij}  
= \bigg[1 + \frac{\as}{\pi} \bigg(D^{(1)}_i(t) 
+D^{(1)}_j(t)\bigg) \bigg]\, 4\pi\alpha_s\, \Mreduced_{ij \to ij}^{(0)} \,, 
\ee
where $D^{(1)}_{i/j}(t)$ are the finite single-Reggeon 
impact factors evaluated at one loop. In turn, the even
reduced amplitude at NLL accuracy 
is given by \cite{Caron-Huot:2013fea}
\beq\label{Even_NLL}
\Mreduced^{(+)}_{ij\to ij}|_{\rm NLL} 
= i \pi \sum_{\ell = 1}^{\infty} \frac{1}{\ell!} \left(\frac{\as}{\pi}\right)^\ell L^{\ell-1} 
\times d_{\ell} \, \times 4\pi\alpha_s\, \Mreduced_{ij \to ij}^{(0)},
\eeq
where the coefficients $d_{\ell}$ contain 
non-diagonal color operators. These coefficients 
are infrared divergent and have been calculated 
in \cite{Caron-Huot:2013fea} up to the fourth 
order. One has, for instance,
\begin{align} \label{dells}
d_{1} &= {\mathbb{d}}_1\Tsu, &
{\mathbb{d}}_1&=\rGamma\,\frac{1}{2\eps}, \nn \\ 
d_{2} &= {\mathbb{d}}_2[\Tt, \Tsu],&
{\mathbb{d}}_2&=(\rGamma)^2\left(-\frac{1}{4\eps^2} - \frac{9}{2}\eps \zeta_{3}
-\frac{27}{4}\eps^2 \zeta_4 + {\cal O}(\eps^3)\right),  \\ \nn
d_{3} &= {\mathbb{d}}_3[\Tt,[\Tt, \Tsu]], &
{\mathbb{d}}_3&=(\rGamma)^3\left(\frac{1}{8\eps^3} - \frac{11}{4}\zeta_{3}
-\frac{33}{8}\eps \zeta_4 -\frac{357}{4}\eps^2 \zeta_{5}
+ {\cal O}(\eps^3)\right)\,.
\end{align}
$\Tsu$ in \eqn{dells} represents a color operator acting 
on the tree-level vector of amplitudes in  \eqn{odd_even_Ms}, 
according to the color-space formalism introduced in 
\cite{Catani:1996jh,Catani:1996vz,Catani:1998bh}.
With this notation, a color operator $\T_i$ 
corresponds to the color generator associated with 
the $i$-th parton in the scattering amplitude, 
which acts as an SU($N_c$) matrix on the color 
indices of that parton. More in details, one assigns 
$(\T_i^a)_{\alpha\beta} = t^a_{\alpha\beta}$ for a 
final-state quark or initial-state anti-quark, 
$(\T_i^a)_{\alpha\beta} = -t^a_{\beta\alpha}$ 
for a final-state anti-quark or initial-state quark, 
and $(\T_i^a)_{bc} = -if^{abc}$ for a gluon. 
We also use the notation $\T_i \cdot \T_j \equiv 
\T_i^a \T_j^a$ summed over $a$. Generators 
associated with different particles trivially commute, 
$\T_i \cdot \T_j = \T_j \cdot \T_i$ for $i\neq j$, while 
$\T_i^2 = C_i$ is given in terms of the quadratic 
Casimir operator of the corresponding color 
representation, i.e $C_g = C_A$ for gluons. 
In the high-energy limit the color factors can be 
simplified considerably, by using the basis of 
Casimirs corresponding to color flow through 
the three channels~\cite{Dokshitzer:2005ig,DelDuca:2011ae}:
\bea\label{TtTsTu}\nn
\T_s &=& \T_1+\T_2=-\T_3-\T_4 \\ \nn
\T_u &=& \T_1+\T_3=-\T_2-\T_4 \\
\T_t  &=& \T_1+\T_4=-\T_2-\T_3
\eea
and using the color conservation identity 
$\left(\T_1+\T_2+\T_3+\T_4\right) \MM=0$ 
to rewrite in terms of signature eigenstates.
One obtains $\T_s^2+\T_u^2+\T_t^2=\sum_{i=1}^4 C_i
\equiv C_{\rm tot}$. One may then define a color 
operator that is \emph{odd} under $s\leftrightarrow u$ 
crossing:
\beq
\Tsu\equiv \tfrac12 \left(\T_s^2-\T_u^2\right),
\eeq
which is the operator used to 
describe the NLL even amplitude in 
\eqn{dells}. Useful relations are given by 
\beqa \label{T_simplifications} \nn 
 \T_1\cdot \T_2+\T_3\cdot \T_4 &=& \T_s^2-\tfrac12 C_{\rm tot}  = \Tsu-\tfrac12 \Tt, \\  
 \T_1\cdot \T_3+\T_2\cdot \T_4 &=&  -\Tsu-\tfrac12 \Tt, \\ \nn
 \T_1\cdot \T_4 + \T_2\cdot \T_3 &=& \Tt-\tfrac12 C_{\rm tot}. 
\eeqa

The goal of this paper is to provide for the first 
time a systematic derivation of the contributions 
arising at the NNLL accuracy. Based on our
discussion so far, we can anticipate that one has 
to consider the following contributions: on the one 
hand, there will be a contribution to the even 
amplitude, in the form of corrections to the 
two-Reggeon exchange. These corrections 
are expected to be of similar origin as the 
ones arising for the single-Reggeon exchange 
at NLL. Namely, there will be a next-to-leading
order correction to the exchange of two Reggeons; 
there will be a correction accounted for by the $i\pi/2$ 
factor included in the expansion parameter $L$; 
and there will be a correction in the form of 
impact factors for the two-Reggeon exchange, 
as indicated by the shaded blobs in 
the diagram at the centre of figure 
\ref{123-Reggeons}.

More interesting, however, are the 
corrections concerning the odd amplitude at 
NNLL accuracy, which, for this reason, are the 
focus of this paper. In this case one has to 
take into account for the first time the exchange 
of three Reggeized gluons, as indicated by the 
right diagram in figure \ref{123-Reggeons}.
This implies that, starting at NNLL, one has 
mixing between one- and three-Reggeons 
exchange. Schematically, this can be encoded 
by writing the full amplitude as  
\beq\label{NNLL_general}
\Mreduced_{ij\to ij}|_{\rm NNLL} = 
\Mreduced^{(-)}_{ij\to ij}|_{\mbox{\scriptsize 1-Reggeon + 3-Reggeon}} 
+ \Mreduced^{(+)}_{ij\to ij}|_{\mbox{\scriptsize 2-Reggeon}}.
\eeq
The mixing between one- and three-Reggeons 
exchange has significant consequences.
First of all, it is at the origin of the breaking of the
simple power law one finds at NLL accuracy in 
\eqn{Odd_NLL}. Such a breaking appears
for the first time at two loops, and has been
singled out for the first time in a perturbative
calculation in \cite{DelDuca:2001gu},
and investigated further from the point of view of 
the infrared factorisation formula in 
\cite{DelDuca:2013ara,DelDuca:2014cya}.
Second, it implies that, starting at three loops, 
there will be a single-logarithmic contribution
originating from the three-Reggeon exchange, 
and from the interference of the one- and 
three-Reggeon exchange as well. As a consequence,
the interpretation of the Regge trajectory at three 
loops, i.e. the coefficient $\alpha_g^{(3)}$, needs to 
be clarified. Understanding these issues requires 
to investigate the structure of the amplitude in the 
context of the BFKL theory, which we are going to
introduce in the next section.

\subsection{BFKL theory abridged}
\label{BFKL_abridged}

The modern approach to high-energy scattering 
can be formulated in terms of Wilson lines:
\be
U(z_\perp) =  \mathcal{P}\exp\left[ig_s\int_{-\infty}^{+\infty}\,
A_+^a(x^+,x^-{=}0,z_\perp)dx^+ T^a\right].
\ee
The Wilson lines follow the paths of color charges 
inside the projectile, and are thus null and labelled 
by transverse coordinates $z_\perp$.
The idea is to approximate, to leading power,
the fast projectile and target by Wilson lines and 
then compute the scattering amplitude between 
Wilson lines. An important feature of this limit is 
that the full transverse structure needs to be 
retained, because the high-energy limit is taken 
with fixed momentum transfer. This has nontrivial implications 
since, due to quantum fluctuations, a projectile 
necessarily contains multiple color charges at 
different transverse positions: the number of Wilson 
lines cannot be held fixed. However, in perturbation 
theory, the unitary matrices $U(z)$ will be close to 
identity and so can be usefully parametrized by a 
field $W$ (from now on we drop the 
$\perp$ subscript):
\be
U(z) = e^{ig_s\,T^aW^a(z)}\,.
\ee
The color-adjoint field $W^a$ sources a BFKL 
Reggeized gluon. A generic projectile, created 
with four-momentum $p_1$ and absorbed with 
$p_4$, can thus be expanded at weak coupling 
as
\bea\label{OPE} \nn
\ket{\psi_i} \equiv \frac{Z_i^{-1}}{2p_1^+} a_i(p_4) a^\dagger_i(p_1)\ket{0} 
&\sim& g_s \, D_{i,1}(t) \, \ket{W} + g_s^2 \,D_{i,2}(t)\, \ket{WW} 
+ g_s^3\, D_{i,3}(t)\, \ket{WWW} + \ldots  \\   
&\equiv& \ket{\psi_{i,1}}+\ket{\psi_{i,2}}+\ket{\psi_{i,3}}+\ldots, 
\eea
where the factor $Z_i^{-1}$ removes 
collinear divergences from the wavefunction 
$\ket{\psi_i}$, and is related to our definition 
of the reduced amplitude in \eqn{Mreduced}.
The factors $D_{i,j}$ depend on the 
transverse coordinates of the $W$ fields, 
suppressed here, but not on the center of mass 
energy. They correspond to the impact factors 
for the exchange of one-, two- and three-Reggeons 
discussed in section~\ref{Regge_limit_perturbative} 
and represented in figure \ref{123-Reggeons}. 
A more precise definition with exact momentum 
dependence will be given in section~\ref{Regge}. 
The energy dependence enters from the fact that 
the Wilson lines have rapidity divergences which 
must be regulated, which leads to a rapidity 
evolution equation:
\be \label{rapidity_evolution}
-\frac{d}{d\eta}\,\ket{\psi_i} = H\, \ket{\psi_i}. 
\ee
The Hamiltonian, known as the Balitsky-JIMWLK 
equation, is given in the next section. A key feature 
for our perturbative purposes is that it is diagonal 
in the leading approximation:
\bea \label{Hamiltonian_schematic_form}
H  \left( 
\begin{array}{c}
  W      \\  WW     \\  WWW  \\  \cdots
\end{array}
\right) &\equiv&
\left(
\begin{array}{cccc}
 H_{1{\to}1} & 0  & H_{3{\to}1} & \ldots \\
 0 & H_{2{\to}2}  & 0  &  \ldots \\
 H_{1{\to}3} & 0  & H_{3{\to}3} & \ldots\\
 \cdots & \cdots  & \cdots & \cdots\\
\end{array}
\right)\left(
\begin{array}{c}
  W      \\  WW     \\  WWW  \\  \cdots\end{array}
\right)
\nn\\&\sim&
\left(
\begin{array}{cccc}
 g_s^2 & 0  & g_s^4 & \ldots \\
 0 & g_s^2  & 0  &  \ldots \\
 g_s^4 & 0  & g_s^2 & \ldots\\
 \cdots & \cdots  & \cdots & \cdots\\
\end{array}
\right) \left(
\begin{array}{c}
  W      \\  WW     \\  WWW  \\  \cdots\end{array}
\right). 
\eea
Notice, moreover, that only even transition
$n \to n \pm 2$ are allowed: odd transition 
of the type $n \to n \pm 1$ are forbidden 
by the signature symmetry, because they 
would originate transitions between even 
and odd parts of the amplitude. 

After using the rapidity evolution equation 
\eqn{rapidity_evolution} to resum all logarithms 
of the energy, the amplitude is obtained from the 
scattering amplitude between equal-rapidity Wilson 
lines, which depends only on the transverse scale $t$:
\be\label{amp_from_inner_product}
\frac{i(Z_iZ_j)^{-1}}{2s} \MM_{ij\to ij} = 
\bra{\psi_{j}} e^{-H \Log} \ket{\psi_i}.
\ee
The prefactor on the left comes simply from the 
terms like $Z_i^{-1}/(2p_1^+)$ in \eqn{OPE}, which 
we have included in order to remove trivial tree-level factors 
and factorized collinear divergences. In fact, we can 
go further and make contact with the reduced amplitude 
$\Mreduced$ of \eqn{Mreduced}, by removing the
Regge trajectory from the evolution:
\be \label{reduced_amp_from_inner_product}
\frac{i}{2s} \Mreduced_{ij\to ij} = 
\bra{\psi_{j}}e^{-\Hhat\Log}\ket{\psi_i},
\qquad \Hhat \equiv H+\T_t^2 \, \alpha_g(t).
\ee
In these expressions we have identified the 
evolution variable, the rapidity $\eta$, with the 
signature-even logarithm appearing in \eqn{L-def}:
\be
\eta= \Log\equiv \log\left|\frac{s}{t}\right|-i\frac{\pi}{2}.
\ee
The essential requirement is that $\eta$ increases 
by one unit under boost of the projectile by one e-fold 
compared to the target, which $L$ can be verified to 
do due to the $\log s$. The $t$ in the denominator is 
arbitrary and could be replaced by any other 
boost-invariant scale, for example $\mu^2$, since 
different choices represent simply different conventions
for the impact factors $\ket{\psi_{i}}$. Choosing $t$ 
however avoids introducing much artificial infrared 
dependence. The $-i\pi/2$ term is a similarly arbitrary 
choice, but it ensures that the coefficients of powers 
of $L$ have simple reality properties, as discussed 
previously, which greatly minimize the number of $i\pi$'s 
appearing in equations. All these conventions, embodied 
in \eqn{reduced_amp_from_inner_product}, will go a long 
way toward simplifying the higher-loop BFKL calculations.

The inner product in 
\eqn{reduced_amp_from_inner_product} 
is by definition the scattering amplitude of 
Wilson lines renormalized to equal rapidity. 
It must be calculated within the full QCD 
theory and therefore cannot be predicted 
within the effective theory of Wilson lines 
that we are working in. For our purposes 
of this paper, however, it will suffice to know 
that it is Gaussian to leading-order:
\be \label{inner_product_oneW}
G_{11'} \equiv \bracket{W_{1}}{W_{1'}} 
= i\,\frac{\delta^{a_1a_1'}}{p_1^2}\,
\delta^{(2-2\eps)}(p_1-p_1') +\ord(g_s^2).
\ee
Multi-Reggeon correlators are obtained 
by Wick contractions, e.g.
\beqa \label{inner_product}\nn
\bracket{W_{1}W_{2}}{W_{1'}W_{2'}} &=& 
G_{11'}G_{22'}+G_{12'}G_{21'} + \ord(g_s^2), \\ \nn
\bracket{W_{1}W_{2}W_3}{W_{1'}W_{2'}W_{3'}} 
&=& G_{11'}G_{22'}G_{33'}+\mbox{(5 permutations)}
+\ord(g_s^2), \\ 
&&\mbox{etc.}
\eeqa
We believe that the $\ord(g_s^2)$ corrections 
could be extracted, if needed, from the results 
of \cite{Babansky:2002my}. There are also 
off-diagonal elements, which can be \emph{defined} to 
have zero overlap:
\beqa\label{3WtoW-Wto3W}
\bracket{W_1W_2W_3}{W_4}  &=&
\bracket{W_4}{W_1W_2W_3} = 0;
\eeqa    
in other words, we assume the 
Reggeons to be free fields. This is 
an implicit assumption in the classic 
BFKL literature. In the Wilson line 
approach it can be justified by noticing 
that, starting from a scheme in which 
the inner products in \eqn{3WtoW-Wto3W}
is different from zero, it is always possible
to perform a scheme transformations 
(redefinition of the $W$ field, for instance 
$WWW\mapsto WWW-g_s^2G \, W$) such as 
to reduce to \eqn{3WtoW-Wto3W}. It is
possible to derive the transformation $G$ only by
calculating the inner product in \eqn{3WtoW-Wto3W}
in full QCD in a given scheme. While we leave this 
calculation to be investigated in future work, we 
notice that the precise form of $G$ is not needed 
in order to obtain quantitative predictions for NNLL 
amplitudes. Indeed, choosing the 1-$W$ and 3-$W$ 
states to be orthogonal, combined with symmetry of 
the Hamiltonian, which in turn is a consequence of 
boost invariance: 
\be
\frac{d}{d\eta}\langle {\cal O}_1| {\cal O}_2 \rangle = 0
\quad \Leftrightarrow \quad   
\langle H {\cal O}_1|{\cal O}_2 \rangle = 
\langle {\cal O}_1| H{\cal O}_2 \rangle\equiv \langle {\cal O}_1|H|{\cal O}_2\rangle,
\label{symmetry-of-H}
\ee
where ${\cal O}_1$, ${\cal O}_2$ represent an arbitrary 
number of $W$ fields, implies that in this scheme one 
has $H_{1\to 3} = H_{3\to 1}$, and more in general 
$H_{k\to k+2} = H_{k+2\to k}$. This relation is 
known as projectile-target duality. As we will see 
in the next section, it is actually essential in order 
to obtain predictions at NNLL accuracy based only 
on the leading order BFKL hamiltonian. As an 
additional comment, we note that in principle one 
could diagonalize the Hamiltonian in 
\eqn{Hamiltonian_schematic_form}, 
given the fact that it is symmetrical with respect to the 
inner product, so there is no invariant meaning to its 
``off-diagonal elements being nonzero''. In practice, 
however, this would require inverting its (complicated) 
diagonal terms, and for this reason we work with the 
undiagonalized Hamiltonian.

We can finally list the ingredients which build up the
amplitude up to three loops. Since the odd and even 
sectors are orthogonal and closed under the action 
of $\Hhat$ (as a consequence of signature 
symmetry), we have
\be \label{Regge-odd-Even-Amplitude}
\frac{i}{2s}\Mreduced_{ij\to ij} \xrightarrow{{\rm Regge}}
\frac{i}{2s}\left(\Mreduced^{(+)}_{ij\to ij} + 
\Mreduced^{(-)}_{ij\to ij} \right) 
\equiv \bra{\psi^{(+)}_j}e^{-\Hhat\Log}\ket{\psi^{(+)}_i}
+\bra{\psi^{(-)}_j}e^{-\Hhat\Log}\ket{\psi^{(-)}_i}.
\ee
Using that multi-Reggeon impact factors are 
coupling-suppressed, $\ket{\psi_{i_k}}\sim g^k$,
and using the suppression \eqn{Hamiltonian_schematic_form} 
of off-diagonal elements in the Hamiltonian, the 
signature even amplitude becomes to three loops:
\begin{subequations}
\label{3loop_even_bfkl}
\begin{align}
\label{3loop_even_bfkl-1}
\frac{i}{2s}\Mreduced_{ij\to ij}^{(+)\, \textrm{1-loop}} &=
\bracket{\psi_{j,2}}{\psi_{i,2}}^{\LO}, \\[0.3cm]  
\label{3loop_even_bfkl-2}
\frac{i}{2s}\Mreduced_{ij\to ij}^{(+)\, \textrm{2-loops}} &=
-\Log \bra{\psi_{j,2}} \Hhat_{2{\to}2}\ket{\psi_{i,2}}^{\LO} +
\bracket{\psi_{j,2}}{\psi_{i,2}}^{\NLO}, \\[0.3cm] \nn
\label{3loop_even_bfkl-3}
\frac{i}{2s}\Mreduced_{ij\to ij}^{(+)\, \textrm{3-loops}} &=
\frac{\Log^2}{2} \bra{\psi_{j,2}} (\Hhat_{2{\to}2})^2\ket{\psi_{i,2}}^{\LO}
-\Log \bra{\psi_{j,2}} \Hhat_{2{\to}2}\ket{\psi_{i,2}}^{\NLO} \\
&+\, \bracket{\psi_{j,4}}{\psi_{i,4}}^{\LO} +
\bracket{\psi_{j,2}}{\psi_{i,2}}^{\NNLO}.
\end{align}
\end{subequations}
Here ``LO'' means that all ingredients are needed 
only to leading nonvanishing order. The first term 
was analyzed in ref.~\cite{Caron-Huot:2013fea} 
and found to be quite powerful: it predicted that 
there should be no $\sim \alpha_s^3\Log^2$ 
corrections to the dipole formula.  At four loops, 
a similar leading-logarithmic computation predicted 
a non-vanishing $\Gamma\sim \alpha_s^4\Log^3$ 
correction to the dipole formula, which hopefully 
will be tested in the future.

In this paper we analyze the similar expansion 
for the signature odd sector:
\begin{subequations}
\label{3loop_odd_bfkl}$H_{1{\to}1} = - C_A \, \alpha_{g}(t)$
\begin{align}
\label{3loop_odd_bfkl-0}
\frac{i}{2s} \Mreduced_{ij\to ij}^{(-)\, \textrm{tree}} &=
\bracket{\psi_{j,1}}{\psi_{i,1}}^{\LO}, \\[0.3cm] 
\label{3loop_odd_bfkl-1}
\frac{i}{2s} \Mreduced_{ij\to ij}^{(-)\, \textrm{1-loop}} &=
-\Log \bra{\psi_{j,1}}\Hhat_{1{\to1}}\ket{\psi_{i,1}}^{\LO}
+\bracket{\psi_{j,1}}{\psi_{i,1}}^{\NLO}, \\[0.3cm] \nn
\label{3loop_odd_bfkl-2}
\frac{i}{2s} \Mreduced_{ij\to ij}^{(-)\, \textrm{2-loops}} &=
+\frac12\Log^2 \bra{\psi_{j,1}}(\Hhat_{1{\to}1})^2\ket{\psi_{i,1}}^{\LO}
-\Log \bra{\psi_{j,1}}\Hhat_{1{\to1}}\ket{\psi_{i,1}}^{\NLO} \\ 
&+\, \bracket{\psi_{j,3}}{\psi_{i,3}}^{\LO} 
+\bracket{\psi_{j,1}}{\psi_{i,1}}^{\NNLO}, \\[0.3cm] \nn
\label{3loop_odd_bfkl-3}
\frac{i}{2s} \Mreduced_{ij\to ij}^{(-)\, \textrm{3-loops}} &=
-\frac16\Log^3 \bra{\psi_{j,1}}(\Hhat_{1{\to}1})^3\ket{\psi_{i,1}}^{\LO}
+\frac12\Log^2 \bra{\psi_{j,1}}(\Hhat_{1{\to}1})^2\ket{\psi_{i,1}}^{\NLO} \\ \nn
& -\,\Log  \Big\{\bra{\psi_{j,1}}\Hhat_{1{\to1}}\ket{\psi_{i,1}}^{\NNLO}
+ \Big[\bra{\psi_{j,3}} \Hhat_{3{\to}3}\ket{\psi_{i,3}} 
+\bra{\psi_{j,3}} \Hhat_{1{\to}3}\ket{\psi_{i,1}} \\ 
&\hspace{1.0cm} 
+\bra{\psi_{j,1}} \Hhat_{3{\to}1}\ket{\psi_{i,3}}
\Big]^{\LO} \Big\} + \bracket{\psi_{j,3}}{\psi_{i,3}}^{\NLO}
+\bracket{\psi_{j,1}}{\psi_{i,1}}^{\rm(N^3LO)},
\end{align}
\end{subequations}
where, for illustrative purposes, we have listed 
all terms that need to be considered by taking into 
account \eqn{3WtoW-Wto3W}, but without any 
specific assumption about the form of $\hat H$. 
Inspecting \eqn{reduced_amp_from_inner_product}, 
we notice now that the $1\to 1$ transition is given, 
according to \eqn{Regge-Pole-General}, by the Regge 
trajectory $H_{1{\to}1} = - C_A \, \alpha_{g}(t)$. As a 
consequence one has $\hat H_{1{\to}1} = 0$, and 
this set to zero all terms of the type 
\be\label{Hhat-LL}
\bra{\psi_{j,1}}(\Hhat_{1{\to}1})^n\ket{\psi_{i,1}}^{(\ldots)} = 0,
\ee
in \eqn{3loop_odd_bfkl}.  Starting from NNLL order, 
the ``gluon Regge trajectory'' is scheme-dependent.
In this paper we \emph{define} it to be $-H_{1\to 1}/\CA$ 
in the scheme defined below \eqn{3WtoW-Wto3W}, so that 
$\Hhat_{1{\to}1}$ identically vanishes. Excluding these terms, 
subleading logarithms in the reduced amplitude arise 
from roughly two mechanisms: corrections to the single-Reggeon
exchange in the form of impact factors, such as for instance the 
term $\bracket{\psi_{j,1}}{\psi_{i,1}}^{\NNLO}$ in \eqn{3loop_odd_bfkl}, 
and exchanges of multiple Reggeized gluons, such as 
terms like $\bracket{\psi_{j,3}}{\psi_{i,3}}^{\LO} $ 
and $\bra{\psi_{j,3}} \Hhat_{1{\to}3}\ket{\psi_{i,1}}^{\LO}$. 

The key observation for us will be that the NLO and NNLO 
effects are strongly constrained by factorization: for example, 
since the elementary Reggeon is color-adjoint, any term in the 
(full) amplitude related to the exchange of a single Reggeon 
vanishes upon projecting the amplitude onto other color 
structures. Due to this, as noted below eq.~(\ref{8a_LL}),
many formally NNLL ($\sim L^1$) 
terms in the three-loop amplitude can be predicted using only 
the LO BFKL theory! In the next section we quantitatively 
work out these predictions.

\section{The Balitsky-JIMWLK equation and the three-loop amplitude}
\label{Regge}

The BFKL prediction \eqn{3loop_odd_bfkl} for 
the three-loop amplitude involves the rapidity evolution 
$H$ and impact factors $\ket{\psi}$. We now describe 
both to the relevant order in perturbation theory.

The evolution equation takes a simple and compact 
form in the planar limit, known as the Balitsky-Kovchegov 
equation~\cite{Balitsky:1995ub,Balitsky:1998kc,Balitsky:1998ya,Kovchegov:1999yj,Kovchegov:1999ua}:
\be\label{BK_equation}
H \, U_{ij} = \frac{\as \CA}{2\pi^2}
\int \frac{d^2z_0\,z_{ij}^2}{z_{i0}^2z_{0j}^2}
\left[ U_{ij}-U_{i0}U_{0j}\right] +\ord(\as^2), 
\ee
where $U_{ij}=\frac{1}{N_c}\TR[U(z_i)U(z_j)^\dagger]$ 
is the trace of a color dipole and $z_{ij}=z_i{-}z_j$ is a 
transverse distance. Physically, this accounts for radiation 
of a gluon at the impact parameter $z_0$ and its effect on 
the perceived color charge density of a projectile.

This form holds for a color singlet projectile, but
a similar equation can also be derived for scattering of 
colored partons. However, since $U_{ij}=1+O(1/N_c^2)$ 
in the planar limit, the equation turns out to linearize and 
its solution for $2\to2$ scattering is essentially trivial: a 
pure Regge pole $\mathcal{M}\propto s^{C_A \, \alpha_g(t)}$ 
to any order in the `t Hooft coupling $g_s^2N_c$. We refer 
to section 3 of~\cite{Caron-Huot:2013fea} for more details.

The effects we focus on in this paper are fundamentally 
non-planar. To describe them we will need the finite $N_c$ 
generalization of \eqn{BK_equation}, known as the 
Balitsky-JIMWLK equation, which involves a sum over 
all possible color attachments of the radiated gluon:
\beqa \label{B-JIMWLK} \nn
H &=& \frac{\as}{2\pi^2}
\int [dz_i][dz_j][dz_0]
K_{ij;0} \Big[T_{i,L}^aT_{j,L}^a+T_{i,R}^aT_{j,R}^a  \\
&&\hspace{4.5cm}-\, U_{\rm ad}^{ab}(z_0)
\left(T_{i,L}^aT_{j,R}^b+T_{j,L}^a T_{i,R}^b \right)\Big] +\ord(\as^2). 
\eeqa
Anticipating infrared divergences, here we have switched 
to dimensional regularization: $[dz]\equiv d^{2{-}2\eps}z$,
where we recall that $z$ parametrizes the transverse 
impact parameter plane. $U_{\rm ad}^{ab}$ is the adjoint 
Wilson line associated with the radiated gluon, and the 
$T_{L/R}$'s are generators for left and right color rotations:
\beq \label{T_LR}
T_{i,L}^a = \left[T^aU(z_i)\right]\frac{\delta}{\delta U(z_i)},\qquad
T_{i,R}^a(z) = \left[U(z_i)T^a\right]\frac{\delta}{\delta U(z_i)}.
\eeq
These act on the projectile and target impact factors $\ket{\psi}$,
which are represented as functionals of Wilson lines $U(z)$.
(In perturbation theory these are just polynomials, so the $i$ 
and $j$ integrals effectively represent discrete sums.)
The $\ord(\as^2)$ correction in \eqn{B-JIMWLK}
has been recently determined by three
groups~\cite{Balitsky:2013fea,Kovner:2013ona,Kovner:2014xia,Kovner:2014lca,Caron-Huot:2015bja,Lublinsky:2016meo}.
In the following, however, we will need only the 
leading-order dimensionally-regulated kernel $K_{ij;0}$,
which turns out to admit a simple, dimension-independent 
expression in momentum space (see ref.~\cite{Caron-Huot:2013fea}):
\be \label{R_kernel}
R(q,p)=\frac{(q+p)^2}{q^2p^2}. 
\ee
The corresponding coordinate space expression is then
\beq\label{Kij_Fourier} 
K_{ij;0} \equiv S_\eps(\mu^2)
\int [\dbar q] [\dbar p] \, 
e^{i q\cdot (z_i {-} z_0)} e^{i p\cdot (z_j {-} z_0)} (-2\pi^2)R(q,p) 
= S_\eps(\mu^2)\frac{\Gamma(1-\eps)^2}{\pi^{-2\eps}} 
\frac{z_{0i}\cdot z_{0j}\phantom{\eps}}{(z^2_{0i}z^2_{0j})^{1-\eps}},
\eeq
where we have defined the integration measure 
$[\dbar q] \equiv \frac{d^{2-2\eps} q}{(2\pi)^{2-2\eps}}$, 
and $S_\eps(\mu^2)=\left(\frac{\mu^2}{4\pi e^{-\gamma_E}}\right)^{\eps}$ 
is the usual $\MS$ loop factor. As $\eps\to 0$ this 
reduces indeed to the well-known four-dimensional formula 
(compare for instance with eq. (2.7) of~\cite{Caron-Huot:2013fea}).
We note that in computing this Fourier transform we have dropped 
contact terms $\delta^{2{-}2\eps}(z_0{-}z_i)$, which vanish in 
\eqn{B-JIMWLK} as a result of the color identities 
$U^{ab}(z_i)T_{i,R}^b=T_{i,L}^a$ and 
$U^{ab}(z_i)T_{i,L}^a =T_{i,R}^b$, see \cite{Caron-Huot:2013fea}.

The corrections to the Balitsky-JIMWLK Hamiltonian 
\eqn{B-JIMWLK} are suppressed by $\as$ in a 
power-counting where the Wilson lines are generic, 
$U\sim 1$. This is more general than the perturbative counting 
of the preceding section, where $1-U\sim g_s W\sim g_s$,
implying that the equation resums infinite towers of 
Reggeon iterations. The relationship will be clarified shortly.
First of all, one expands the Wilson line $U$ in terms of 
the Reggeon field $W$:
\bea\label{UexpansionW} \nn
U = e^{ig_s \,W^a T^a} &=& 1+ ig_s \, W^a \, T^a - \frac{g^2_s}{2} W^a W^b \, T^a T^b
-i\frac{g^3_s}{6}W^a W^b W^c\, T^a T^b T^c \\
&& + \, \frac{g^4_s}{24}W^a W^bW^cW^d\, T^aT^b T^c T^d + {\cal O}(g_s^5 \,W^5).
\eea
Then, to extract the interactions efficiently, we simply use the 
Campbell-Baker-Hausdorf formula to convert the rotations defined by
\eqn{T_LR} to derivatives with respect to $W$:
\bea\label{CBH_formula}
iT_{j,L/R}^{a} &=& \frac{1}{g_s}\frac{\delta}{\delta W_j^a} 
\pm\frac{1}{2} f^{abx} W_j^x \frac{\delta}{\delta W_j^b} 
-\frac{g_s}{12} W_j^x W_j^y (F^xF^y)^{a}{}_{b}
\frac{\delta}{\delta W_j^b} \nn\\ 
&&\hspace{0.0cm}-\,\frac{g_s^3}{720} W_j^x W_j^y W_j^z W_j^t 
(F^xF^yF^zF^t)^{a}{}_b\frac{\delta}{\delta W_j^b} + \ldots,
\eea
where we have introduced the Hermitian color matrix 
$(F^x)^{a}{}_b\equiv if^{axb}$. It is then a straightforward, 
if lengthy, exercise in algebra to expand the Hamiltonian 
\eqn{B-JIMWLK} in powers of $g_s$:
\be
H = H_{k{\to}k} + H_{k{\to}k{+}2} +\ldots
\ee
For the diagonal terms, commuting $\delta/\delta W$'s to 
the right of $W$'s by using 
\beq
\frac{\delta W^b(z')}{\delta W^a(z)}\equiv
\delta^{ab}\delta^{2{-}2\eps}(z{-}z'),
\eeq
one finds \cite{Caron-Huot:2013fea}:
\bea \label{hamiltonian_diag_position} \nn
H_{k{\to}k}&=&
\frac{\as\CA}{2\pi^2} \int [dz_i][dz_0] K_{ii;0}\, (W_i {-} W_0)^a 
\frac{\delta}{\delta W_i^a} \\
&& -\frac{\as}{2\pi^2} \int[dz_i][dz_j][dz_0] K_{ij;0} 
(W_i {-} W_0)^x (W_j {-} W_0)^y (F^xF^y)^{ab} 
\frac{\delta^2}{\delta W_i^a\delta W_j^b}.
\eea
For the first nonlinear corrections, not previously 
written in the literature, we find:
\bea \label{non_linear_coordinate_space} 
H_{k{\to}k{+}2} &=& \frac{\as^2}{3\pi}
\int [dz_i][dz_0]\, K_{ii;0}\, (W_i {-} W_0)^x 
W_0^y (W_i {-} W_0)^z\,\TR\big[F^xF^yF^zF^a\big] 
\frac{\delta}{\delta W_i^a} \\  \nn
&+& \frac{\as^2}{6\pi} \int [dz_i][dz_j][dz_0] \, K_{ij;0}\, 
(F^xF^yF^zF^t)^{ab} \Big[(W_i{-}W_0)^xW_0^yW_0^z(W_j{-}W_0)^t \\ \nn
&&\hspace{0.5cm}-\, W_i^x(W_i{-}W_0)^yW_0^z(W_j{-}W_0)^t
-(W_i{-}W_0)^xW_0^y(W_j{-}W_0)^zW_j^t\Big]
\frac{\delta^2}{\delta W_i^a\delta W_j^b}. 
\eea
We have included the second term for future reference 
only, since in this paper we will only need the $1\to 3$ 
transition, contained in the first line. (We observe, a 
posteriori, that the two terms are not completely 
independent: the first can be obtained from the 
second by moving $\delta/\delta W_j$ 
to the left and letting it act on $W_i$.)

Finally, let us explain the relationship between the 
Balitsky-JIMWLK power counting ($U\sim 1$) and 
the BFKL power-counting ($W\sim 1$), and how it 
justifies our extraction of the multi-Reggeon vertices.
The key is to substitute eqs.~(\ref{UexpansionW}) and 
(\ref{CBH_formula}) into (\ref{B-JIMWLK}), which show 
that an  $m{\to}m{+}k$ transition taken from the $\ell$-loop 
Balitsky-JIMWLK equation is proportional to 
$g_s^{2\ell+k}$. Thus for $k\geq0$, all the leading 
interactions can be extracted from just the leading-order 
equation. On the other hand, because of the symmetry 
of $H$ (\ref{symmetry-of-H}), interactions with $k<0$ are 
suppressed by at least $g_s^{2+|k|}$, which means that 
they can first appear in the $(|k|{+}1)$-loop Balitsky-JIMWLK 
Hamiltonian. Thus to obtain the $m{\to}m{-}2$ transition by 
direct calculation of the Hamiltonian would require a rather 
formidable three-loop non-planar computation. However, this 
is unnecessary, since the symmetry of $H$ predicts the result;
this is carried out explicitly in the following subsection 
(see eq.~(\ref{Ham31})).

\subsection{Evolution in momentum space}
\label{W-momentum-space}

Due to the simple form \eqn{R_kernel} of the kernel 
in momentum space, the perturbative calculation will 
be easier in this space. Let us thus introduce 
the Fourier transform:
\be\label{Wtransform}
W^a(p) = \int [dz] \,e^{-ipz}\,W^a(z), \qquad
W^a(z) = \int [\dbar p] \,e^{ipz}\, W^a(p).
\ee
Substituting into \eqn{hamiltonian_diag_position}, 
and using the Fourier representation of the kernel 
\eqn{Kij_Fourier}, one finds, after a bit of algebra 
again,
\bea \label{ktok_momentum_space} 
H_{k{\to}k} &=& -\int [dp] \, C_A \, \alpha_g(p) \,
W^a(p)\frac{\delta}{\delta W^a(p)} \\  \nn
&&\hspace{-1.2cm}+\, \as\int [\dbar q][dp_1][dp_2]\, 
H_{22}(q;p_1,p_2) \,W^x(p_1{+}q)W^y(p_2{-}q)
(F^xF^y)^{ab}\frac{\delta}{\delta W^a(p_1)}
\frac{\delta}{\delta W^b(p_2)},
\eea
where the gluon Regge trajectory 
and pairwise interactions come out as some specific
combinations of the momentum space kernel $R$ 
of \eqn{R_kernel} (see \cite{Caron-Huot:2013fea} 
for more details). Given that we consider here
only the leading order contribution to the kernel 
$K_{ij;0}$ in \eqn{B-JIMWLK}, the gluon Regge 
trajectory in \eqn{ktok_momentum_space} is 
actually the leading-order trajectory defined in 
\eqn{GluonRegge}, that we recall here for 
the reader's convenience:
\beqa\label{GluonRegge-bis}\nn
\alpha_g(p) &=& \frac{\as}{\pi} \, \alpha^{(1)}_g(p^2) +{\cal O}(\as^2) \\
&=& -\as(\mu) S_\eps(\mu^2)\int[\dbar q] \frac{p^2}{q^2(p-q)^2} +{\cal O}(\as^2) 
= \frac{\as(\mu) \, \rGamma}{2\pi\eps} \left(\frac{\mu^2}{p^2}\right)^{\eps} 
+{\cal O}(\as^2).
\eeqa
The solution to the single-Reggeon 
part of the evolution equation above, 
in which one consider the LO
Regge trajectory, is responsible for the 
leading-logarithmic behaviour of the amplitude. 
Below we will analyse the structure of the scattering
amplitude up to NNLL accuracy, which means that 
we will need also the first two corrections to 
$\alpha_g(p^2)$, namely $\alpha^{(2)}_g(p^2)$
and $\alpha^{(3)}_g(p^2)$. The NLO Regge trajectory 
$\alpha^{(2)}_g(p^2)$ has been calculated in 
\cite{Fadin:1995xg,Fadin:1996tb,Fadin:1995km,Blumlein:1998ib};
it can also be extracted from two-loop calculations
of $2\to 2$ scattering amplitudes, see 
\cite{DelDuca:2001gu}. The NNLO correction
to the Regge Trajectory $\alpha^{(3)}_g(p^2)$ is 
instead not yet known in full QCD, though it will be possible to extract it
below at least in ${\cal N} = 4$ SYM from 
a recent three-loop calculation \cite{Henn:2016jdu}. 
As we will discuss below, it is not even possible to 
define it precisely, beyond the planar limit, without 
taking into account the mixing in the evolution 
between one- and three-Reggeon exchange
given by $H_{1\to 3}$ and $H_{3\to 1}$. The other 
ingredient appearing in \eqn{ktok_momentum_space}
is then the leading-order momentum kernel for the 
evolution of two Reggeon states, 
\cite{Caron-Huot:2013fea}, i.e.
\beq \label{Ham22}
H_{22}(q;p_1,p_2) = \frac{(p_1+p_2)^2}{p_1^2p_2^2}
-\frac{(p_1+q)^2}{p_1^2q^2}-\frac{(p_2 - q)^2}{q^2p_2^2}. 
\eeq
These ingredients are of course precisely as in the 
classic BFKL equation~\cite{Kuraev:1977fs,Balitsky:1978ic}, 
and \eqn{ktok_momentum_space} encapsulates in a concise 
way its generalization to multi-Reggeon 
states~\cite{Bartels:1980pe,Jaroszewicz:1980mq,Kwiecinski:1980wb}.
Here they has been obtained in a systematic and straightforward 
way by linearizing the non-planar version of our starting point,
the Balitsky-Kovchegov equation (\ref{BK_equation}).

\begin{figure}[t]
\begin{center}
  \includegraphics[width=0.70\textwidth]{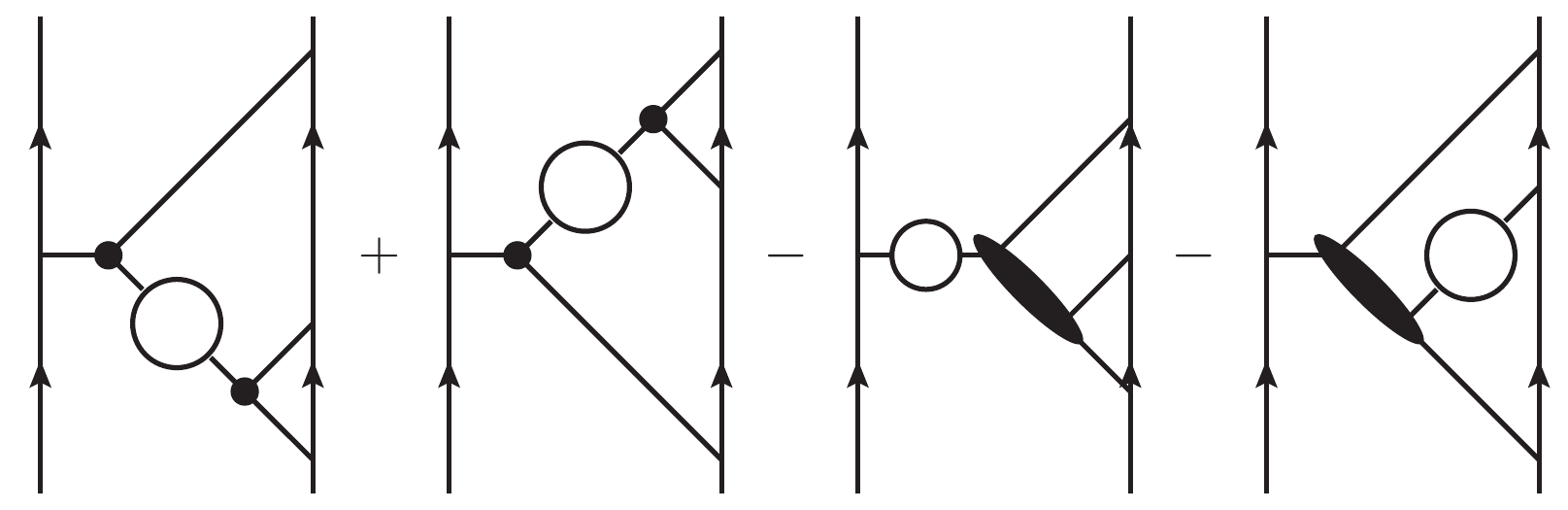}
  \end{center}
  \caption{Diagrams representing the 
  kinematical structure of the  $1\to 3$ and $3\to 1$ evolution, 
  i.e. the factor $H_{13}(p_1,p_2,p_3)$ in \eqn{H13kin}.
  The hamiltonian $H_{13}(p_1,p_2,p_3)$ is derived 
  in the context of an effective field theory in $2-2\eps$ dimensions, 
  therefore the vertices indicated by black dots must be thought as
  effective vertices. The actual color structure associated to the $1\to 3$ 
  and $3\to 1$ evolution is given by the diagrams in figure \ref{1to3-3to1col}.}
\label{1to3-3to1}
\end{figure}

The less familiar ingredient we will need is the $1\to 3$ 
transition, obtained again as the Fourier transform of 
\eqn{non_linear_coordinate_space}:
\be \label{Ham13}
H_{1{\to}3} = \as^2\int [\dbar p_1][\dbar p_2][dp] \,
\TR[F^aF^bF^cF^d] \, W^b(p_1)W^c(p_2)W^d(p_3) 
\, H_{13}(p_1,p_2,p_3)\, \frac{\delta}{\delta W^a(p)},
\ee
where $p_3=p-p_1-p_2$ and the kernel is
\bea \label{H13kin} \nn
H_{13}(p_1,p_2,p_3) &=&
\frac{2\pi}{3}S_\eps(\mu^2)\int [\dbar q]\bigg[
\frac{(p_1{+}p_2)^2}{q^2(p_1{+}p_2{-}q)^2} 
+\frac{(p_2{+}p_3)^2}{q^2(p_2{+}p_3-q)^2} \\
&&\hspace{3.0cm}-\, 
\frac{(p_1{+}p_2{+}p_3)^2}{q^2(p_1{+}p_2{+}p_3{-}q)^2}
-\frac{p_2^2}{q^2 (p_2{-}q)^2} \bigg] \\ \nn
&=& \frac{\rGamma}{3\eps} \left[
\left(\frac{\mu^2}{(p_1{+}p_2{+}p_3)^2}\right)^{\eps}
+\left(\frac{\mu^2}{p_2^2}\right)^{\eps} 
 -\left(\frac{\mu^2}{(p_1{+}p_2)^2}\right)^{\eps}
-\left(\frac{\mu^2}{(p_2{+}p_3)^2}\right)^{\eps} 
\right].
\eea
Taking its transpose with respect to the inner 
product \eqn{inner_product_oneW} then gives 
the conjugate vertex:
\bea\label{Ham31}\nn
H_{3{\to}1} &=& \as^2\int [dp_1][dp_2][dp_3] \,
\TR[F^aF^bF^cF^d]\, W^d(p_1{+}p_2{+}p_3)\, 
\frac{\delta}{\delta W^a(p_1)}\frac{\delta}{\delta W^b(p_2)}
\frac{\delta}{\delta W^c(p_3)} \\ 
&&\hspace{5.0cm}\times (-1)\frac{(p_1{+}p_2{+}p_3)^2}{p_1^2p_2^2p_3^2}
\, H_{13}(p_1,p_2,p_3). 
\eea
This was obtained simply by equating the matrix 
elements 
$$ 
\big(\bra{WWW}H\big)\ket{W}=\bra{WWW}\big(H\ket{W}\big),
$$
taking into account the mismatching propagators, 
$\frac{i}{(p_1{+}p_2{+}p_3)^2}$ compared with 
$\frac{i^3}{p_1^2p_2^2p_3^2}$.

\Eqn{non_linear_coordinate_space} describes
not only the $1\to 3$,  but also $2\to 4$ transitions
in position space. The latter are not necessary for 
the calculation of the odd contribution to the amplitude 
at three loops: $2\to 4$ transitions start contributing 
only at four loops. It is however straightforward to 
derive their representation in momentum space, 
and we list it here for future reference. One has
\bea \nn
H_{2\to 4} &=& \frac{\pi\as^2}{3}S_\eps(\mu^2)
\int [\dbar p_1][\dbar p_2][\dbar p_3][\dbar p_4][dp_a][dp_b]
(2\pi)^{2{-}2\eps}\delta^{2{-}2\eps}(p_1{+}p_2{+}p_3{+}p_4{-}p_a{-}p_b) \\ 
&&\hspace{-1.0cm} \times\, H_{24}(p_{i}) \,(F^xF^yF^zF^t)^{ab} \,
W^x(p_1)W^y(p_2)W^z(p_3)W^t(p_4) \,
\frac{\delta}{\delta W^a(p_a)}\frac{\delta}{\delta W^b(p_b)},
\eea
where:
\bea \nn
H_{24}(p_{i}) &=&
2R(p_a,p_b{-}p_4)+2R(p_a{-}p_1,p_b)-R(p_a,p_b) \\ \nn 
&&-\,3R(p_a{-}p_1,p_b{-}p_4)+R(p_a{-}p_1,p_b{-}p_4{-}p_3)
-R(p_a,p_b{-}p_4{-}p_3) \\
&&+\, R(p_a{-}p_1{-}p_2,p_b{-}p_4)-R(p_a{-}p_1{-}p_2,p_b),
\eea
and we recall that $R(p,q)=\frac{(p+q)^2}{p^2q^2}$
from \eqn{R_kernel}. Similarly, taking its transpose,
\bea\nn
H_{4\to 2} &=& \frac{\pi\as^2}{3}
\int [\dbar p_a][\dbar p_b][d p_1][d p_2][d p_3][d p_4]
(2\pi)^{2{-}2\eps}\delta^{2{-}2\eps}(p_a{+}p_b{-}p_1{-}p_2{-}p_3{-}p_4) \\ \nn
&&\hspace{2.0cm} \times\, (-1) \,\frac{p_a^2\,p_b^2}{p_1^2 \,p_2^2\, p_3^2 \,p_4^2}
\,H_{24}(p_{i}) \,(F^xF^yF^zF^t)^{ab} \,W^a(p_a)W^b(p_b) \\ 
&&\hspace{4.0cm}  \times\, 
\frac{\delta}{\delta W^x(p_1)}\frac{\delta}{\delta W^y(p_2)}
\frac{\delta}{\delta W^z(p_3)}\frac{\delta}{\delta W^t(p_4)}.
\eea

\subsection{Impact factors}

Given the Hamiltonian, all one needs to compute
the amplitude are the target and projectile impact 
factors. At leading order these follow simply from 
the naive eikonal approximation:
\be \label{simple_eikonal}
\ket{\psi_i}^{\LO} = \int [dz]e^{ip{\cdot}z} U_i(z), 
\ee
where the Wilson line is in the representation of 
particle $i$, and $p$ in the transferred momentum, $p^2=-t$.
Expanding in powers of the Reggeon field according
to \eqn{UexpansionW}, and going to momentum 
space, this can also be written to NNLL accuracy as
\beqa \label{LO_wavefunction} \nn
\ket{\psi_i}^{\LO} &=& ig\,\T_i^a W^a(p) 
- \frac{g^2}{2} \T_i^a\T_i^b \int [\dbar q] \, W^a(q)W^b(p{-}q) \\ 
&&\hspace{-1.0cm} -\, \frac{ig^3}{6} \T_i^a\T_i^b\T_i^c 
\int [\dbar q_1] [\dbar q_2] \, W^a(q_1)W^b(q_2)W^c(p{-}q_1{-}q_2) 
+ \ord({\rm N^3LL}),
\eeqa
where we have dropped the coefficient of the unit operator.

At higher orders in the coupling, the color charge 
of the projectile is no longer concentrated in a single 
point, which leads to a nontrivial momentum dependence 
for multi-Reggeon impact factors. Restricting again to NNLL 
accuracy, the relevant corrections at relative order $\as$ reads
\beqa \label{NLO_wavefunction} \nn
\ket{\psi_i}^{\NLO} &=& \frac{\as}{\pi} 
\bigg[ig\, \T_i^a W^a(p)D_i^{(1)}(p)  \\ 
&&\hspace{-1.0cm} -\, \frac{g^2}{2}\T_i^a\T_i^b 
\int[\dbar q] \, \psi^{(1)}_i(p,q) \, W^a(q)W^b(p{-}q) 
+ \ord({\rm N^3LL}) \bigg],
\eeqa
and at the next order one has :
\be \label{NNLO_wavefunction}
\ket{\psi_i}^{\NNLO} = \left(\frac{\as}{\pi}\right)^2 
\left[ig\, \T_i^a W^a(p)D_i^{(2)}(p) + O({\rm N^3LL})\right].
\ee

\subsection{Odd amplitude up to two loops}

According to \eqn{3loop_odd_bfkl}, to get the 
signature-odd amplitude to two loops we need
exchanges of one and three Reggeons, the latter 
first appearing at two loops. Let us consider first
the single Reggeon exchange.

\subsubsection*{$W\to W$ amplitude}

Concerning the reduced amplitude, 
the one-Reggeon exchange is rather simple, 
since the Regge trajectory is subtracted to all 
loop, see \eqn{Hhat-LL}. As a consequence, 
the $1\to 1$ transitions involves only the impact 
factors, and is given by a generalisation of 
\eqn{Odd_NLL} to include NNLL effects. In terms 
of transitions between Wilson lines it is given by 
\be \label{reduced_11_NNLL}
\bra{\psi_{j,1}}e^{-\Hhat_{1{\to}1} \Log}\ket{\psi_{i,1}}
= D_i(t)\,D_j(t)\, \frac{i}{2s} \, 4\pi\alpha_s\, \Mreduced_{ij\to ij}^{(0)}, 
\ee
where $\Mreduced_{ij\to ij}^{(0)} = \MM_{ij\to ij}^{(0)}$ 
has been defined in \eqn{treeConvention}. Effects up 
to NNLL are retained by considering impact factors 
$D_{i/j}$ up to NNLO. At tree level one trivially has
\be\label{M11_LO}
\bracket{\psi_{j,1}}{\psi_{i,1}}^{\LO} = \frac{i}{2s}\, 4\pi\alpha_s\, \Mreduced_{ij\to ij}^{(0)},
\ee
while at one and two loops one 
obtains
\bea \label{M11_NLO} 
\bracket{\psi_{j,1}}{\psi_{i,1}}^{\NLO} &=&
\frac{\as}{\pi}\left(D^{(1)}_i(t) + D^{(1)}_j(t)\right) 
\frac{i}{2s} \, 4\pi\alpha_s\,\Mreduced_{ij\to ij}^{(0)}, \\
\label{M11_NNLO} 
\bracket{\psi_{j,1}}{\psi_{i,1}}^{\NNLO} &=&
\left( \frac{\as}{\pi} \right)^2 \left(D_i^{(2)}+D_j^{(2)}+D_i^{(1)} D_j^{(1)}\right)
\frac{i}{2s} \, 4\pi\alpha_s\, \Mreduced_{ij\to ij}^{(0)}.
\eea

\subsubsection*{$3W\to 3W$ amplitude}

The exchange of three Reggeons contributes 
to the amplitude starting at two-loops, and is 
given according to \eqn{3loop_odd_bfkl-2} by 
a simple Wick contraction of free propagators:
\be\label{2loop_3W3W}
\bracket{\psi_{j,3}}{\psi_{i,3}}^{\LO} =   
-i \pi^2\,(\rGamma)^2\,\II[1] \,\frac{g^2}{t}
\left( \frac{\as}{\pi} \right)^2 \,C_{33}^{(2)} 
\ee
where $C_{33}^{(2)} $ represents the color factor, 
to be discussed below, and we have defined the 
basic two-loop integral
\be\label{two-loop-int}
\II[N] \equiv  \left(\frac{4\pi S_\eps(p^2)}{\rGamma}\right)^2
\int [\dbar p_1][\dbar p_2] \, 
\frac{p^2 }{p_1^2p_2^2(p{-}p_1{-}p_2)^2} \, N\,
\ee
where $N$ should be understood to be a function of the momenta $p_1$, $p_2$ and $p$.
Integrals of the type $\II[N]$ are trivial to 
calculate, because they correspond to bubble 
integrals of the type 
\begin{align}\label{bubbleint}
\begin{split}
&\int \frac{d^{2-2\eps}k}{(2\pi)^{2-2\eps}} \frac{1}{[k^2]^{\alpha}[(p-k)^2]^{\beta}} = 
\frac{B_{\alpha,\beta}(\eps)}
{(4\pi)^{1-\eps} } 
\,(p^2)^{1-\eps-\alpha-\beta}
\\
&
\\
\qquad&\text{with}\qquad \qquad
B_{\alpha,\beta}(\eps)\equiv\frac{\Gamma(1-\alpha-\eps)\Gamma(1-\beta-\eps)\Gamma(\alpha+\beta-1+\eps)}
{\Gamma(\alpha)\Gamma(\beta)\Gamma(2-2\eps-\alpha-\beta)}\,.
\end{split}
\end{align}
In particular, in case of \eqn{2loop_3W3W} we 
need the case $N=1$, for which we get 
\be\label{I1}
 \II[1] = \frac{4}{\epsilon^2}\frac{B_{1,1+\eps}(\eps)}{B_{1,1}(\eps)}=\frac{3}{\eps^2}-18\eps\zeta_3 
 -27\eps^2\zeta_4 +\ldots 
\ee
This is a nice feature of the Regge limit: a 
two-loop amplitude has been reduced to 
essentially a free theory computation in 
the effective Reggeon theory. The more 
difficult aspect is to deal with the color 
factor:
\be
C_{33}^{(2)} = \frac{1}{36}\sum_{\sigma\in \mathcal{S}_3}
\left(\T_i^{\sigma(a)}\T_i^{\sigma(b)}\T_i^{\sigma(c)}\right)_{a_1a_4}
\left(\T_j^{a}\T_j^b\T_j^c\right)_{a_2a_3}.
\ee
Our strategy, keeping in mind our goal to 
compare the infrared divergent part, is to 
express this as some kind of operator acting 
on the tree color factor. Fortunately, there is 
a systematic way to do so: we iteratively 
peel off contracted indices, starting from 
the outermost ones, and re-express them 
in terms of Casimirs, for example
\be
\left[\left( \T_i^a \cdots\right)_{a_1a_4}
\left( \T_j^a \cdots\right)_{a_2a_3} \right] =
\frac12(\T_s^2-C_i-C_j)
\left[\left(\cdots\right)_{a_1a_4}
\left(\cdots\right)_{a_2a_3} \right].
\ee
With the help of the identities used in 
\eqn{T_simplifications}, 
the Casimirs can be further decomposed into 
signature even and odd combinations, which 
gives us the following two useful formulas:
\beqa \label{peeled_contractions} \nn
\left[\left( \T_i^a \cdots\right)_{a_1a_4}
\left(\T_j^a\cdots \right)_{a_2a_3} \right]
&=& \tfrac12\left(\Tsu - \tfrac12 \Tt\right)
\left[\left(\cdots\right)_{a_1a_4}
\left(\cdots\right)_{a_2a_3} \right], \\
\left[\left( \T_i^a \cdots\right)_{a_1a_4}
\left(\cdots \T_j^a\right)_{a_2a_3} \right]
&=& \tfrac12\left(\Tsu + \tfrac12 \Tt\right)
\left[\left(\cdots\right)_{a_1a_4}
\left(\cdots\right)_{a_2a_3} \right].
\eeqa
By repeatedly applying these formulas 
it is now a simple exercise to obtain that
\be\label{C33_value}
C_{33}^{(2)} = \frac1{24}
\left[(\Tsu)^2-\frac{1}{12}(\CA)^2\right] 
(T_i^b)_{a_1a_4} (T_j^b)_{a_2a_3} , 
\ee
and substituting into (\ref{2loop_3W3W}) 
gives the two-loop amplitude:
\be\label{M33}
\bracket{\psi_{j,3}}{\psi_{i,3}}^{\LO}  = -\frac{\pi^2}{24} \,
\left( \frac{\as}{\pi} \right)^2  (\rGamma)^2\, \II[1]\, 
\left[(\Tsu)^2- \tfrac1{12}(\CA)^2\right] 
\frac{i}{2s}\, 4\pi\alpha_s\, \Mreduced_{ij\to ij}^{(0)}.
\ee

\subsubsection*{Total to two loops}

Adding the results of 
eqs.~(\ref{M11_LO}),~(\ref{M11_NLO}),~(\ref{M11_NNLO})
and~(\ref{M33}) as indicated in \eqn{3loop_odd_bfkl} 
we get the total contribution to the odd amplitude 
at one and two loops. Explicitly, expanding the reduced
amplitude in powers of $\as/\pi$ as defined for the 
complete amplitude in \eqn{Mpower_expansion}, 
we have 
\begin{subequations}
\label{one_two_loops_regge} 
\begin{align}
\label{one_two_loops_regge-1} 
\Mreduced_{ij\to ij}^{(-,1)} &=
\left(D_i^{(1)}+D_j^{(1)}\right) 
 \Mreduced_{ij\to ij}^{(0)}, \\ 
\label{one_two_loops_regge-2} 
\Mreduced_{ij\to ij}^{(-,2)} &= \bigg[
D_i^{(2)}+D_j^{(2)}+D_i^{(1)} D_j^{(1)} 
+ \pi^2 R^{(2)} \Big( (\Tsu)^2 - 
\tfrac{1}{12}(\CA)^2\Big) \bigg]  \Mreduced_{ij\to ij}^{(0)},
\end{align}
\end{subequations}
where we have introduced the function
\be\label{R2def}
R^{(2)} \equiv-\frac{1}{24} \,(\rGamma)^2 \,\II[1]\, =\, 
-\frac{(\rGamma)^2}{6 \epsilon^2} \,\frac{B_{1,1+\eps}(\eps)}{B_{1,1}(\eps)}=
(\rGamma)^2 \left(-\frac{1}{8\eps^2}
+\frac{3}{4}\eps\zeta_3 +\frac{9}{8}\eps^2\zeta_4 +\ldots\right)\,,
\ee
where $B_{\alpha,\beta}(\eps)$ is given in eq.~(\ref{bubbleint}).
Here we have factored out $\pi^2$ to emphasize that this 
term originates as a Regge cut proportional to $(i\pi)^2$.
This formula, in particular the fact that $R^{(2)}$ 
multiplies the nontrivial color factor $(\Tsu)^2$, is 
responsible for the breakdown of Regge pole 
factorization as will be discussed in section~\ref{dip_comparison}.
The fact that with two unknown impact factors 
$D_g^{(2)}$, $D_q^{(2)}$, this formula can describe 
the three processes of gluon-gluon, gluon-quark 
and quark-quark scattering is highly nontrivial.

\subsection{Odd amplitude at three loops}
\label{ReggeOddTreeLoops}

The calculation of the three-loop amplitude
through NNLL requires the evaluation of the 
triple, double and single $L$ coefficients in 
\eqn{3loop_odd_bfkl-3}. 

\subsubsection*{$W\to W$ amplitude}

Once again, given \eqn{Hhat-LL}, 
the contribution of the $1\to 1$ transition
to the reduced amplitude is given by the
higher-orders corrections to the impact factors, 
according to \eqn{reduced_11_NNLL}. This 
equation does not involve evolution, and therefore
at three loops it contributes only at N$^3$LO:
\be
\bracket{\psi_{j,1}}{\psi_{i,1}}^{\rm(N^3LO)} = 
\left( \frac{\as}{\pi} \right)^3 
\left(D_i^{(3)} + D_j^{(3)} + D_i^{(2)} D_j^{(1)} + D_i^{(1)} D_j^{(2)} \right) 
\frac{i}{2s} \, 4\pi\alpha_s\,\Mreduced_{ij\to ij}^{(0)}.
\ee
This is beyond the logarithmic 
accuracy which is the target of this paper, 
and therefore we will not consider this 
contribution further.

\subsubsection*{$3W\to 3W$ amplitude}

\begin{figure}[t]
\begin{center}
  \includegraphics[width=0.20\textwidth]{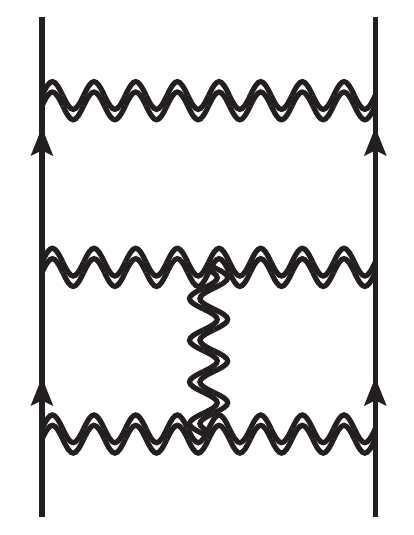}
  \end{center}
  \caption{Example of a diagram involved in the calculation 
  of the three-Reggeon cut at three loops. This diagram, together 
  with all the other diagrams obtained by inserting a rung 
  in all possible ways between the three Reggeons, and considering 
  all possible permutation of the three Reggeons themselves, arises 
  from the insertion of a single factor of $\hat{H}_{3\to 3}$, as discussed 
  below \eqn{H33}.}
\label{3-Reggeons-1-Rung}
\end{figure}

We start by considering the single logarithmic 
term originating by applying the diagonal term 
$H_{k\to k}$ given in (\ref{ktok_momentum_space}) 
to the wavefunction $| \psi_{i,3} \rangle$. A major 
simplification is that only the leading order wavefunction 
\eqn{LO_wavefunction} is required, whose momentum 
and color dependence are separately 
permutation invariant. This allows the sum 
over pairwise color factors in the Hamiltonian 
(\ref{ktok_momentum_space}) to be simplified 
in terms of the total Casimir in the $t$-channel 
(a typical graph is shown in fig.~\ref{3-Reggeons-1-Rung}). 
After a computation we find
\beqa\label{H33} \nn
\hat{H}_{3\to 3}\, 
W^a(p_1)W^b(p_2)W^c(p_3)\Big|_{\mathcal{S}^3} && \\ \nn
&&\hspace{-6.0cm} \simeq \, \frac{\as\rGamma}{2\pi\eps}
\left[\Tt-3\CA\left(\frac{p^2}{p_1^2}\right)^{\eps}\right]
W^a(p_1)W^b(p_2)W^c(p_3) \\ 
&&\hspace{-5.0cm}-\as\left(\Tt-3\CA\right)S_\eps\int [\dbar q] 
H_{22}(q;p_1,p_2)\, W^a(p_1{+}q)W^b(p_2{-}q)W^c(p_3),
\eeqa
where $H_{22}$ is the BFKL kernel in eq.~(\ref{Ham22}).
We emphasize that the simplification of 
the Hamiltonian is only valid for permutation 
invariant momentum dependence.
Contracting the $W$'s against the target
then gives the color factor derived in \eqn{C33_value},
times three propagators, which produce simple 
two-dimensional integral:
\beqa \label{result_33}\nn
\bra{\psi_{j,3}} \Hhat_{3{\to}3}\ket{\psi_{i,3}} &=&
\frac{\pi^2}{48}\,\left( \frac{\as}{\pi} \right)^3 (\rGamma)^3
\Big[\Tt\, (2\II_b{-}\II_a{-}\II_c) + 3\CA\, (\II_c-\II_b)\Big]  \\ 
&&\hspace{3.0cm}\cdot \, 
\Big[(\Tsu)^2-\tfrac1{12}(\CA)^2\Big] 
\frac{i}{2s}\, 4\pi\alpha_s\,\Mreduced_{ij\to ij}^{(0)}.
\eeqa
Here, using the elementary bubble integral in 
eq.~(\ref{two-loop-int}), we have expressed all 
integrals in terms of three basic ones:
\begin{subequations}
 \label{integrals} 
\begin{align}
\II_a &\equiv  \II\!\left[\frac1\eps\right]\,=\, 
\frac{4}{\epsilon^3}\frac{B_{1,1+\eps}(\eps)}{B_{1,1}(\eps)}
=
\frac{3}{\eps^3}-18\zeta_3-27\eps \zeta_4+\ldots \\ 
\II_b &\equiv \II\!\left[\frac1\eps\left(\frac{p^2}{p_1^2}\right)^\eps\right]
= \frac{4}{\epsilon^3}\frac{B_{1+\eps,1+\eps}(\eps)}{B_{1,1}(\eps)}
\,=\,\frac{2}{\eps^3}-44\zeta_3-66\eps\zeta_4+\ldots \\
\II_c &\equiv \II\!\left[\frac1\eps\left(\frac{p^2}{(p_1+p_2)^2}\right)^\eps\right]
= \frac{4}{\epsilon^3}\frac{B_{1,1+2\eps}(\eps)}{B_{1,1}(\eps)}
\,=\, \frac{8}{3\eps^3}-\frac{128}{3}\zeta_3-64\eps\zeta_4+\ldots.
\end{align}
\end{subequations}
While the integrals $\II_{a,b,c}$ are readily available in terms  
of $B_{\alpha,\beta}(\eps)$ of~\eqn{bubbleint} to all orders in~$\eps$, 
here we chose to display the first few orders in their expansion, 
which will be used below.

\subsubsection*{$3W\to W$ and $W\to 3W$ amplitudes: transition vertices}

\begin{figure}[t]
\begin{center}
  \includegraphics[width=0.45\textwidth]{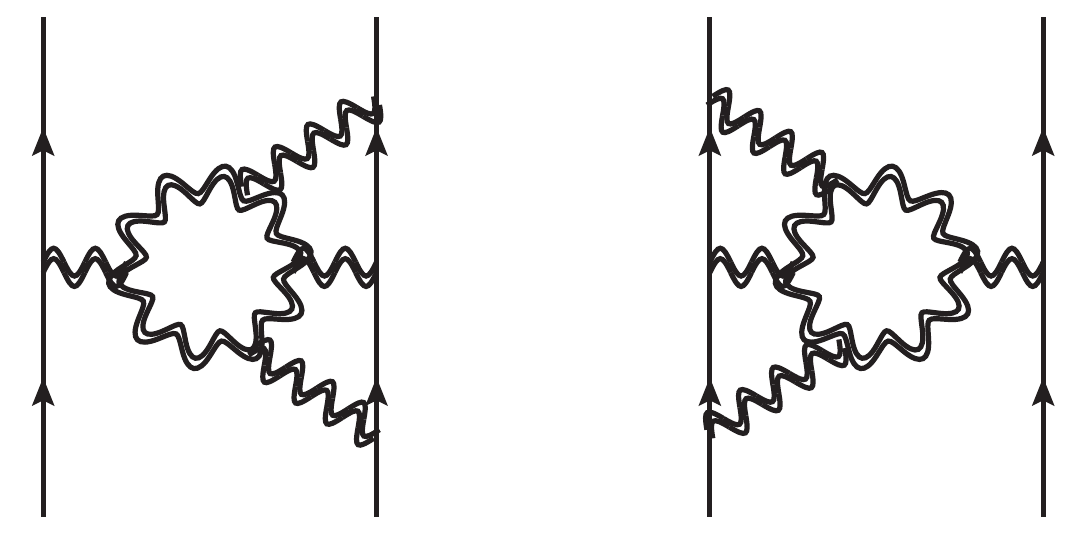}
  \end{center}
  \caption{Diagrams representing the color structure 
  of the $1\to 3$ and $3\to 1$ transitions. Notice that these 
  diagrams are different from the ones representing the 
  kinematical structure of the  $1\to 3$ and $3\to 1$ transitions, 
  i.e. $H_{13}(p_1,p_2,p_3)$ in \eqn{H13kin}. This is a 
  consequence of the fact that the BFKL evolution 
  derived in section \ref{W-momentum-space} represents 
  an effective field theory in $2-2\eps$ dimensions, in which 
  the longitudinal degrees of freedom have been integrated out.}
\label{1to3-3to1col}
\end{figure}

The next contribution comes from the 
off-diagonal $1\to 3$ and $3\to 1$ terms 
in the Hamiltonian, given in \eqns{Ham13}{Ham31}.
These produce the color factor (represented 
by the graphs in fig.~\ref{1to3-3to1col}):
\be
C^{(3)}_{13+31} \equiv \frac16 \sum_{\sigma\in\mathcal{S}^3}
\TR\big[F^aF^{\sigma(b)}F^{\sigma(c)}F^{\sigma(d)}\big]\left[
  (T_i^a)_{a_1a_4}(T_j^bT_j^cT_j^d)_{a_2a_3} 
+(T_i^bT_i^cT_i^d)_{a_1a_4} (T_j^a)_{a_2a_3} 
\right].
\ee
Multiplying with the propagators according 
to our master equation (\ref{3loop_odd_bfkl-3}), 
and collecting the integrals, this contribution to 
the reduced amplitude is again written in terms 
of the same elementary integrals:
\be
\bra{\psi_{j,3}} \Hhat_{1{\to}3}\ket{\psi_{i,1}} +\bra{\psi_{j,1}} \Hhat_{3{\to}1}\ket{\psi_{i,3}} 
=\frac{i}{12} \, \left( \frac{\as}{\pi} \right)^3 \, \pi^2\, (\rGamma)^3 \,
\Big[2\II_c{-}\II_a{-}\II_b\Big] \,\frac{g^2}{t} \, C^{(3)}_{13+31}.
\ee
The main nontrivial task is to simplify the 
color factor. Again we would like to obtain 
a color operator acting on the tree amplitude.
This can be achieved by a simple systematic 
algorithm: move all $f^{abc}$'s onto the external 
states by using the Jacobi identity:
\be
f^{abc}T_i^c = -i [T_i^a,T_i^b].
\ee
In fact this can be done in multiple distinct ways, 
since one can applies this on the $i$ or $j$ leg.
This makes it possible to arrange to get 4 color 
generators to act on each of the $i$ and $j$ legs, 
which then enable to use \eqn{peeled_contractions} 
to read off the result in terms of quadratic Casimirs. 
In fact, we find that for the $1\to 3$ and $3\to 1$ 
transitions separately, the quadratic Casimir 
operators do not provide a sufficient basis 
since the nesting for some terms does not 
allow to extract any generator acting from 
the outside. However, the obstruction is 
odd under interchange of $i$ and $j$, and 
upon adding the two diagrams we do find 
a compact expression:
\bea \nn
C^{(3)}_{13+31} &=& \frac14
\Big(2\Tsu[\Tt,\Tsu]-[\Tt,\Tsu]\Tsu \\ 
&&\hspace{2,0cm}-\, (\Tsu)^2\CA 
-\tfrac{1}{12}(\CA)^{3}\Big) (T_i^b)_{a_1a_4} (T_j^b)_{a_2a_3} ,
\eea
thus leading to 
\bea\label{result_13} \nn
&&\bra{\psi_{j,3}} \Hhat_{1{\to}3}\ket{\psi_{i,1}}+\bra{\psi_{j,1}} \Hhat_{3{\to}1}\ket{\psi_{i,3}}  \\ 
&&\hspace{10mm}=\frac{\pi^2}{48}\,\left( \frac{\as}{\pi} \right)^3\, (\rGamma)^3\,
\left(2\II_c{-}\II_a{-}\II_b\right) \Big(2\Tsu[\Tt,\Tsu] \\ \nn
&&\hspace{20mm}-\, [\Tt,\Tsu]\Tsu-(\Tsu)^2\CA
-\tfrac{1}{12}(\CA)^3\Big)
\frac{i}{2s} \, 4\pi\alpha_s\, \Mreduced_{ij\to ij}^{(0)}.
\eea
Adding the results in \eqns{result_33}{result_13}, 
and expressing the color operators in a common basis, we get:
\begin{align}  \label{result_using_ints}
\begin{split}
&\bra{\psi_{j,3}} \Hhat_{3{\to}3}\ket{\psi_{i,3}} +
\bra{\psi_{j,3}} \Hhat_{1{\to}3}\ket{\psi_{i,1}} +
\bra{\psi_{j,1}} \Hhat_{3{\to}1}\ket{\psi_{i,3}}  \\ 
&\hspace{2.0cm}=\, \frac{\pi^2}{48}\, 
\left( \frac{\as}{\pi} \right)^3 \, (\rGamma)^3 
\bigg[ 3(\II_c{-}\II_a)\, \Tsu[\Tt,\Tsu]+3(\II_b{-}\II_c)\, [\Tt,\Tsu]\Tsu \\ 
&\hspace{6.0cm}-\tfrac1{6}(2\II_c{-}\II_a{-}\II_b)\, (\CA)^3\bigg]
\frac{i}{2s} \, 4\pi\alpha_s\, \Mreduced_{ij\to ij}^{(0)}.
\end{split}
\end{align}

\subsection{Result: the three-loop reduced amplitude to NNLL accuracy}

To summarize, in this section we used BFKL theory
to calculate the signature odd part of the $2\to 2$ 
amplitude to NNLL accuracy. The result at one- and 
two-loop is recorded in \eqn{one_two_loops_regge},
while the three-loop result is obtained by multiplying
the preceding equation with the appropriate minus 
sign and factor from eq.~(\ref{3loop_odd_bfkl}):
\be \label{three_loops_regge}
\Mreduced_{ij\to ij}^{(-,3,1)} =
\pi^2\Big( R_A^{(3)}\, \Tsu[\Tt,\Tsu] +\, R_B^{(3)}\, [\Tt,\Tsu]\Tsu
+R_C^{(3)}\, (\CA)^3 \Big) \Mreduced_{ij\to ij}^{(0)}\,,
\ee
where we have introduced the functions
\bea\nn\label{R3def}
R_A^{(3)} &=& \frac{1}{16}\, (\rGamma)^3 (\II_a{-}\II_c)
\,=\, (\rGamma)^3 \left(\frac{1}{48 \eps^3} + \frac{37}{24}\zeta_3 +\ldots\right), \\ \nn
R_B^{(3)} &=& \frac{1}{16}\, (\rGamma)^3 (\II_c{-}\II_b)
\,=\, (\rGamma)^3 \left(\frac{1}{24 \eps^3} +\frac{1}{12}\zeta_3  +\ldots\right), \\
R_C^{(3)} &=& \frac{1}{288}\, (\rGamma)^3 (2\II_c{-}\II_a{-}\II_b) 
\,=\, (\rGamma)^3 \left(\frac{1}{864 \eps^3} - \frac{35}{432}\zeta_3 +\ldots\right).
\eea
This equation is the main result of this section.
The integrals $\II_{a,b,c}$ are defined in \eqn{integrals} 
where they are evaluated, using the bubble integral 
(\ref{bubbleint}), to all orders in $\epsilon$ in terms 
of~$\Gamma$ functions. Here we will be interested 
in particular in their $\epsilon\to 0$ limit, hence we 
quote their expansion through finite terms.

We note that all the integrals entering 
$\Mreduced_{ij\to ij}^{(-,3,1)}$ in eq.~(\ref{three_loops_regge}) 
are of uniform polylogarithmic weight 3 (as usual in this context, 
$\epsilon$ is assigned weight $-1$). Given that 
$\Mreduced_{ij\to ij}^{(-,3,1)}$ is itself the coefficient 
of a single (high-energy) logarithm, and taking into 
account the overall factor of $\pi^2$ in eq.~(\ref{three_loops_regge}), 
we see that the weight adds up to $6$, which is 
the maximal weight at three loops. Such a uniform 
maximal weight structure is expected in ${\cal N}=4$ 
SYM theory, while in general not in QCD. However, 
as we have seen, $\Mreduced_{ij\to ij}^{(-,3,1)}$ is 
fully determined by gluon interactions, and therefore 
entirely independent of the matter contents of the theory. 
Thus, it is indeed expected that the result, which is valid 
for any gauge theory, should retain the uniform maximal 
weight nature characteristic of ${\cal N}=4$ SYM. 

We further emphasise that these results are valid for 
arbitrary projectiles (quarks or gluons) in arbitrary 
representation of the gauge group; only the impact 
factors $D_i^{(1)}$ and $D_i^{(2)}$ in eq.~(\ref{one_two_loops_regge}) 
depend upon this choice. In the next section we 
discuss our predictions for the amplitude itself, 
and discuss its nontrivial consistency with 
infrared exponentiation theorems.

Finally note that the gluon Regge trajectory does 
not enter the above formulae, because it is subtracted 
in the definition of the reduced amplitude, 
eq.~(\ref{Mreduced}). This definition is also the 
reason why terms with more logarithms are 
absent: $\Mreduced_{ij\to ij}^{(-,1,1)}
=\Mreduced_{ij\to ij}^{(-,2,2)}=\Mreduced_{ij\to ij}^{(-,3,3)}=0$ 
and well as $\Mreduced_{ij\to ij}^{(-,2,1)}=\Mreduced_{ij\to ij}^{(-,3,2)}=0$.
The logarithm-free term at three loops, $\Mreduced_{ij\to ij}^{(-,3,0)}$, 
is beyond our current NNLL accuracy. The presently known results 
from BFKL theory in the even sector, which hold to NLL accuracy,
have been reviewed in eq.~(\ref{Even_NLL}).

\section{Comparison between Regge and 
infrared factorisation}
\label{dip_comparison}

As mentioned in the introduction, the structure of 
infrared divergences in massless scattering 
amplitudes is known in full to three-loop 
order~\cite{Almelid:2015jia}.
The prediction for the reduced 
amplitude presented in the previous 
section is based solely on evolution 
equations of the Regge limit, and has 
taken no input from the theory of 
infrared divergences. It is therefore 
a highly nontrivial consistency test that 
this prediction is consistent with the known
exponentiation pattern and the anomalous 
dimensions governing infrared divergences. 
Conversely, the prediction of the previous 
section can also be seen as a constraint on 
the soft anomalous dimension: the high-energy limit 
of the latter has a very special structure, which may 
ultimately help in determining it beyond three loops. 

The possibility of performing a 
systematic comparison between 
results obtained in the context of 
Regge theory and the infrared 
factorisation theorem has been 
considered in the past~\cite{Bret:2011xm,DelDuca:2011ae,Caron-Huot:2013fea,DelDuca:2013ara,DelDuca:2014cya}.
Given our calculation of the reduced 
amplitude up to NNLL within the Regge 
theory, we are now able to extend this 
analysis systematically to this logarithmic
accuracy. In the following section we 
exploit this possibility by performing a 
comparison up to three loops: 
this will allow us to check consistency
with the structure of infrared divergences 
in the first place; moreover, we will be 
able to use our result obtained in the
context of Regge theory to extract the
infrared renormalised amplitudes, i.e. 
the so-called hard functions, up to three
loops. 

We start this discussion by reviewing 
the structure of infrared divergences 
in the high-energy limit. In particular, 
the expansion of the quadrupole 
correction at three loops in this limit 
has not been presented elsewhere.

\subsection{Infrared renormalization and the soft anomalous dimension}
\label{IRfact}

The infrared divergences of scattering amplitudes 
are controlled by a renormalization group equation, 
whose integrated version takes the form
\beq \label{IRfacteq}
\MM_n \left(\{p_i\},\mu, \as (\mu^2) \right) \, = \, 
{\bf Z}_n \left(\{p_i\},\mu, \as (\mu^2) \right)
\Hhard_n \left(\{p_i\},\mu, \as (\mu^2) \right),
\eeq
where $\MM_n $ represents now an $n$-point 
scattering amplitude, and ${\bf Z}_n$ is given as a 
path-ordered exponential of the soft-anomalous 
dimension:
\beq \label{RGsol}
{\bf Z}_n \left(\{p_i\},\mu, \as (\mu^2) \right) \, = \,  
{\cal P} \exp \left\{ -\frac{1}{2}\int_0^{\mu^2} \frac{d \lambda^2}{\lambda^2}\,
{\bf \Gamma}_n \left(\{p_i\},\lambda, \as(\lambda^2) \right) \right\}\,,
\eeq
where the dependence on the scale is both explicit
and via the $4-2\eps$ dimensional coupling, which obeys 
the renormalization group equation
\beq\label{betagamma}
\beta(\as,\eps) \equiv \frac{d\as}{d\ln \mu}= 
-2\eps \,\as - \frac{\as^2}{2\pi} \sum_{n = 0}^{\infty} 
b_n \, \left(\frac{\as}{\pi}\right)^n\,,
\eeq 
with $b_0=\frac{11}{3}C_A-\frac{2}{3}\nf$.
The soft anomalous dimension for scattering of massless 
partons ($p_i^2=0$) is an operator in color space given, through three loops, 
by~\cite{Becher:2009cu,Gardi:2009qi,Becher:2009qa,Gardi:2009zv,Almelid:2015jia}
\begin{eqnarray} 
\label{gammaSoft1}
&&{\bf \Gamma}_{n}\left(\{p_i\},\lambda, \as(\lambda^2) \right) \,=\,
{\bf \Gamma}_{n}^{\rm dip.}\left(\{p_i\},\lambda, \as(\lambda^2) \right)
\,+\,{\bf \Delta}_{n}\left(\{\rho_{ijkl}\}\right)
\\ && \nn\\
&\qquad\text{with}&\qquad \nn
{\bf \Gamma}_{n}^{\rm dip.}\left(\{p_i\},\lambda, \as(\lambda^2) \right)=
-\frac{\gamma_{K}  (\as)}{2} \, \, \sum_{i<j} 
\log \left(\frac{-s_{ij}}{\lambda^2}\right) \, \T_i \cdot \T_j \,+\, \sum_i \gamma_i (\as) \,,
\end{eqnarray}
where ${\bf \Gamma}_{n}^{\rm dip.}$ involves only 
pairwise interactions amongst the hard partons, 
and is therefore referred to as the ``dipole formula'' 
\cite{Becher:2009cu,Gardi:2009qi,Becher:2009qa,Gardi:2009zv}, 
while the term ${\bf \Delta}_{n}\left(\{\rho_{ijkl}\}\right)$ 
involves interactions of up to four partons, and is called
the ``quadrupole correction''.
In eq.~(\ref{gammaSoft1}) one defines the kinematic 
variables $-s_{ij} = 2 |p_i \cdot p_j | e^{-i \pi \lambda_{ij}}$ with 
$\lambda_{ij} = 1$ if partons $i$ and $j$ both belong to either 
the initial or the final state and $\lambda_{ij} = 0$ otherwise; 
$\T_i$ represent color change operators~\cite{Catani:1998bh} 
in an arbitrary representation, according to the 
notation introduced in section \ref{Regge_limit_perturbative}. 
The function ${\gamma}_{K}(\alpha_s)$ in \eqn{gammaSoft1} 
is the (lightlike) cusp anomalous dimension
\cite{Korchemsky:1985xj,Korchemsky:1985xu,Korchemsky:1987wg}, 
\emph{normalised} by the quadratic Casimir of the corresponding 
Wilson lines. The universality of ${\gamma}_{K}$ 
(so-called \emph{Casimir scaling}) may be broken 
at four loops and beyond. Corresponding corrections 
may be induced in ${\bf \Gamma}_n$ in \eqn{gammaSoft1}, 
but these will not be discussed here, since we restrict explicit 
computations to three loops. In turn, $\gamma_i (\as)$ 
represent the field anomalous dimension corresponding 
to the parton $i$, which governs hard collinear singularities. 
The coefficients of both ${\gamma}_{K}$ and $\gamma_i$ 
are known through three loops and are summarized in 
Appendix \ref{GammaAndZ}.  

The quadrupole correction ${\bf \Delta}_{n}\left(\{\rho_{ijkl}\}\right)$, 
which appears first at three loops, depends on the cross ratios 
$\rho_{ijkl} = \frac{(-s_{ij})(-s_{kl})}{(-s_{ik})(-s_{jl})}$,
which are invariant under rescaling of any of 
the momenta. The quadrupole correction is 
expanded in powers of $\as/\pi$ as follows:
\be\label{quadrupole1}
{\bf \Delta}_{n}(\{\rho_{ijkl}\}) = \sum_{i = 3}^{\infty}
\left(\frac{\alpha_s}{\pi}\right)^{i}{\bf \Delta}^{(i)}_{n}(\{\rho_{ijkl}\}).
\ee
The leading contribution has been 
computed for the first time only recently~\cite{Almelid:2015jia}, 
and is given by
\bea\label{quadrupole2} \nn
{\bf \Delta}^{(3)}_{n}(\{\rho_{ijkl}\}) &=& \frac{1}{4}f^{abe} f^{cde} 
\sum_{1\leq i<j<k<l \leq n} \Big[ 
{\bf T}_i^{a} {\bf T}_j^{b} {\bf T}_k^{c} {\bf T}_l^{d}
\,{\cal F}(\rho_{ikjl},\rho_{iljk}) \\ \nn
&&+\,
{\bf T}_i^{a} {\bf T}_k^{b} {\bf T}_j^{c} {\bf T}_l^{d}
\,{\cal F}(\rho_{ijkl},\rho_{ilkj}) +\,  
{\bf T}_i^{a} {\bf T}_l^{b} {\bf T}_j^{c} {\bf T}_k^{d}
\,{\cal F}(\rho_{ijlk},\rho_{iklj}) \Big]   \\ 
&&-\, \frac{C}{4} f^{abe} f^{cde} 
\sum_{i=1}^n \sum_{\substack{1\leq j < k \leq n,\\ j,k\neq i}} 
\{{\bf T}_i^{a}, {\bf T}_i^{d} \} {\bf T}_j^{b} {\bf T}_k^{c},
\eea
where ${\cal F}$ is a 
function of two cross-ratios and $C$ is a constant:
\bea\nn 
{\cal F}(\rho_{ikjl},\rho_{ilkj}) &=& F(1-z_{ijkl}) - F(z_{ijkl}), \\ 
C &=& \zeta_5 + 2 \zeta_2 \zeta_3, 
\eea
with $z_{ijkl} \, \bar z_{ijkl} = \rho_{ijkl}$
and $(1-z_{ijkl})(1-\bar z_{ijkl}) = \rho_{ilkj}$. In turn one has 
\be
F(z) = {\cal L}_{10101}(z) +2 \zeta_2 \Big({\cal L}_{001}(z) + {\cal L}_{100}(z) \Big),
\ee
where the functions ${\cal L}_w(z)$ are Brown's 
single-valued harmonic polylogarithms 
\cite{BrownSVHPLs} (see also \cite{Dixon:2012yy}) 
in which $w$ is a word made out of 0's and 1's.
The function $F$ implicitly depends on $\bar z$ as well, 
but it is initially defined in the part of the Euclidean region 
where $\bar z = z^*$, where it is single valued. One may 
then analytically continue the function beyond this region, 
treating $z$ and $\bar{z}$ as independent variables. It 
can then be seen that $F$ develops discontinuities, 
with three branch points for $z$ and $\bar{z}$ equals 
$\{0,1,\infty\}$ corresponding to forward or backward 
scattering.

Focusing now on the case of 2 to 2 scattering 
amplitudes, we restrict the index $n$ in \eqn{IRfacteq} 
above to $n = 4$, and drop the index $n$ from now on. 
The dipole contributions to the anomalous dimension 
for $2\to 2$ scattering with timelike $s=s_{12}>0$ and 
spacelike $t=s_{14}<0$ and $u=s_{13}<0$, is
\bea\label{Gammadip1234}
{\bf \Gamma}^{\rm dip.}\left(\{p\},\lambda, \as(\lambda^2)\right) &=&  
-\frac{\gamma_{K} (\as)}{2} \bigg[ \left(\T_1 \cdot \T_2 + \T_3 \cdot \T_4 \right) 
\log \frac{ s \,e^{-i \pi}}{\lambda^2} \\ 
&&\hspace*{-1.4cm}+ \, \left(\T_1 \cdot \T_3 + \T_2 \cdot \T_4 \right) 
\log \frac{-u}{\lambda^2} 
+\left(\T_1 \cdot \T_4 + \T_2 \cdot \T_3 \right) 
\log \frac{-t}{\lambda^2}  \bigg] 
+ \sum_{i=1}^4 \gamma_{i} (\as)\,.\nn
\eea
In the high-energy limit $u\approx -s$ this expression 
simplifies significantly. In particular, by expressing it in 
terms of the color operators introduced in \eqn{TtTsTu}
one obtains
\beqa \label{GammaDipHigh} 
\Gamma^{\rm dip.}\left(\{p_i\},\lambda, \as(\lambda^2)\right) 
\,&\xrightarrow{{\rm Regge}}&\, \frac{\gamma_{K} (\as)}{2} 
\bigg[ L\, \Tt +i \pi \, \Tsu+\frac{C_{\rm tot}}{2} \log \frac{-t}{\lambda^2} \bigg]  
\\&&\hspace{4.0cm}
+\, \sum_{i=1}^4 \gamma_{i} (\as) + \ord \left(\frac{t}{s}\right),\nn
\eeqa
where $L=\log\left|\frac{s}{t}\right| - i \frac{\pi}{2}$ is the natural 
signature-even combination of logarithms introduced in \eqn{L-def}.

Obtaining ${\bf \Delta}^{(3)}$ in the high-energy limit 
requires some more work~\cite{Almelid:2015jia,longpaper}. 
This function is initially defined in Euclidean kinematics 
where the invariants are all spacelike, and the momenta 
of the colored partons $p_i$ are not required to admit 
momentum conservation.  One therefore needs to first 
analytically continue the functions ${\cal F}$ in 
\eqn{quadrupole2} across the cut to the region where 
$p_1$ and $p_2$ are incoming while $p_3$ and $p_4$ 
outgoing. Once this is done, one imposes the momentum 
conserving limit where one identifies $s=s_{12}=s_{34}>0$ 
and $t=s_{14}=s_{23}<0$, and the variables $z$ and $\bar{z}$ 
approach the real axis from opposite sides and coincide, 
such that $z,\bar{z}\to s/(s+t)$. At the final stage one takes 
the high-energy limit where $s\gg -t$. Details of these 
calculations will be presented elsewhere. One obtains 
\bea\label{quadrupole4} \nn
{\bf \Delta}^{(3)} &=& i \pi \, [\Tt,[\Tt, \Tsu]] \frac{1}{4} 
\bigg[ \zeta_3 L + 11 \zeta_4 \bigg] 
+  \frac{1}{4}  [\Tsu,[\Tt, \Tsu]] \bigg[\zeta_5  - 4 \zeta_2 \zeta_3 \bigg] \\ \nn
&&-\frac{\zeta_5+2\zeta_2\zeta_3}{8} \bigg\{ f^{abe}f^{cde} 
\bigg[ \{\T_t^a, \T_t^d \}  \Big(\{\T_{s-u}^b, \T_{s-u}^c \} 
+ \{\T_{s+u}^b, \T_{s+u}^c \} \Big) \\ 
&&\hspace{4.0cm}+\,   \{\T_{s-u}^a, \T_{s-u}^d \}
 \{\T_{s+u}^b, \T_{s+u}^c \} \bigg] -\frac{5}{8} C_A^2 \Tt \bigg\},
\eea
where we introduced the color operators
\beq
\T^a_{s-u} \equiv \tfrac{1}{\sqrt{2}} \left(\T^a_s-\T^a_u\right), \qquad 
\T^a_{s+u} \equiv \tfrac{1}{\sqrt{2}} \left(\T^a_s+\T^a_u\right). 
\eeq
Note that the second and third lines in (\ref{quadrupole4}) 
correspond to the kinematics-independent term $C$ in the 
quadrupole correction of \eqn{quadrupole2}; it appears that 
it cannot be written in terms of quadratic invariants. The 
symmetry properties of (\ref{quadrupole4}) under $s$ to $u$ 
exchange, are nevertheless clear, and as expected 
(recall that the hard function on which this operator will act 
is color odd) the imaginary part is color odd while the real 
part is color even. We observe that this expression contains 
only a single factor of $L$, with an imaginary coefficient. 
Therefore the quadrupole contribution to the even amplitude 
${\cal M}^{(+)}$ starts at NNLL while for the odd amplitude 
${\cal M}^{(-)}$ it starts only  at N$^3$LL. The evaluation 
of the color operator in the second and third line of 
\eqn{quadrupole4} in an explicit color basis is provided 
in the appendices. More specifically, in appendix~\ref{3LoopPredictOrtho}  
we provide it in an orthonormal color basis in the $t$-channel, 
while in appendix~\ref{3LoopPredictTrace} we give it 
in a ``trace'' color basis. 

The anomalous dimension would be straightforward to 
exponentiate according to eq.~(\ref{RGsol}), were it not 
for the fact that $\Tt$, $\Tsu$ and the color operators in 
${\bf \Delta}^{(3)}$ do not commute.  This non-commutativity 
by itself implies that the amplitude projected on the tree-level 
color factor cannot be written as a simple power law, that is, it 
cannot be interpreted as exchange of a single Reggeized 
gluon~\cite{DelDuca:2011ae,Caron-Huot:2013fea}, as 
discussed in section~\ref{Regge_limit_perturbative}.

The last two terms in the dipole formula
\Eqn{GammaDipHigh} do not depend on colors nor 
or the total energy $s$, which suggests to attribute 
them to the projectile and target separately and write
$Z$ factor in \eqn{RGsol} in the following factorized form:
\beqa \label{RGsolFact}
{\bf Z} \left(\{p_i\},\mu, \as (\mu^2) \right)  &=&  
{\bf \tilde Z} \left(\frac{s}{t},\mu, \as (\mu^2) \right)
Z_i \left(t,\mu, \as (\mu^2) \right) Z_j \left(t,\mu, \as (\mu^2) \right),
\eeqa
where the $Z_{i/j}$ are just scalar factors that depend only 
on either the projectile or target:
\be\label{Zidef}
 Z_i = \exp\left\{- \int_0^{\mu^2} \frac{d\lambda^2}{\lambda^2} \left[
 \frac{{\gamma}_K\left(\as(\lambda^2)\right)}{4}\, C_i \, \log \frac{-t}{\lambda^2}  
 + \gamma_{i}\left(\as(\lambda^2)\right)\right]\right\}\,.
\ee
The more interesting factor is ${\bf \tilde Z}$ which 
is a color operator given to three-loop accuracy as:
\beq\label{ZtildeDef}
{\bf \tilde Z} \left(\frac{s}{t},\mu, \as (\mu^2) \right) = 
\exp \Big\{ K\left(\as(\mu^2)\right) \left[L\, \Tt +i \pi \, \Tsu \right] 
+ Q^{(3)}_{\bf \Delta} \Big\}
\eeq
with $K\left(\as(\mu^2)\right)$ defined as the integral 
over the cusp anomalous dimension: 
\be\label{Kdef}
K\left(\as(\mu^2)\right) = -\frac{1}{4}\int_0^{\mu^2} 
\frac{d \lambda^2}{\lambda^2} \,
\gamma_{K} \left(\as(\lambda^2)\right)\,
= \,\frac{1}{2\eps}\, \frac{\as(\mu^2)}{\pi} + \ldots,
\ee
while $Q^{(3)}_{\bf \Delta}$ represent the contribution 
of the quadrupole correction at three loops,
\be\label{Ztilde-quadrupole}
Q^{(3)}_{\bf \Delta} = -\frac{{\bf \Delta}^{(3)}}{2} \int_0^{\mu^2} 
\frac{d\lambda^2}{\lambda^2} 
\left(\frac{\alpha_s(\lambda^2)}{\pi}\right)^{3}  =  
\frac{{\bf \Delta}^{(3)}}{6\eps} 
\left(\frac{\alpha_s(\mu^2)}{\pi}\right)^{3}\,.
\ee

The extra logarithm of $\lambda$ in the integration in 
\eqn{Zidef} is  responsible for double poles combining 
infrared and collinear singularities. Thus we see that 
all double poles are included in the factors $Z_{i/j}$, 
while the factor $K$ (and consequently ${\bf \tilde Z}$) 
contains at most a single infrared pole per loop order. 
To three loops one has 
\bea\label{Kexplicit} \nn 
K(\as) &=& \frac{\as}{\pi} \, \frac{{\gamma}_K^{(1)}}{4 \epsilon} 
+ \left(\frac{\as}{\pi}\right)^2 \left( \frac{{\gamma}_K^{(2)}}{8 \epsilon} -
\frac{b_0 \, {\gamma}_K^{(1)}}{32 \epsilon^2}\right) , \\
&&\hspace{1.0cm} +\, \left(\frac{\as}{\pi}\right)^3
\left(\frac{{\gamma}_K^{(3)}}{12\epsilon} 
-\frac{b_0 \, {\gamma}_K^{(2)} 
+ b_1 \,{\gamma}_K^{(1)}}{48 \epsilon^2} 
+ \frac{b_0^2 \, {\gamma}_K^{(1)}}{192 \epsilon^3}\right) 
+\ord(\as^4)
\eea
where explicit expressions for the $\as$ expansion 
of the cusp anomalous dimensions $\gamma_{K}$, 
as well as the quark and gluon anomalous dimension 
$\gamma_{i}$ and the scalar factor $Z_{i/j}$ are provided 
in appendix \ref{GammaAndZ}.

The scalar factors $Z_i$ removed in \eqn{RGsolFact}
are the same as those we removed from the reduced 
amplitude \eqn{Mreduced} in the BFKL context, and 
in fact, at leading log accuracy the exponent of 
\eqn{ZtildeDef} is also very similar to the gluon 
Regge trajectory subtracted in the reduced 
amplitude. This makes the relation between 
the ``infrared-renormalized'' amplitude (hard 
function) $\Hhard_{ij\to ij}$ and reduced matrix 
element particularly simple. Comparing 
\eqn{IRfacteq} with \eqn{Mreduced} and 
using \eqns{RGsolFact}{ZtildeDef}, we 
indeed find
\bea\label{Mred_vs_H}\nn
\Hhard_{ij\to ij} \left(\{p_i\},\mu, \as (\mu^2) \right) &=&
\exp^{-1} \Big\{ K\left(\as(\mu^2)\right) \left[L\, \Tt +i \pi \, \Tsu \right] 
+ Q^{(3)}_{\bf \Delta} \Big\}   \\ 
&&\hspace{1.0cm} \cdot \,
\exp \bigg\{ \alpha_g(t) L \, \Tt \bigg\}\, 
\Mreduced_{ij\to ij}\left(\{p_i\},\mu, \as (\mu^2) \right).
\eea
This equation allows us to pass from directly from 
the reduced amplitude $\Mreduced_{ij\to ij}$, predicted 
in the previous section using BFKL theory, to the more 
conventional scattering amplitude or hard function. 
In particular, the statement that the left-hand-side
$\Hhard_{ij\to ij}$ is finite, which is equivalent to 
the exponentiation of infrared divergences,
is a highly nontrivial constraint on our result.

\subsection{Expansion of the hard amplitude}

Similarly to \eqn{Mpower_expansion},
we introduce a power expansion for the 
hard function: 
\be \label{HExpansion}
\Hhard_{ij\to ij} \left(\{p_i\},\mu, \as (\mu^2) \right) 
= 4\pi \alpha_s \sum_{n = 0}^\infty \sum_{k = 0}^n
\left( \frac{\alpha_s}{\pi} \right)^n L^k\,
\Hhard^{(n,k)} \left( \frac{-t}{\mu^2} \right).
\ee
In the rest of this section we 
derive the coefficients $\Hhard^{(n,k)}$
order by order in perturbation theory,
applying \eqn{Mred_vs_H} to the results 
of the preceding section. The color 
factors in the exponent do not commute, 
but the formula can be expanded in 
perturbation theory by repeatedly applying 
the Baker-Campbell-Hausdorff formula. 
Up to three loops we find explicitly:
\bea\label{Mred_vs_H2}\nn
\Hhard_{ij\to ij} \left(\{p_i\},\mu, \as (\mu^2) \right) &=&
\Bigg( 1
+\tfrac{K^3(\as) }{3!} \Big(2 \pi^2 \, L\, [\T_{s-u}^2, [\T_t^2, \T_{s-u}^2] ] 
-  i \, \pi L^2 \,  [\T_t^2, [\T_t^2, \T_{s-u}^2 ] ] \Big)
\\ \nn && \hspace{5mm}+i \pi \tfrac{K^2(\as)}{2} L \, [\T_t^2, \T_{s-u}^2]-Q^{(3)}_{\bf \Delta}\Bigg)
\cdot \exp \Big\{- i \pi  \, K(\as)\, \T_{s-u}^2\Big\}
 \\
&&\hspace{0mm}
\cdot\, \exp \Big\{ \Big( \alpha_g(t) - K(\as) \Big) \,L\, \T_t^2 \Big\} 
\, \Mreduced_{ij\to ij}\left(\{p_i\},\mu, \as (\mu^2) \right).\,
\eea
Notice that we have combined the exponent 
containing the Regge trajectory with the $\T_t^2$ term
in the infrared factorisation formula, since they 
have the same color structure.  Because of 
the structure of this exponent, the combination 
$\alpha_g(t) - K(\as)$ frequently appears 
in the following. For this reason, it proves useful 
to introduce the short-hand notation 
\be\label{alphaghat}
 \hat \alpha_g(t) = \alpha_g(t) - K(\as) ,
\ee 
to indicate the ``finite'' Regge trajectory divided by $N_c$. 
Expanding in the coupling, we write
\be
\hat \alpha_g(t)=\hat\alpha_g^{(n)}\,\,\left(\frac{\alpha_s(-t)}{\pi}\right)^n\,.
\ee
The fact that the combination entering the hard function 
is the difference between the Regge trajectory and $K(\as)$ 
is a manifestation of the relation between the divergent part 
of the gluon Regge trajectory and the cusp anomalous dimension 
discovered in Refs.~\cite{Korchemskaya:1994qp,Korchemskaya:1996je}.
Below we will see that this relation breaks down by the Regge cut, 
and in our scheme $\hat{\alpha}_g(t)$ will not be finite at three loops.

At one- and two-loops, using the known trajectory in 
\eqn{two_loop_gluon_trajectory} and cusp anomalous 
dimension in \eqn{gammaK} we get $\hat \alpha_g(t)$ 
with
\begin{subequations}
\label{alphahat}
\begin{align}
\hat\alpha_g^{(1)} &= \frac{1}{2\eps}(\rGamma - 1) = 
-\frac14\zeta_2\, \eps -\frac76\zeta_3 \,\eps^2 + {\cal O}(\eps^3),
\label{alpha1hat} \\
\hat\alpha_g^{(2)}&=C_A \left(\frac{101}{108}  
- \frac{\zeta_3}{8}\right) -\frac{7\nf}{54} +O(\eps)\,.
\label{alpha2hat}
\end{align}
\end{subequations}
This is nicely infrared finite.  The first term would in fact 
vanish if we worked in a scheme where the coupling is 
$\alpha_s \rGamma$ instead of $\alpha_s$, which would 
simplify many of our predictions. However, to simplify 
comparisons with the literature, we will stick with the 
standard $\MS$ coupling $\alpha_s$.

At leading logarithmic accuracy, and to any order in the 
coupling, only the rightmost exponential factor in 
(\ref{Mred_vs_H2}) is relevant, and we obtain 
\begin{align}
\label{Hnn}
\Hhard_{ij\to ij} ^{(n,n)} &= \frac{1}{n!}\left(\hat \alpha_g^{(1)}\right)^n
\left(\Tt\right)^n \, \Mreduced^{(0)}_{ij\to ij} , 
\end{align}
which is of course finite.
At NLL accuracy and beyond the expansion of 
(\ref{Mred_vs_H2}) requires input with regards 
to the coefficients of $\Mreduced_{ij\to ij}$.
The computation is significantly simplified here 
by working with the reduced amplitude, whose 
leading logarithms $\Mreduced_{ij\to ij}^{(n,n)}$ 
vanish, and whose next-to-leading logarithms 
beyond one loop are purely imaginary and are 
given by eq.~(\ref{Even_NLL}) above (the real 
part of $\Mreduced_{ij\to ij}^{(n,n-1)}$ for $n\geq 2$ 
vanishes by construction). Next-to-next-to-leading 
logarithms in the real part of $\Mreduced_{ij\to ij}$ 
are determined by the BFKL analysis of the previous section. 

An important feature visible in eq.~(\ref{Mred_vs_H2}) 
is that the conversion to hard function does not commute 
with the projection onto even and odd signatures.
Specifically, the odd part of the hard function at NNLL 
receives some contamination from the even reduced 
amplitude at NLL, multiplied by $i\pi K(\as) \T_{s-u}^2$ 
or $i\pi \tfrac{K^2(\as)}{2} [\T_t^2, \T_{s-u}^2]$. This is 
not going to pose a problem, because these ingredients 
are already known.

Our comparison between Regge and infrared factorization 
below follows closely the analysis in \cite{DelDuca:2014cya} 
(see also \cite{Caron-Huot:2013fea}). Nevertheless, there 
are several new elements allowing us to make a significant 
step forward: first, our present analysis makes a clear and 
transparent separation between signature odd and even, 
corresponding respectively to real and imaginary parts of 
the amplitude expressed in terms of $L\equiv \log\left|\frac{s}{t}\right| 
- i \frac{\pi}{2}$; second is the possibility to compare the infrared 
factorisation formula with the contribution originating from 
three Reggeon exchange at two and three loops, which we 
have calculated here for the first time; third is the availability 
of the complete infrared structure at three loops, i.e. 
eq.~(\ref{quadrupole4}) based on~\cite{Almelid:2015jia,longpaper}, 
which implies, in particular, that the odd amplitude receives no 
new NNLL high-energy corrections beyond the dipole formula 
through three loops, while the even amplitude does; a final 
new ingredient is the availability of the ${\cal N}=4$ SYM 
result for $2\to2$ gluon-gluon scattering amplitude~\cite{Henn:2016jdu}, 
which beyond consistency checks, also provides new information 
on the odd amplitude at NNLL:  together with the computation 
of the three-Reggeon cut performed here, it allows us to fix 
the three-loop gluon Regge trajectory in this theory.

\subsection{Comparison at one loop}

At tree level one has $\Hhard^{(0)} 
= \Mreduced^{(0)} =\MM^{(0)}$. 
Comparison at one loop is simple, 
and completely equivalent to the 
discussion in \cite{Caron-Huot:2013fea,DelDuca:2014cya}.
We repeat it here in order to adapt it
to the conventions used in this paper, in particular, 
the fact that we expand the amplitude in powers of $L
= \log |s/t| - i \pi /2$ instead of 
powers of  $\log |s/t|$. 

Expanding \eqn{Mred_vs_H2} to one loop, 
and suppressing the indices $ij\to ij$ for brevity, we get 
\begin{subequations}
\begin{align}
\label{H11}
\Hhard^{(1,1)} &= \hat \alpha_g^{(1)} \, \T_t^2 \, \Mreduced^{(0)} ,\\
\label{H10}
\Hhard^{(1,0)} &= \Mreduced^{(1,0)} 
- i  \pi \, K^{(1)} \, \Tsu \, \Mreduced^{(0)}.
\end{align}
\end{subequations}
As anticipated (see the discussion regarding 
\eqn{alphaghat}) the fact that the hard function 
must be finite relates to the connection between the 
divergent part of  $\alpha_g^{(1)}$ and the cusp 
anomalous dimension~\cite{Korchemskaya:1994qp,Korchemskaya:1996je}. 
The vanishing of $\hat \alpha_g^{(1)}$ in the four-dimensional 
limit, as shown in \eqn{alpha1hat}, reflects the fact that gluon 
Reggeisation at this order is determined entirely by soft corrections, 
hence no high-energy logarithms arise in the hard 
function at one loop in the $\epsilon\to 0$ limit.

The finite part in \eqn{H10} contains informations 
both its in real and imaginary parts. Using the direct 
correspondence between the real and imaginary parts 
of the amplitude, respectively, and its odd and even 
signature parts, we get
\begin{subequations}
\begin{align}
\RE[\Hhard^{(1,0)}] =& \Mreduced^{(-,1,0)} , \\
i \, \IM[\Hhard^{(1,0)}] =& \Mreduced^{(+,1,0)} 
-i \pi \, K^{(1)} \, \Tsu \, \Mreduced^{(0)}.
\end{align}
\end{subequations}
Using the results for $\Mreduced^{(-,1,0)}$ and 
$\Mreduced^{(+,1,0)}$ in \eqns{one_two_loops_regge-1}{Even_NLL} 
one explicitly gets 
\begin{subequations}
\label{H10b}
\begin{align}
\label{H10Real}
\RE[\Hhard^{(1,0)}] &= \Big(D_i^{(1)} + D_j^{(1)}\Big)\, \Mreduced^{(0)}, \\
\label{H10Img}
i\,\IM[\Hhard^{(1,0)}] &= i \pi \left({\mathbb{d}}_{1}  - K^{(1)}\right)\Tsu  \, \Mreduced^{(0)}
\,=\,  i \pi \,\hat\alpha_g^{(1)}\, \Tsu  \, \Mreduced^{(0)}\,,
\end{align}
\end{subequations}
where ${\mathbb{d}}_1=\frac{\rGamma}{2\eps}$ is the 
one-loop coefficient in \eqn{dells}; in the last expression 
in \eqn{H10Img} we used (\ref{alphahat}) to replace the 
difference of divergent coefficients ${\mathbb{d}}_{1}  - K^{(1)}$ 
by the ${\cal O}(\epsilon)$ coefficient~$\hat\alpha_g^{(1)}$. 
This replacement will be used in what follows to obtain simpler 
expressions at higher orders.

Infrared factorization tells us that both of the equations 
in \eqn{H10b} are finite as $\eps\to 0$. This is evidently 
satisfied for the imaginary part, \eqn{H10Img}. 
Finiteness of the real part in \eqn{H10Real} in turn 
implies that the impact factors $D_i$ must also be finite 
-- indeed they are, as we have already extracted the 
divergences into the factors $Z_i$ of \eqn{Zidef} 
(see \eqn{Regge-Pole-General}). A systematic 
way to extract these, which will work to higher orders 
as well, is to consider the fixed-order hard functions 
projected onto the color octet (see e.g. 
\eqn{odd_even_Ms}). Then we have simply
\be\label{impactfactorOneLoop}
D_i^{(1)} 
= \frac{1}{2}\, \frac{\RE[\Hhard_{ii\to ii}^{(1,0)[8_a]}]}{\Hhard_{ii\to ii}^{(0)[8_a]}}.
\ee
Explicitly, using the one-loop gluon-gluon and 
quark-quark octet hard function from Ref.~\cite{DelDuca:2014cya},
converting to the convention where the amplitude is expanded
in powers of $L = \log |s/t| - i \pi /2$ instead of powers of 
$\log |s/t|$, we extract the results for the one-loop impact 
factors, which are indeed finite:
\begin{subequations}
\begin{align}
 \begin{split}
D_g^{(1)} & = - N_c \left(\frac{67}{72} - \zeta_2 \right) + \frac{5}{36} n_f 
+ \eps \bigg[ N_c \left(-\frac{101}{54} + \frac{11}{48}\zeta_2 + \frac{17}{12} \zeta_3 \right) 
+ n_f  \left( \frac{7}{27} - \frac{\zeta_2}{24} \right) \bigg] \\ 
&+ \eps^2 \bigg[ N_c \left(- \frac{607}{162} +\frac{67}{144}\zeta_2 
+ \frac{77}{72}\zeta_3 +\frac{41}{32}\zeta_4\right) 
+ n_f  \left( \frac{41}{81} - \frac{5}{72}\zeta_2 -\frac{7}{36}\zeta_3 \right) \bigg]
+ {\cal O}(\eps^3) \,,  \label{D1g}
\end{split}
\\
\begin{split}
D_q^{(1)} & = N_c \left(\frac{13}{72}  + \frac{7}{8} \zeta_2 \right)
+ \frac{1}{N_c} \left( 1 - \frac{1}{8} \zeta_2 \right) - \frac{5}{36} n_f  
+ \eps \bigg[ N_c \left(\frac{10}{27} - \frac{\zeta_2}{24} + \frac{5}{6} \zeta_3 \right)  \\ 
&\hspace{-0.7cm}+ \frac{1}{N_c} \left(2 - \frac{3}{16}\zeta_2 - \frac{7}{12} \zeta_3 \right) 
+ n_f  \left( -\frac{7}{27} + \frac{\zeta_2}{24} \right) \bigg]
+ \eps^2 \bigg[ N_c \left(\frac{121}{162} - \frac{13}{144}\zeta_2 
- \frac{7}{36} \zeta_3 + \frac{35}{64}\zeta_4 \right)  \\
&\hspace{-0.7cm}+ \frac{1}{N_c} \left(4 - \frac{\zeta_2}{2} - \frac{7}{8} \zeta_3 - \frac{47}{64}\zeta_4 \right) 
+ n_f  \left( -\frac{41}{81} + \frac{5}{72} \zeta_2 + \frac{7}{36}\zeta_3 \right) \bigg]
+ {\cal O}(\eps^3) \,. \label{D1q}
\end{split}
\end{align}
\end{subequations}
Note that, with these two coefficients extracted, the 
quark-gluon amplitude is then predicted unambiguously 
and correctly, as explicitly shown in Ref.~\cite{DelDuca:2014cya}
(see eq. (4.17) there). 

\subsection{Comparison at two loops}

At two-loops, the expansion of 
\eqn{Mred_vs_H2} gives
\begin{subequations}
\begin{align}
\label{H22}
\Hhard^{(2,2)} &= \frac{1}{2}(\hat \alpha_g^{(1)})^2
(\Tt)^2 \, \Mreduced^{(0)} , \\ \nn
\label{H21}
 \Hhard^{(2,1)} &=
\Mreduced^{(2,1)} 
+\hat \alpha_g^{(1)}  \Tt \,\Mreduced^{(1,0)} 
+\hat \alpha_g^{(2)}  \Tt \, \Mreduced^{(0)}  \\ 
&\hspace{1.0cm}+\, i\pi K^{(1)} \Big[ 
\tfrac12K^{(1)} [\Tt, \Tsu]-  \hat \alpha_g^{(1)} \Tsu \Tt \,  \Big] \Mreduced^{(0)},  \\
\label{H20}
\Hhard^{(2,0)} &=
\Mreduced^{(2,0)}  -\frac{\pi^2}{2} (K^{(1)})^2 (\Tsu)^2 \Mreduced^{(0)} 
- i \pi \left[K^{(2)} \Tsu \Mreduced^{(0)} 
+ K^{(1)} \Tsu \Mreduced^{(1,0)} \right].
\end{align}
\end{subequations}
Note that the leading-log term of \eqn{H22} 
is a simple exponentiation of \eqn{H11}.
More interesting are the lower-logarithmic terms 
of the amplitude. Using explicitly the information 
for $\Mreduced^{(1,0)}$  and $\Mreduced^{(2,1)}$ 
in eq.~(\ref{Odd_NLL}),~(\ref{Even_NLL}) and 
(\ref{one_two_loops_regge-2}) we obtain 
\begin{subequations}
\begin{align}
\label{H21Real}
\RE[\Hhard^{(2,1)}] &= \left[\hat \alpha_g^{(2)} 
+ \hat \alpha_g^{(1)} \left(D_i^{(1)} + D_j^{(1)}\right) \right] \Tt \, \Mreduced^{(0)}, \\
\label{H21Img}
i\, \IM[\Hhard^{(2,1)}] &= i \pi 
\left[\Big(\tfrac12{\mathbb{d}}_2+\tfrac12 (K^{(1)})^2  + K^{(1)}\hat \alpha_g^{(1)} \Big) [\Tt,\Tsu] 
+ \left(\hat \alpha_g^{(1)}\right)^2  \Tt \Tsu \right] \Mreduced^{(0)}.
\end{align}
\end{subequations}
Finiteness of the first line is manifest, and 
finiteness of the second line is a constraint 
on the divergent part of ${\mathbb{d}}_2$, 
which again is satisfied by the explicit expression
in \eqn{dells} (with $K^{(1)}=1/(2\epsilon)$, see 
eq.~(\ref{Kexplicit})); this was also verified in 
Ref.~\cite{Caron-Huot:2013fea}.

Considering finally the coefficient of the zero-th 
order logarithm, i.e. \eqn{H20}, the operator 
$\left(\Tsu\right)^2$ makes its first appearance. 
We focus on the odd component, i.e. 
$\Mreduced^{(-,2,0)}$, which we have 
calculated in \eqn{one_two_loops_regge-2}. 
Inserting this result along with the previous
result for the one-loop even amplitude we obtain 
\bea \label{H20Odd} \nn
\RE[\Hhard^{(2,0)}]  &=&
\bigg[ D_i^{(2)}+D_j^{(2)}+D_i^{(1)} D_j^{(1)} 
- \pi^2R^{(2)} \tfrac{1}{12}(\CA)^2  \\
&&\hspace{1.0cm}+\pi^2 \,\Big(R^{(2)} + \tfrac{1}{2}  (K^{(1)})^2 
+ \, K^{(1)} \hat \alpha_g^{(1)}  \Big) (\Tsu)^2 \bigg]
\Mreduced^{(0)}. 
\eea
It is clear at this point that the term
proportional to  $(\Tsu)^2$ in the 
infrared factorisation formula can be attributed to
multi-Reggeon exchange, and 
this is confirmed by the fact that the 
quantity in squared brackets in \eqn{H20Odd} 
proportional to $(\Tsu)^2$ is finite. Upon 
explicit substitution of $R^{(2)}$ in \eqn{R2def},
we get
\be\label{R2hat}
{\hat{R}}^{(2)} \equiv
R^{(2)} +\tfrac{1}{2}  (K^{(1)})^2 + K^{(1)}  \hat \alpha_g^{(1)}
= \frac{3}{4}\eps\zeta_3+ \frac{67}{64}\eps^2\zeta_4+\ldots
\ee
which is indeed finite,
as required for the infrared renormalized 
amplitude $\RE[\Hhard^{(2,0)}]$.

This equation can thus be used to extract 
the impact factors at two loops from the 
known two-loop fixed-order amplitudes.
As before, it suffices to consider the 
projection of the amplitude onto the 
adjoint channel, but the projection of 
the color factor $(\Tsu)^2$ needs to be 
carried out on a case-by-case basis.  
This can be done using the matrices 
given in appendix \ref{3LoopPredictOrtho}.
For gluon-gluon scattering with SU($N_c$) 
gauge group, we get: 
\be\label{ImpactFactorTwoLoops}
2D_g^{(2)} = \frac{\Hhard_{gg\to gg}^{(2,0)[8_a]}}{\Hhard_{gg\to gg}^{(0)[8_a]}}  
-(D_g^{(1)})^2 + \pi^2R^{(2)}\frac{N_c^2}{12}-\pi^2{\hat{R}}^{(2)}\frac{N_c^2+24}{4},
\ee
where in turn $D_g^{(1)}$ can be found in eq.~(\ref{D1g}).
The impact 
factor $D_g^{(2)}$ would be finite, were 
it not for the double pole originating from the 
$R^{(2)}\approx -\frac{1}{8\eps^2}$ term.
For quark-gluon scattering we find: 
\be
\label{qplusg}
D_q^{(2)}+D_g^{(2)} = \frac{\Hhard_{qg\to qg}^{(2,0)[8_a]}}{\Hhard_{qg\to qg}^{(0)[8_a]}}  
-D_q^{(1)}D_g^{(1)} + \pi^2R^{(2)}\frac{N_c^2}{12}-\pi^2{\hat{R}}^{(2)}\frac{N_c^2+4}{4}\,.
\ee
Finally, for quark-quark scattering, we find instead:
\be
\label{twoq}
2D_q^{(2)} = \frac{\RE[\Hhard_{qq\to qq}^{(2,0)[8_a]}]}{\Hhard_{qq\to qq}^{(0)[8_a]}}  
-(D_q^{(1)})^2 + \pi^2R^{(2)}\frac{N_c^2}{12}-\pi^2{\hat{R}}^{(2)}\frac{N_c^4-4N_c^2+12}{4N_c^2}.
\ee
The important thing to notice is that the coefficient 
of the ${\hat{R}}^{(2)}$ term, which represents the
color structure $(\Tsu)^2$ attributed to three-Reggeon 
exchange, is different in each case. This contribution 
(in addition to the extra factors of $(\Tsu)^2$ coming 
from the infrared renormalization) explains why the 
amplitude does not take a simple factorized form, as 
was first observed in ref.~\cite{DelDuca:2001gu} 
based on explicit computations of two-loop amplitudes 
for gluon-gluon, quark-gluon and quark-quark scattering 
(for the departure from simple Regge-pole factorization 
see also~\cite{Bret:2011xm,DelDuca:2011ae,DelDuca:2013ara,DelDuca:2014cya,Caron-Huot:2013fea}).
Quantitatively, it is a highly nontrivial check on the 
BFKL formalism that the three equations 
(\ref{ImpactFactorTwoLoops}) through (\ref{twoq}) 
can be solved for the two unknowns $D_g^{(2)}$ 
and $D_q^{(2)}$. By using the explicit result for the 
two-loop gluon-gluon and quark-quark hard functions $\Hhard_{ij\to ij}^{(2,0)[8_a]}$
provided in Ref.~\cite{DelDuca:2014cya}, and 
expanding them in powers of  $L = \log |s/t| - i \pi /2$
we find:
\begin{subequations}
\label{D2s}
\begin{align}
\begin{split}
\label{D2g} 
D_g^{(2)} & = - \frac{\zeta_2}{32\eps^2} N_c^2 + 
N_c^2 \bigg(-\frac{26675}{10368} + \frac{335}{288} \zeta_2 
+ \frac{11}{18}\zeta_3 - \frac{\zeta_4}{64} \bigg) \\ 
& + N_c n_f \bigg(\frac{2063}{3456} - \frac{25}{144} \zeta_2+ \frac{\zeta_3}{72} \bigg) 
+ \frac{n_f}{N_c} \bigg(-\frac{55}{384} + \frac{\zeta_3}{8} \bigg) -\frac{25}{2592} n_f^2
+ {\cal O}(\eps) \,,  
\end{split}
\\ 
\begin{split}
\label{D2q}
D_q^{(2)} & =  - \frac{\zeta_2}{32\eps^2} N_c^2 + 
N_c^2 \bigg(\frac{22537}{41472} + \frac{87}{64} \zeta_2 
+ \frac{41}{144}\zeta_3 - \frac{15}{256} \zeta_4 \bigg) +\frac{28787}{10368} +\frac{19}{32}\zeta_2  \\ 
& - \frac{205}{288}\zeta_3
-\frac{47}{128}\zeta_4 + \frac{1}{N_c^2} \bigg(\frac{255}{512} + \frac{21}{64} \zeta_2 
- \frac{15}{32}\zeta_3 - \frac{83}{256} \zeta_4 \bigg) 
\\
& + N_c n_f \bigg(-\frac{325}{648} - \frac{\zeta_2}{4} - \frac{23}{144} \zeta_3\bigg) 
+ \frac{n_f}{N_c} \bigg(-\frac{505}{1296} - \frac{\zeta_2}{16} - \frac{19}{144}\zeta_3 \bigg) 
+\frac{25}{864} n_f^2 + {\cal O}(\eps) \,.
\end{split}
\end{align}
\end{subequations}
Remarkably, these impact factors then correctly predict the quark-gluon amplitude
according to \eqn{qplusg}, as it should!

Finally, we comment on the infrared divergences 
in $D_i^{(2)}$, which contrast with the finite~$D_i^{(1)}$.
We believe one should not be overly concerned about 
this, because of the arbitrary basis choice in 
\eqn{3WtoW-Wto3W} which has forced the physics 
into a very specific basis, where one- and three-Reggeon 
states are orthogonal to each other, therefore removing 
$1\to 3$ and $3\to 1$ correlators of Wilson lines.
In practice, the color factors $\sim (\CA)^2$ of such 
correlators would not be distinguishable from 
single-Reggeon exchange at this order. It  
seems plausible that, in a more natural basis, 
the infrared divergences would  appear only in these
off-diagonal contributions rather than being pushed 
into the $1\to 1$ single-Reggeon transition, thus 
leaving finite impact factors which may be closer 
to the ones defined empirically in Ref.~\cite{DelDuca:2014cya}.
We leave this for future investigation: since the 
choice used in this paper corresponds to a 
well-defined basis, it should always be possible 
to convert the result to other schemes.

\subsection{Comparison at three loops}
\label{ssec:comp3}

Let us now turn to the comparison between the 
BFKL results and infrared factorization at three loops. 
Because the expressions are rather lengthy, we will 
discuss the various logarithmic orders in turn. At leading 
logarithmic accuracy the expansion of the infrared 
factorisation formula in eq.~(\ref{Mred_vs_H2}) gives 
\be \label{AmpCoeff33tilde}
\Hhard^{(3,3)} =
\frac{1}{6}  \left(\hat{\alpha}_g^{(1)}\right)^3  \left(\Tt\right)^3 \Mreduced^{(0)},
\ee
as anticipated in \eqn{Hnn}. 
At next-to-leading logarithmic accuracy, 
the infrared factorisation in formula in 
\eqn{Mred_vs_H2} gives 
\bea \label{H32} \nn
\Hhard^{(3,2)} &=&   
\Mreduced^{(3,2)} +\hat \alpha_g^{(1)} \, \Tt \, \Mreduced^{(2,1)}
+\tfrac12(\hat \alpha_g^{(1)})^2 (\Tt)^2 \Mreduced^{(1,0)}
+ \hat \alpha_g^{(1)} \hat \alpha_g^{(2)}  \, (\Tt)^2 \, \Mreduced^{(0)} \\ 
&&+\, i\pi \Big( -\tfrac12 (\hat \alpha_g^{(1)})^2 K^{(1)} \Tsu (\Tt)^2
+ \tfrac12  \hat\alpha_g^{(1)} (K^{(1)})^2 [\Tt,\Tsu]\Tt \\ \nn
&&\hspace{6.0cm}-\, \tfrac16 (K^{(1)})^3  [\Tt,[\Tt,\Tsu]] \Big) \Mreduced^{(0)} .
\eea
Inserting the explicit results for the 
$\Mreduced^{(n,n-1)}$ terms as given 
in eqs.~(\ref{one_two_loops_regge}) and (\ref{Even_NLL}) the 
hard function can be brought 
into the form 
\begin{subequations}
\label{H32explicit1}
\begin{align}
\label{H32Real}
\RE[\Hhard^{(3,2)}] &= \hat \alpha_g^{(1)}  \Big[\hat \alpha_g^{(2)} 
+ \tfrac12 \hat \alpha_g^{(1)} \left(D_i^{(1)} + D_j^{(1)}\right) \Big] (\Tt)^2 \, \Mreduced^{(0)}, \\ \nn
\label{H32Img}
i\, \IM[\Hhard^{(3,2)}] &= i \pi 
\Big[ \tfrac16\Big(\mathbb{d}_3 - (K^{(1)})^3  
- 3 K^{(1)} (\hat \alpha_g^{(1)})^2 - 3 (K^{(1)})^2 \hat \alpha_g^{(1)} \Big) [\Tt,[\Tt,\Tsu]] \\ \nn
&\hspace{1,0cm}+\, \tfrac12 \hat \alpha_g^{(1)}   \Big(\mathbb{d}_2 + (K^{(1)})^2 
+ 2 K^{(1)} \hat \alpha_g^{(1)} \Big) \Tt [\Tt,\Tsu] \\
&\hspace{1,5cm}+\, \tfrac12 (\hat \alpha_g^{(1)})^3 (\Tt)^2 \Tsu \Big] \Mreduced^{(0)}.
\end{align}
\end{subequations}

It is easy to check by explicit substitution 
of the functions involved in \eqn{H32explicit1}
that $\Hhard^{(3,2)}$ is indeed finite. The only 
${\cal O}(\eps^0)$ contribution is given by 
$\IM[\Hhard^{(3,2)}]$, i.e. one has 
\bea\nn \label{H32explicit2}
\RE[\Hhard^{(3,2)}] &=& {\cal O}(\eps), \\
i\, \IM[\Hhard^{(3,2)}] &=& i \pi \left( - \frac{11}{24}\zeta_3 
+ {\cal O}(\eps) \right)  [\Tt,[\Tt,\Tsu]]   + {\cal O}(\eps).
\eea
More interesting is the amplitude at NNLL, 
since at this logarithmic accuracy we can 
confront our new predictions concerning the 
three-Reggeon exchange to the infrared 
factorisation formula. Starting from the latter, 
\eqn{Mred_vs_H2} gives 
\bea \nn
\Hhard^{(3,1)} &=& \Mreduced^{(3,1)}
+ \hat \alpha_g^{(1)} \Tt \Mreduced^{(2,0)}
+ \hat \alpha_g^{(2)} \Tt \Mreduced^{(1,0)}
+ \hat \alpha_g^{(3)} \Tt \Mreduced^{(0)}  \\ \nn
&&\hspace{-1.0cm}+\, \frac{\pi^2}{6} \Big[ - 3 \hat \alpha_g^{(1)} (K^{(1)})^2 (\Tsu)^2 \Tt
+\, (K^{(1)})^3 \Big( 2 \Tsu [\Tt,\Tsu] + [\Tt,\Tsu]\Tsu \Big) \Big]\Mreduced^{(0)} \\ \nn
&&\hspace{-1.0cm} +\, i \pi \Big[ - K^{(1)} \Tsu \Mreduced^{(2,1)}
+ \Big( \tfrac12 (K^{(1)})^2 [\Tt,\Tsu] 
- K^{(1)} \hat \alpha_g^{(1)}  \Tsu \Tt \Big) \Mreduced^{(1,0)} \\ \nn
&& +\, \Big( K^{(1)} K^{(2)} [\Tt,\Tsu] 
-K^{(2)} \hat \alpha_g^{(1)}  \Tsu \Tt 
-K^{(1)} \hat \alpha_g^{(2)}  \Tsu \Tt \\
&&\hspace{1.0cm}-\, 
\frac{\zeta_3}{24\eps} [\Tt,[\Tt,\Tsu]] \Big) \Mreduced^{(0)} \Big]\, .
\label{H31} 
\eea
Note that the last term this equation, proportional 
to $\zeta_3/\eps$, originates from the recently-computed 
quadrupole correction~\cite{Almelid:2015jia}, as shown 
in section~\ref{IRfact} (see \eqns{quadrupole4}{Ztilde-quadrupole} 
above), while all other terms in eq.~(\ref{H31}), which 
involve $K^{(n)}$, originate in the dipole formula 
$\Gamma^{\rm dip.}$. \Eqn{H31} shows explicitly 
that, in the high-energy limit, the quadrupole 
correction contributes first at NNLL, and it only 
contributes at this logarithmic order to the 
\emph{even part} of the amplitude.

Our prediction from BFKL
theory concerns the odd amplitude, 
hence we focus now on the real part of 
\eqn{H31}. Inserting results for the amplitude 
coefficients $ \Mreduced^{(n,k)}$ determined 
in the previous section, we get 
\bea \label{H31Real} \nn
\RE[\Hhard^{(3,1)}] &=& 
\Big[\hat \alpha_g^{(3)}
+ \hat \alpha_g^{(2)}\Big(D_i^{(1)} + D_j^{(1)}\Big) 
+ \hat \alpha_g^{(1)}\Big(D_i^{(2)} + D_j^{(2)}+D_i^{(1)} D_j^{(1)} \Big)  
 \Big] \Tt \, \Mreduced^{(0)} \\ 
&&\hspace{-0.0cm}+\, \pi^2 \Big[R_C^{(3)} 
-\tfrac{1}{12}\hat \alpha_g^{(1)} R^{(2)} \Big]   (\Tt)^3 \, \Mreduced^{(0)} 
+ \pi^2  \,  \hat \alpha_g^{(1)} \,\hat R^{(2)}\,  \Tt (\Tsu)^2 \, \Mreduced^{(0)} \\ \nn 
&&+\,\pi^2 \Big[R_A^{(3)} \,+ \,\tfrac{1}{6}\, K^{(1)}\Big(2(K^{(1)})^2 
 +3 \hat \alpha_g^{(1)} K^{(1)} + 3   \mathbb{d}_2 \Big) 
 \Big] \Tsu [\Tt,  \Tsu] \, \Mreduced^{(0)} \\ \nn
&& +\, 
\pi^2 \Big[R_B^{(3)}-\tfrac{1}{3}\,K^{(1)} \,\Big((K^{(1)})^2 
+3\hat \alpha_g^{(1)} K^{(1)}   +\, 3 (\hat \alpha_g^{(1)})^2  \Big)  
\Big] [\Tt, \Tsu] \Tsu \, \Mreduced^{(0)} .
\eea
In this equation, the parameters $\hat \alpha_g^{(i)}$
are related to the perturbative expansion of the Regge trajectory, 
representing the one-Reggeon evolution, according to the definition 
in \eqn{alphaghat}. As already discussed, these parameters are 
unknown in our formulation of the Regge theory, beyond 
$\hat \alpha_g^{(1)}$. However, $\hat \alpha_g^{(2)}$ can 
be determined from the two loop analysis, see \eqn{alpha2hat}, 
which means that only $\alpha_g^{(3)}$ is unknown in \eqn{H31Real}. 
We discuss below how this parameter can be extracted from 
a three-loop calculation. Similarly, the impact factors 
$D_{i/j}^{(n)}$ represent corrections to the one-Reggeon 
wavefunction, and can be determined by matching with 
explicit calculations. \Eqn{H31Real} depends on the impact
factors up to two loops, and these have all been determined
through the one- and two-loop analysis, see 
eqs.~(\ref{impactfactorOneLoop}) and 
(\ref{ImpactFactorTwoLoops}) through (\ref{twoq}).
The two terms proportional to $\Tsu [\Tt, \Tsu]$ 
and $[\Tt, \Tsu] \Tsu$ depend only on quantities 
which have been calculatated explicitly: the loop functions 
$R_{A,B,C}^{(3)}$ originating from the BFKL evolution of 
the $1\to3$, $3\to1$ and $3\to3$ Reggeon exchange, 
terms from the one-loop soft anomalous dimension cubed,
plus the signature-odd log part of the quadrupole correction 
\eqn{quadrupole4} (which turns out to be zero).
The fact that these terms add up to something finite is 
therefore a highly non-trivial check of both BFKL theory 
and of the specific form of the quadrupole correction.
Indeed, expanding explicitly to ${\cal O}(\eps^0)$ 
one finds  
\bea \label{H31RealEps0} \nn
\RE[\Hhard^{(3,1)}] &=& 
\Bigg[\hat \alpha_g^{(3)}
+ \hat \alpha_g^{(2)}\Big(D_i^{(1)} + D_j^{(1)}\Big) 
+ \hat \alpha_g^{(1)}\Big(D_i^{(2)} + D_j^{(2)}+D_i^{(1)}D_j^{(1)}\Big)  
\\ &&\quad +\CA^2
\frac{\pi^2}{864}\Big(\frac{1}{\epsilon^3} 
- \frac{15 \zeta_2}{4 \eps} - \frac{175 \zeta_3}{2}\Big)
\Bigg] \CA \, \Mreduced^{(0)} \\ \nn
&&
+ \pi^2 \frac{5 \zeta_3}{12} \Tsu [\Tt,  \Tsu] \, \Mreduced^{(0)} 
+ \pi^2 \frac{ \zeta_3}{12} [\Tt, \Tsu] \Tsu \, \Mreduced^{(0)} +{\cal O}(\eps).
\eea
where the term proportional to $C_A^2$ originates in the combination $\pi^2 \Big[R_C^{(3)} 
-\tfrac{1}{12}\hat \alpha_g^{(1)} R^{(2)} \Big]$ in eq.~(\ref{H31Real}).

We stress that the color operators 
$\Tsu [\Tt, \Tsu]$ and $[\Tt, \Tsu] \Tsu$
originate, within the infrared factorisation 
approach, only from the expansion of the 
``dipole term'' in \eqn{GammaDipHigh},
since, as discussed after \eqn{H31}, the 
quadrupole correction turns out to contribute at 
NNLL only to the even amplitude. 
The fact that the calculation of the odd 
amplitude at NNLL within the Regge theory 
matches exactly the poles originating from 
the dipole contribution can be seen as an 
indirect confirmation of the result in 
Ref.~\cite{Almelid:2015jia}; in the computation 
of the previous section, the fact that the quadrupole 
contribution to the odd amplitude vanishes can be 
seen to be a reflection of the absence of $1/\eps$ 
single poles in the bubble integrals of eq.~(\ref{integrals}).
Finiteness of the left-hand-side also predicts the 
infrared poles of the presently unknown 
``trajectory'' $\hat{\alpha}_3$.

\Eqn{H31Real} represents not only 
a check of the infrared factorisation formula, 
but also a prediction for the real 
part of the infrared-finite amplitude, in the 
high-energy limit, up to three loops. In order 
to show what parts of the amplitude 
are predicted, we focus now on gluon-gluon
scattering. Recalling our discussion in section 
\ref{Regge_limit_perturbative}, in particular
\eqn{odd_even_Ms}, we see that the real part of the
hard function corresponds to the antisymmetric 
octet $8_a$ and the $10+\overline{10}$ 
components of the amplitude. Evaluating the 
color operators in \eqn{H31RealEps0} in the 
orthonormal color basis in the $t$-channel defined 
in appendix \ref{3LoopPredictOrtho} we find 
\begin{subequations} 
\label{H31RealEps0Components} 
\begin{align}
\label{H31RealEps0Components8}  \nn
\RE[\Hhard^{(3,1),[8_a]}] &= 
\bigg\{ \CA \Big[\hat \alpha_g^{(3)}
+ \hat \alpha_g^{(2)}\Big(D_i^{(1)} + D_j^{(1)}\Big) 
+ \hat \alpha_g^{(1)}\Big(D_i^{(2)} + D_j^{(2)}+D_i^{(1)}D_j^{(1)}\Big)  
 \Big]  \\ 
&\hspace{0.5cm}+ \CA^3
\frac{\pi^2}{864}\Big(\frac{1}{\epsilon^3} - \frac{15 \zeta_2}{4 \eps} - \frac{175 \zeta_3}{2}\Big)
-  C_A \pi^2 \frac{2\zeta_3}{3}  +{\cal O}(\eps) \bigg\} \Mreduced^{(0),[8_a]}, \\ 
\label{H31RealEps0Components10} 
\RE[\Hhard^{(3,1),[10+\overline{10}]}] &=
\sqrt{2}C_A \sqrt{C_A^2-4} \bigg\{\frac{11\pi^2 \zeta_3}{24} 
+{\cal O}(\eps) \bigg\}\Mreduced^{(0),[8_a]}.
\end{align}
\end{subequations}
Concerning the antisymmetric octet
component, we see that it involves the Regge trajectory at three loops, 
$\hat \alpha_g^{(3)}$, which is unknown 
within our formalism. Given that the impact factors 
up to two loops are known from our previous 
analysis, see in particular \eqn{ImpactFactorTwoLoops}, 
this means that, 
knowing $\RE[\Hhard^{(3,1),[8_a]}]$, \eqn{H31RealEps0Components8}
can be used to extract $\hat \alpha_g^{(3)}$.
We will take this point of view below. Before, 
however, we note that $\hat \alpha_g^{(3)}$ does
not contribute to the $10+\overline{10}$ component
of the amplitude. Therefore, in our formalism we are able to 
predict this term unambiguously, and in \eqn{H31RealEps0Components10}
we have provided the explicit result up to three loops. 
As already mentioned, this result does not depend on the matter 
contents of the theory. Indeed, we find that our prediction agrees 
perfectly with a recent calculation~\cite{Henn:2016jdu} of $2\to 2$ 
gluon-gluon scattering amplitude at three loops in ${\cal N}=4$ SYM!

In appendix \ref{3LoopPredictOrtho} we provide 
an explicit prediction for the gluon-gluon hard amplitude
up to three loops in perturbation theory, which is based 
on the combination of the BFKL theory developed in section 
\ref{Regge} and the comparison with the infrared factorisation
formula discussed in this section. The hard function is provided
in appendix \ref{3LoopPredictOrtho} in an orthonormal color basis
in the $t$-channel, while in appendix \ref{3LoopPredictTrace} 
we provide the same quantity in the ``trace'' basis commonly 
used in literature, see Ref.~\cite{Naculich:2011ep}. 

For completeness, we end this section 
by quoting the infrared factorisation result 
for the N$^3$LL coefficient of the hard function, 
namely, $\Hhard^{(3,0)}$.
This result relies on the 3-loop soft anomalous dimension described in \eqn{quadrupole4} but not on BFKL theory.
One has 
\bea \label{AmpCoeff30tilde} \nn
\Hhard^{(3,0)} &=&  \Mreduced^{(3,0)} 
-  \frac{\pi^2}{2} (K^{(1)})^2 (\Tsu)^2  \Mreduced^{(1,0)}  
- \pi^2 K^{(1)} K^{(2)} (\Tsu)^2  \Mreduced^{(0)}   \\ \nn
&&-\, \frac{1}{6\eps}
\bigg[ [\Tsu,[\Tt, \Tsu]] \frac{\zeta_5 -4 \zeta_2 \zeta_3}{4} \\ \nn
&&\hspace{1.0cm}-\frac{\zeta_5+2\zeta_2\zeta_3}{8} \Big\{ f^{abe}f^{cde} 
\Big[ \{\T_t^a, \T_t^d \}  \Big(\{\T_{s-u}^b, \T_{s-u}^c \} 
+ \{\T_{s+u}^b, \T_{s+u}^c \} \Big) \\ \nn
&&\hspace{4.0cm}+\,   \{\T_{s-u}^b, \T_{s-u}^c \}
\{\T_{s+u}^b, \T_{s+u}^c \} \Big] 
-\frac{5}{8} C_A^2 \Tt \Big\} \bigg] \Mreduced^{(0)}    \\ \nn
&& - i\pi \Big\{ K^{(1)} \Tsu \Mreduced^{(2,0)} 
+ K^{(2)} \Tsu \Mreduced^{(1,0)}  
+ 6 K^{(3)} \Tsu \Mreduced^{(0)}  \\ 
&&\hspace{1.0cm}
- \pi^2 (K^{(1)})^3 (\Tsu)^3 \, \Mreduced^{(0)}
-\frac{11\zeta_4}{24\eps} [\Tt,[\Tt,\Tsu]] \, \Mreduced^{(0)} \Big\}.
\eea
This result is interesting on its 
own, because it provides the structure of infrared 
divergences at three loops, for a $2\to 2$ scattering 
amplitude in the high energy limit, including the 
quadrupole correction calculated in 
\cite{Almelid:2015jia}. The explicit structure 
can be obtained in the orthonormal color basis 
in the $t$-channel defined in \eqn{cgggg}, or in 
the ``trace'' basis defined in \eqn{cTrace}, by 
substituting the color operators with their explicit 
matrix representations in that basis, which are 
also provided in the appendices~\ref{3LoopPredictOrtho}
and~\ref{3LoopPredictTrace}. The structure of 
infrared singularities in \eqn{AmpCoeff30tilde} 
agrees with the calculation of gluon gluon 
scattering at three loops in ${\cal N}=4$ SYM
presented in \cite{Henn:2016jdu}.
\Eqn{AmpCoeff30tilde} is however more 
general, as it predicts the infrared structure 
for any $2\to2$ scattering amplitude in QCD, 
thus including also quark-quark and quark-gluon 
scattering.

\subsubsection*{Three-loop gluon Regge trajectory}

Finally, let us state the precise relation between the 
three-loop ``gluon Regge trajectory'' and the logarithmic
terms in the three-loop amplitude. Starting from three loops 
the ``gluon Regge trajectory'' is scheme-dependent. In 
this paper we pragmatically defined it to be the one-to-one 
matrix element of the Hamiltonian, $\alpha_{g}(t) = -H_{1\to 1}/\CA$,
in the scheme defined by \eqn{3WtoW-Wto3W} where 
states corresponding to a different number of Reggeon are 
orthogonal, as discussed following \eqn{3loop_odd_bfkl}. 
This can be related to fixed-order amplitudes by taking the 
logarithm of the reduced amplitude projected onto the 
signature-odd adjoint channel. When projected onto 
that channel, the full amplitude and reduced amplitude 
defined in \eqn{Mreduced} differ by a simple multiplicative 
factor whose logarithm is linear in $L$. Therefore, evaluating 
the prediction \eqn{three_loops_regge} in the adjoint 
representation using the matrices given appendix 
\ref{3LoopPredictOrtho}, we find
\be
\label{traj_relation}
 \log\frac{\mathcal{M}^{[8_a]}_{gg\to gg}}{\mathcal{M}^{(0)[8_a]}_{gg\to gg}}
= L \bigg\{- H_{1\to 1}(t) + \left(\frac{\alpha_s}{\pi}\right)^3\pi^2
\Big[N_c\Big(-2R^{(3)}_A+2 R^{(3)}_B\Big)+N_c^3 R^{(3)}_C\Big]\bigg\}
+ {\cal O} (L^0,\alpha_s^4),
\ee
where the constants $R_A^{(3)}$, $R_B^{(3)}$, $R_C^{(3)}$ 
are given in \eqn{R3def}.

While this paper was in preparation, a remarkable calculation 
of the non-planar three-loop gluon-gluon amplitude in 
$\mathcal{N}=4$ SYM appeared~\cite{Henn:2016jdu}, 
which yields, in terms of the $\MS$ coupling at scale $-t$, 
\bea\label{Henn_result} 
&&\log\frac{\mathcal{M}^{[8_a],\, \mathcal{N}=4}_{gg\to gg}}{\mathcal{M}^{(0)[8_a]}_{gg\to gg}}\bigg|_{L} =N_c\left[
\frac{\as}{\pi} k_1 +  \left(\frac{\as}{\pi}\right)^2 k_2 +  \left(\frac{\as}{\pi}\right)^3 k_3 +\cdots\right]
\eea
with
\bea\label{Henn_result_coef} 
k_1&=&\,  
\frac{1}{2 \eps} - \eps \frac{\zeta_2}{4} - \eps^2 \frac{7}{6}\zeta_3 
- \eps^3  \frac{47}{32}\zeta_4 +\eps^4\bigg(\frac{7}{12}\zeta_2\zeta_3 
- \frac{31}{10} \zeta_5\bigg) +{\cal O}(\eps^5)  \\ \nn  
k_2&=& \, N_c \bigg[-\frac{\zeta_2}{8}\frac{1}{\eps} -\frac{\zeta_3}{8} 
- \eps \frac{3}{16}\zeta_4 +\eps^2 \bigg(\frac{71}{24}\zeta_2 \zeta_3 + \frac{41}{8}\zeta_5 \bigg) 
+{\cal O}(\eps^3) \bigg] \\ \nn
k_3&=&\, 
 N_c^2 \bigg[\frac{11\zeta_4}{48}\frac{1}{\eps}+\frac{5}{24}\zeta_2\zeta_3
+\frac14\zeta_5+{\cal O}(\eps)\bigg] +\,  \bigg[\frac{\zeta_2}{4}\frac{1}{\eps^3}
-\frac{15\zeta_4}{16}\frac{1}{\eps} -\frac{77}{4}\zeta_2\zeta_3+{\cal O}(\eps) \bigg]\,.
\eea
Using (\ref{traj_relation}) we are therefore 
able to obtain, in this theory, the ``trajectory'' 
$\alpha_g(t)N_c = - H_{1\to 1}$ to three loop:
\be\begin{aligned} \label{3loop_traj}
-H_{1\to 1}^{{\cal N}=4 \, \rm SYM} &= N_c\left[
\frac{\as}{\pi} \alpha_g^{(1)}|_{{\cal N}=4 \, \rm SYM} +  \left(\frac{\as}{\pi}\right)^2 \alpha_g^{(2)}|_{{\cal N}=4 \, \rm SYM}
+ \left(\frac{\as}{\pi}\right)^3 \alpha_g^{(3)}|_{{\cal N}=4 \, \rm SYM} +\cdots\right]
\end{aligned}\ee
with the first two coefficients, $\alpha_g^{(1)}|_{{\cal N}=4 \, \rm SYM}=k_1$ and $\alpha_g^{(2)}|_{{\cal N}=4 \, \rm SYM}=k_2$ 
given in eq.~(\ref{Henn_result_coef}),  while the 
three-loop one given instead by 
\be \label{ReggeTrajectory3N4SYM}
\alpha_g^{(3)}|_{{\cal N}=4 \, \rm SYM}=  N_c^2 \bigg[-\frac{\zeta_2}{144}\frac{1}{\eps^3}
+\frac{49\zeta_4}{192}\frac{1}{\eps}+\frac{107}{144}\zeta_2\zeta_3
+\frac{\zeta_5}{4} + {\cal O}(\eps)\bigg] +  N_c^0\bigg[0+{\cal O}(\eps)\bigg]\, .
\ee

It is important to stress that, even though to three loop 
accuracy the adjoint amplitude may \emph{look} like
a Regge pole, e.g. a pure power-law, it is actually not: 
starting from two-loops it is really a sum of multiple powers.
Simply exponentiating the exponent defined by \eqn{Henn_result} 
would predict a definitely incorrect four-loop amplitude.
The correct, predictive,  procedure is to exponentiate 
the action of the Hamiltonian following 
\eqn{amp_from_inner_product}.  With the 
``trajectory'' \eqn{3loop_traj} now fixed, this 
procedure will not require any new parameter 
for the odd amplitude at NNLL to all loop orders.

Finally, we comment on the fact that the trajectory 
of \eqn{3loop_traj}, minus single-poles from the 
cusp anomalous dimension, is not finite. Superficially, 
this would seem to contradict the prediction of 
Ref.~\cite{Korchemskaya:1996je}. However, it is important 
to stress that $\alpha_g^{(3)}$ is not physically 
observable by itself and in the present BFKL framework, 
it depends on an arbitrary choice of scheme used to 
separate one- and three-Reggeon contributions.  
As explained below \eqn{D2s}, it is likely that our 
(arbitrary) choice to force the physics into a somewhat peculiar 
basis, in which multi-Reggeon states are orthogonal, 
is causing these spurious divergences in the 
intermediate quantity $H_{1\to 1}$. In fact this 
can be seen clearly in the planar limit, where 
general arguments show that in the $U$-basis 
the evolution is trivial and the amplitude is a pure 
Regge pole \cite{Caron-Huot:2013fea}, whereas 
in the present $W$-basis this pole is split between 
$1\to 1$, $1\to 3$ and $3\to 3$ transitions. Thus our
$H_{1\to 1}$, even in the planar limit, is \emph{not} 
equal to the position of this pure Regge pole.

Despite the not entirely satisfactory properties of the 
basis we used with regards to the simplicity of the large-$N_c$ limit, nor 
the relation between the singularities of the trajectory and the
cusp anomalous dimension, it is important to stress 
that the basis is well-defined and sufficient to provide a fully 
predictive framework to all loop orders.  A non-trivial 
confirmation is the fact that the $10+\overline{10}$ amplitude 
component \eqn{H31RealEps0Components10} is predicted 
correctly.  Furthermore, one would expect the ambiguities 
from the choice of basis described below in 
\eqn{3WtoW-Wto3W} to be proportional to $\CA^3$, which
is completely consistent with the fact that the $N_c^0$ part 
of \eqn{ReggeTrajectory3N4SYM} is finite. In fact, we see 
that the subleading color term proportional to $N_c^0$ is zero, 
up to ${\cal O}(\eps)$. This is an interesting result, which would 
be important to understand further, especially in light of the integrability 
properties of the planar amplitude~\cite{Lipatov:1993yb,Faddeev:1994zg}.

\section{Conclusions} 
\label{conclusion}

In this paper we have analyzed parton-parton scattering 
in gauge theories in the high-energy limit (Regge limit), 
pushing the accuracy to the next-to-next-to-leading 
logarithmic order. Our main tool has been BFKL theory, 
or more precisely its modern formulation as an effective 
theory of Wilson lines reviewed in section 2. An important 
observation is that many terms at this order can be 
fully predicted using only leading-order ingredients.
These terms are distinguished, for example, by their color factors, 
and this paper has focused on such terms. Our predictions 
provide stringent constraints that the Regge limit of three-loop 
$2\to 2$   QCD amplitudes must satisfy. Specifically, the odd 
reduced amplitude is predicted in \eqn{three_loops_regge} 
to all order in $\epsilon$.

An interesting feature of the Regge limit is the reduction to 
a two-dimensional effective theory. Technically, this dramatically 
simplifies the loop integrals, and indeed the most complicated 
integral we needed in this paper is the standard bubble integral
in \eqn{bubbleint}.  The main work is reduced to the bookkeeping 
of color factors.

The NNLL amplitude is conceptually interesting from the 
BFKL perspective because it exhibits a new phenomenon: 
the mixing between one- and three- Reggeon states, both 
contributing to the odd part of the amplitude. To deal with 
this we used the symmetry property of the Hamiltonian, 
\eqn{symmetry-of-H}, also known as target-projective duality, 
to obtain the $3\to 1$ terms in the Hamiltonian from the 
${1\to 3}$ terms. This is the first time that this symmetry 
property is tested quantitatively. The tests described below 
can therefore be viewed as a nontrivial check of this symmetry.

As a consequence of the mixing between one- and three- 
Reggeon states, starting at NNLL the gluon Regge pole is 
not physically distinct from the Regge cut. In particular, in 
the $t$-channel colour flow basis, the antisymmetric octet 
color component receives contributions from both the pole 
corresponding to $1\to 1$ Reggeon transition and the $3\to 3$ 
as well as the $1\to 3$ and $3\to 1$ cut components.
In general, using this formalism one may compute the signature odd NNLL 
$2\to 2$ amplitude in QCD to all loop orders up to a single presently 
unknown parameter, the three-loop gluon Regge trajectory.
The other color component of the odd amplitude, $10+\overline{10}$, 
is entirely determined by the cut contributions, and hence it 
is fully predicted already, see  \eqn{H31RealEps0Components10}.    
Because of the mentioned mixing, the result in either channel does \emph{not} 
take the form of a single exponential (except in the planar limit), rather what exponentiates 
is the Hamiltonian in \eqn{Regge-odd-Even-Amplitude}.

Our results have been tested in two ways. First, the infrared 
divergent part of the result is in agreement with predictions 
from the general theory, including the recently computed 
three-loop soft anomalous dimension \cite{Almelid:2015jia,Gardi:2016ttq}.
Conversely, our results provide a valuable test of the latter.
Second, our predictions, which are general and valid in any 
theory, turn out to agree with a recent explicit three-loop 
calculation in $\mathcal{N}=4$ super Yang-Mills. This 
comparison also allows us to fix in this theory the one 
free parameter we have left, the three-loop gluon Regge 
trajectory in \eqn{3loop_traj}, thereby making the formalism 
fully predictive at higher loop orders.
Our predictions for the odd part of the three-loop amplitude 
are summarised in appendices~\ref{3LoopPredictOrtho} 
and~\ref{3LoopPredictTrace} in a $t-$channel orthonormal 
basis and in a trace basis, respectively. These explicit 
results may be used as a stringent test of future multiloop 
amplitude computations.

To complete the NNLL description of $2\to 2$ amplitudes, 
the only missing ingredient is in the even sector, namely 
the NLO impact factor to two gluons, which would thus 
be interesting to compute in the future. More generally, 
we have seen that the BFKL theory is consistent with 
infrared exponentiation, such that the hard function
$\mathcal{H}$ (see \eqn{IRfacteq}) is finite; it would 
thus be interesting to understand how to setup the 
BFKL calculation of $\mathcal{H}$ in a manifestly 
finite way, which would alleviate the need to 
$\eps$-expand all intermediate quantities.
This would make it possible to exploit the 
integrability of the Hamiltonian in two dimensions 
\cite{Lipatov:1993yb,Faddeev:1994zg}.

\vspace{30pt}
\noindent
{\bf Note Added:} While this paper was being 
completed, partially overlapping results were 
announced in Ref.~\cite{Fadin:2016wso}.

\vspace{20pt}
\acknowledgments

SCH's research in its early stages was partly funded by the 
Danish National Research Foundation (DNRF91). EG's 
research is supported by the STFC Consolidated Grant 
``Particle Physics at the Higgs Centre.'' LV's research is 
supported by the People Programme (Marie Curie Actions) 
of the European Union's Horizon 2020 Framework Programme 
H2020-MSCA-IF-2014 under REA grant No.~656463 -- ``Soft Gluons''. 
SCH thanks the Higgs Centre for Theoretical Physics for 
hospitality in the early stages this work. LV thanks the Niels 
Bohr Institute, Copenhagen, for hospitality during the completion 
of this work. In addition, EG and LV would like to thank the Erwin Schr\"odinger 
International Institute for Mathematics and Physics (ESI) in Vienna for hospitality 
during the program ``Challenges and Concepts for Field Theory and Applications 
in the Era of LHC Run-2'', during which part of this work was done. 


\appendix

\section{Anomalous dimensions, 
renormalization group factors and Regge trajectory}
\label{GammaAndZ}

We write the $\as$ expansion of the 
anomalous dimension (in the $\MS$ scheme) as
\be
\gamma_i(\as) = \sum_{k=1}^{\infty} 
\gamma_i^{(k)} \left(\frac{\as}{\pi}\right)^k.
\ee
With this notation, the coefficients of the 
cusp anomalous dimension (with the quadratic 
Casimir factor $C_i$ removed) 
read~\cite{Korchemsky:1985xj,Moch:2004pa,Grozin:2014hna}
\bea\nn \label{gammaK}
{\gamma}_K^{(1)} & = & 2, \\ \nn
{\gamma}_K^{(2)} & = & 
\left( \frac{67}{18} - \zeta_2 \right) C_A 
- \frac{10}{9} T_R \nf,  \\ \nn
{\gamma}_K^{(3)} & = & 
\frac{C_A^2}{96} \left( 490 - \frac{1072}{3} \zeta_2 
+ 88 \zeta_3 + 264 \zeta_4 \right) 
+ \frac{C_F T_R \nf}{32} \left( - \frac{220}{3} + 64 \zeta_3 \right) \\ 
&&\hspace{0.0cm} + \, \frac{C_A T_R \nf}{96} 
\left(-\frac{1672}{9} + \frac{320}{3} \zeta_2 - 224 \zeta_3 \right) 
- \frac{2 T_R^2 \nf^2}{27},
\eea 
where $T_R = 1/2$. The coefficients of the quark 
and gluon anomalous dimension are given 
by~\cite{Moch:2005tm,Gehrmann:2010ue}
\beqa \label{gammaq} \nn
\gamma_{q}^{(1)} & = & - \frac{3}{4} \, C_F,   \\ \nn
\gamma_{q}^{(2)} & = & \frac{C_F^2}{16} 
\left( - \frac{3}{2} + 12 \zeta_2 - 24 \zeta_3 \right)  \\
&& + \, \frac{C_A C_F}{16} \left( - \frac{961}{54} - 11 \zeta_2 + 26 \zeta_3 \right)
+\frac{C_F T_R \nf}{16} \left( \frac{130}{27} + 4 \zeta_2 \right),  
\eea
and
\beqa \label{gammag} \nn
\gamma_{g}^{(1)} & = & -  \frac{b_0}{4}, \\
\gamma_{g}^{(2)} & = & \frac{C_A^2}{16} 
\left( - \frac{692}{27}+ \frac{11}{3} \zeta_2 + 2 \zeta_3 \right) 
+  \frac{C_A T_R \nf}{16} \left( \frac{256}{27} - \frac{4}{3} \zeta_2 \right) 
+ \frac{C_F T_R \nf}{4}.
\eeqa
The scalar factors in \eqn{Zidef} start at ${\cal O}\left(\as^0\right)$. 
In terms of the coefficients in \eqns{gammaq}{gammag}, 
and setting $\mu^2=-t$, they are:
\bea \nn 
Z_i^{(0)} & = & 1,  \\ \nn
Z_i^{(1)} & = & -\, C_i\, {\gamma}_K^{(1)} \frac{1}{4\eps^2} 
+ \frac{\gamma_i^{(1)}}{\eps}, \\ 
Z_i^{(2)} & = & C^2_i \left({\gamma}_K^{(1)}\right)^2 \frac{1}{32\eps^4}\, 
+C_i\, \Bigg[ \frac{1}{\eps^3}\frac{{\gamma}_K^{(1)} }{4}\left(\frac{3b_0 }{16} 
- \gamma_i^{(1)}\right) -\,\frac{1}{\eps^2}\frac{ {\gamma}_K^{(2)}}{16} 
\Bigg] 
\\ \nn &&\hspace{0.5cm}
+ \frac{1}{\eps^2} 
\frac{\gamma_i^{(1)}}{2} \left(\gamma_i^{(1)}-\frac{b_0}{4}\right)
+\frac{\gamma_i^{(2)}}{2\eps}. \nn
\eea
Finally, we quote the one- and two-loop gluon Regge trajectory
(divided by $\CA$) in terms of the coupling at scale $\mu^2=-t$ 
\cite{Fadin:1995xg,Fadin:1996tb,Fadin:1995km,Blumlein:1998ib}:
\begin{align}
\label{two_loop_gluon_trajectory}
\begin{split}
\alpha^{(1)}_g(t) &= \frac{\rGamma}{2\eps}, \\ 
\alpha^{(2)}_g(t) &= -\frac{b_0}{16\eps^2} +\frac{1}{8\eps}
\left[\left(\frac{67}{18}-\zeta_2\right)\CA-\frac{10T_R\nf}{9}\right] \\ 
&\hspace{0.5cm}+ \CA\left(\frac{101}{108}-\frac{\zeta_3}{8}\right) 
- \frac{7\, T_R\nf}{27} + {\cal O}(\eps)\,. 
\end{split}
\end{align}    

\section{The hard function for gluon-gluon scattering in an orthonormal $t$-channel color basis}
\label{3LoopPredictOrtho}

Predictions for the infrared renormalised amplitude
(hard function) based on the Regge theory developed 
in this paper have been presented in section \ref{dip_comparison}.
These predictions have been given in color space notation, 
i.e., writing the amplitude in terms of color operators 
acting on a vector amplitude. Predictions for the 
single components can be obtained by choosing a 
specific color basis. In this appendix and the next we 
provide explicit results within two color basis widely considered 
in literature. Here we focus on the orthonormal color basis in 
the $t$-channel, which, as discussed in the main text, is 
particularly useful to highlight the factorisation properties 
of the amplitude in the high-energy limit. In the next 
appendix we will focus on a ``trace'' basis, which 
has been typically used in the context of multi-loop 
calculations. 

Before proceeding, we stress once more that
the calculations performed on this paper are based solely on the
BFKL evolution at leading order. The corrections $D_i$ to the impact factors,
defined in \eqns{NLO_wavefunction}{NNLO_wavefunction}, as well as the
higher-loop corrections to the gluon Regge trajectory $\alpha_g$ 
(more precisely $H_{1\to 1}$ in the scheme \eqn{3WtoW-Wto3W})
are therefore kept in this appendix as free parameters.
Their values can be obtained by matching with fixed-order
amplitudes and are listed in appendix \ref{TrajectoryImpact}. 

\subsection*{Definition of the $t$-channel color basis}

We consider gluon-gluon scattering with external legs 
labelled as in figure~\ref{tree-t}. Within SU$(N_c)$, an 
orthonormal color basis in the $t$-channel can be 
obtained decomposing the color representations 
$8\otimes 8$ of legs one and four into the direct sum 
$1\oplus 8_s \oplus 8_a \oplus 10+ \overline{10} \oplus 27 \oplus 0$.
Such basis has been provided in \cite{DelDuca:2014cya}, 
and we repeat it here for the reader convenience:
\beqa \label{cgggg}
  c^{[1]} & = & \frac{1}{N_c^2 - 1} \, {\delta^{a_4}}_{a_1} \, {\delta^{a_3}}_{a_2} \, ,
  \nonumber \\
  c^{[8_s]} & = & \frac{N_c}{N_c^2 - 4} \, \frac{1}{\sqrt{N_c^2 - 1}} \, 
  d^{\, a_1 a_4 b} \, d^{\, a_2 a_3}_{\phantom{\, a_2 a_3} b} \, , \nonumber \\
  c^{[8_a]} & = & \frac{1}{N_c} \, \frac{1}{\sqrt{N_c^2 - 1}} \, 
  f^{\, a_1 a_4 b} \, f^{\, a_2 a_3}_{\phantom{\, a_2 a_3} b} \, , \nonumber \\
  c^{[10 + \overline{10}]} & = & \sqrt{\frac{2}{(N_c^2 - 4)(N_c^2 - 1)}}
  \left[ \frac{1}{2} \left({\delta^{a_1}}_{a_2} \, {\delta^{a_3}}_{a_4} - 
  {\delta^{a_3}}_{a_1} \, {\delta^{a_4}}_{a_2} \right) - \frac{1}{N_c}
  f^{\, a_1 a_4 b} \, f^{\,a_2a_3}_{\phantom{\, a_2 a_3} b} \right] \, , \nonumber \\
  c^{[27]} & = & \frac{2}{N_c \sqrt{(N_c + 3)(N_c - 1)}} \, \bigg[ - 
  \frac{N_c + 2}{2 N_c (N_c + 1)} \, {\delta^{a_4}}_{a_1} \, {\delta^{a_3}}_{a_2}
  \nonumber \\
  & & + \, \frac{N_c + 2}{4 N_c} \, \big( {\delta^{a_1}}_{a_2} \, {\delta^{a_3}}_{a_4}
  + {\delta^{a_3}}_{a_1} \, {\delta^{a_4}}_{a_2} \big) - \frac{N_c + 4}{4( N_c + 2)} \,
  d^{\, a_1 a_4 b} \, d^{\, a_2 a_3}_{\phantom{\, a_2 a_3} b} \nonumber \\
  & & + \, \frac{1}{4} \, \big( d^{\, a_1 a_2 b} \, d^{\, a_3 a_4}_{\phantom{\, a_2 a_3} b}
  + d^{\, a_1 a_3 b} \, d^{\, a_2 a_4}_{\phantom{\, a_2 a_3} b} \big) \bigg] \, , \\
  c^{[0]} & = & \frac{2}{N_c \sqrt{(N_c - 3)(N_c + 1)}} \, \bigg[ 
  \frac{N_c - 2}{2 N_c (N_c - 1)} \, {\delta^{a_4}}_{a_1} \, {\delta^{a_3}}_{a_2} \nonumber \\
  & & + \, \frac{N_c - 2}{4 N_c} \, \big( {\delta^{a_1}}_{a_2} \, {\delta^{a_3}}_{a_4}
  + {\delta^{a_3}}_{a_1} \, {\delta^{a_4}}_{a_2} \big) + \frac{N_c - 4}{4 (N_c - 2)} \,
  d^{\, a_1 a_4 b} \, d^{\, a_2 a_3}_{\phantom{\, a_2 a_3} b} \nonumber \\
  & & - \, \frac{1}{4} \big( d^{\, a_1 a_2 b} \, d^{\, a_3 a_4}_{\phantom{\, a_2 a_3} b}  
  + d^{\, a_1 a_3 b} \, d^{\, a_2 a_4}_{\phantom{\, a_2 a_3} b} \big) \bigg] \, .
  \nonumber
\eeqa
We treat the two decuplet representations together, since they 
always contribute to the amplitude with the same coefficients. The 
tensors $c^{[8_a]}$ and $c^{[10 + \overline{10}]}$ are odd under the 
exchanges $a_1 \leftrightarrow a_4$ and $a_2 \leftrightarrow a_3$, 
while $c^{[1]}$, $c^{[8_s]}$, $c^{[27]}$ and $c^{[0]}$ are even. 
The last representation does not contribute for $N_c = 3$, 
since its dimensionality is given by
\beq
  {\rm dim} \left[ \, {\bf 0} \, \right] \, = \, \frac{N_c^2 (N_c - 3)(N_c + 1)}{4} \, ,
\label{dim0}
\eeq 
and it vanishes for SU(3). In the orthormal basis defined by \eqn{cgggg} (in that order), 
the diagonal matrix ${\bf T}^2_t$ evaluates to 
\beq
(\Tt)_{gg} \, = \, {\rm diag}
\Big[ 0, \, N_c , \, N_c, \, 2 N_c, \, 2 (N_c + 1), \, 
2 (N_c - 1) \Big] \, ,
\label{Ttgggg}
\eeq
while ${\bf T}_{s-u}$ can be calculated starting from 
${\bf T}_{s}$ provided in \cite{DelDuca:2014cya}, 
by exploiting the relation $\Tt+\Ts+\Tu = C_{\rm tot}$. 
${\bf T}_{s-u}$ is symmetric and traceless, 
and reads 
\beq
  (\Tsu)_{gg} \, = \, \left[
    \begin{array}{cccccc}
    0 & 0 & {\cal T}_{1, 8_a} & 0 & 0 & 0 \\
    0 & 0 & {\cal T}_{8_s, 8_a} & {\cal T}_{8_s, 10} & 0 & 0 \\
    {\cal T}_{1, 8_a} & {\cal T}_{8_s, 8_a} & 0 & 0 & {\cal T}_{8_a, 27} & {\cal T}_{8_a, 0} \\
    0 & {\cal T}_ {8_s, 10} & 0 & 0 & {\cal T}_{10, 27} & {\cal T}_{10, 0} \\
    0 & 0 & {\cal T}_{8_a, 27} & {\cal T}_{10, 27} & 0 & 0 \\
    0 & 0 & {\cal T}_{8_a, 0} & {\cal T}_{10, 0} & 0 & 0
    \end{array}
  \right] \, ,
\label{Tsgggg}
\eeq
where
\beqa
\label{elmatgggg}
  {\cal T}_ {1, 8_a} & = & - \frac{2 N_c}{\sqrt{N_c^2 - 1}} \, , \qquad 
  {\cal T}_ {8_s, 8_a} \, = \, - \frac{N_c}{2} \, , \qquad {\cal T}_ {8_s, 10} \, = \, 
  - N_c \sqrt{\frac{2}{N_c^2 - 4}} \, , \nonumber \\  
  {\cal T}_ {8_a, 27} & = & - \sqrt{\frac{N_c + 3}{N_c + 1}} \, , \qquad 
  {\cal T}_ {8_a, 0} \, = \, - \sqrt{\frac{N_c - 3}{N_c - 1}} \, ,  \\
  {\cal T}_ {10, 27} & = & - \sqrt{\frac{(N_c + 3)(N_c + 1)(N_c - 2)}{2 (N_c + 2)}} \, , \qquad 
  {\cal T}_ {10, 0}  =  - \sqrt{\frac{(N_c - 3)(N_c - 1)(N_c + 2)}{2 (N_c - 2)}} \, .
  \nonumber
\eeqa
Similarly, for the gluon-gluon amplitude we obtain 
also the color matrix representing the color operator 
associated to the constant term $\zeta_5 + 2\zeta_2\zeta_3$ 
in \eqn{quadrupole4}:
\beqa \label{TConst} \nn
\frac{1}{2}\bigg\{ f^{abe}f^{cde} 
\bigg[ \{\T_t^a, \T_t^d \}  \Big(\{\T_{s-u}^b, \T_{s-u}^c \} 
+ \{\T_{s+u}^b, \T_{s+u}^c \} \Big)  \\ 
&&\hspace{-6.0cm}+\,   \{\T_{s-u}^a, \T_{s-u}^d \}
 \{\T_{s+u}^b, \T_{s+u}^c \} \bigg] -\frac{5}{8} C_A^2 \Tt \bigg\}_{gg} \\ \nn
&&\hspace{-9.0cm}= \, \left[
\begin{array}{cccccc}
 2 N_c {\cal T}_{1, 8_a}^2 & 
 N_c {\cal T}_{1, 8_a} {\cal T}_{8_s, 8_a} & 
 0 & 
 0 & 
 -2 {\cal T}_{1, 8_a} {\cal T}_{8_a, 27} & 
 2 {\cal T}_{1, 8_a} {\cal T}_{8_a, 0} \\
 N_c {\cal T}_{1, 8_a} {\cal T}_{8_s, 8_a} & 
 2 N_c {\cal T}_{8_s, 10}^2 & 
 0 & 
 0 & 
 {\cal T}_{8_a, 27} {\cal T}_{A} & 
 {\cal T}_{8_a, 0} {\cal T}_{B} \\
 0 & 
 0 & 
 14 N_c & 
 -\frac{N_c^2}{{\cal T}_{8_s, 10}} & 
 0 & 
 0 \\
 0 & 
 0 & 
 -\frac{N_c^2}{{\cal T}_{8_s, 10}} &
 5 N_c & 
 0 & 
 0 \\
 -2 {\cal T}_{1, 8_a} {\cal T}_{8_a, 27} & 
 {\cal T}_{8_a, 27} {\cal T}_{A} & 
 0 & 
 0 & 
 {\cal T}_{C} & 
 -2 N_c {\cal T}_{8_a, 27} {\cal T}_{8_a, 0} \\
  2 {\cal T}_{1, 8_a} {\cal T}_{8_a, 0} & 
  {\cal T}_{8_a, 0} {\cal T}_{B} &
  0 &
  0 & 
  -2 N_c {\cal T}_{8_a, 27} {\cal T}_{8_a, 0} & 
  {\cal T}_{D}
\end{array}
\right] \, ,
\eeqa
where we have introduced the 
color factors 
\bea \nn
{\cal T}_{A}  = \frac{8N_c + 6 N_c^2 - N_c^3}{2(2+N_c)} && \quad 
{\cal T}_{B}  = \frac{8N_c - 6 N_c^2 - N_c^3}{2(2+N_c)}, \\ 
{\cal T}_{C}  = \frac{26N_c + 31 N_c^2 +10 N_c^3 + N_c^4}{(1+N_c)(2+N_c)} && \quad 
{\cal T}_{D}  = \frac{26N_c - 31 N_c^2 +10 N_c^3 - N_c^4}{(1+N_c)(2+N_c)}.
\eea

For quark-quark scattering, the representations 
are more limited and we let similarly \cite{DelDuca:2014cya}
\be
c^{[1]}_{qq} =  \frac{1}{N_c}\delta^{a_4}{}_{a_1}\delta^{a_3}{}_{a_2}\,,\qquad
c^{[8_a]}_{qq} = \frac{2}{\sqrt{N_c^2-1}}(\T^b)^{a_4}{}_{a_1}(\T^b)^{a_3}{}_{a_2}\,,
\ee
and one finds
\be
 (\Tt)_{qq} = \, {\rm diag} \Big[ 0, \, N_c\Big] \,,\qquad
 (\Tsu)_{qq} = \,\left[\begin{array}{c@{{\hspace{5mm}}}c} 0&
 \frac{\sqrt{N_c^2-1}}{N_c}\\ \frac{\sqrt{N_c^2-1}}{N_c}&\frac{N_c^2-4}{2N_c}\end{array}\right].
\label{Tqq}
\ee
Finally for quark-gluon scattering we have
\bea
c^{[1]}_{qg} &=& \frac{1}{\sqrt{N_c(N_c^2 - 1)}}\delta^{a_4}{}_{a_1} \, {\delta^{a_3}}_{a_2}\,,\qquad
c^{[8_s]}_{qg} = \sqrt{\frac{2N_c}{(N_c^2-4)(N_c^2 - 1)}}(T^b)^{a_4}{}_{a_1} \,d^{\, a_2 a_3b}\,, \nn\\
c^{[8_a]}_{qg} &=& \sqrt{\frac{2}{N_c(N_c^2 - 1)}}(T^b)^{a_4}{}_{a_1} \,i f^{\, a_2 a_3b}\,,
\eea
with
\be
 (\Tt)_{qg} = \, {\rm diag} \Big[ 0, \, N_c,\,N_c \Big] \,,\qquad
 (\Tsu)_{qg} = \,\left[\begin{array}{c@{{\hspace{5mm}}}c@{{\hspace{5mm}}}c} 
 0&0&-\sqrt{2}\\0&0&-\tfrac12\sqrt{N_c^2-4}\\-\sqrt{2}&-\tfrac12\sqrt{N_c^2-4}&0\end{array}\right].
\label{Tqg}
\ee
For antiquark scattering we define the same color structures.
Note that in the quark-quark case the signature in the adjoint 
channel is not determined by the color projection and can only 
be determined by comparing the quark and antiquark amplitudes.
In the quark-gluon case the structures have definite signatures 
(respectively even, even, odd) due to Bose symmetry on the 
gluon side.

\subsection*{The hard function for gluon-gluon scattering}

Let us now present explicit results for the hard function 
components in the orthonormal $t$-channel basis defined 
above.  We restrict the discussion here to the gluon-gluon 
amplitude, since it should hopefully be clear how the 
formulas presented below are obtained from the formulas 
in section \ref{dip_comparison} by evaluating the color 
operators. We decompose the hard function according 
to \eqn{AmpColor}, namely
\be\label{HardColor}
{\cal H}_{gg\to gg}(s,t) =\sum_i \, c^{[i]} \, {\cal H}^{[i]}(s,t). 
\ee 
Within the orthonormal basis \eqn{cgggg}
the tree-level hard function in \eqn{treeConvention}
reads 
\bea \label{H0Components}\nn
\Hhard^{(0),[1]} &=& \Hhard^{(0),[8_s]} = \Hhard^{(0),[10+\overline{10}]}
= \Hhard^{(0),[27]} = \Hhard^{(0),[0]} = 0, \\
\Hhard^{(0),[8_a]} &=& -2\frac{s}{t} N_c \sqrt{N_c^2 -1}.  
\eea

In section \ref{dip_comparison} we have 
presented results up to three loops, but the 
Regge theory develop in section \ref{Regge}
allows one to calculate higher orders, too. 
For feature reference, therefore, we expand 
here the amplitude in powers of $\eps$, consistently
as it would be needed for a four loop calculation. Namely, 
we expand the one loop functions up to power $\eps^6$, 
the two loop ones up to $\eps^4$, and the three loops 
functions up to power $\eps^2$. 

The one loop amplitude, and more in general 
the leading logarithmic contribution can be 
expressed entirely in terms of the one-loop 
function defined in \eqn{alpha1hat}.
Up to $\eps^6$ one has 
\bea\label{alphag1hat-eps6} \nn
\hat \alpha_g^{(1)} &=& \frac{1}{2\eps}(\rGamma - 1)  =
-\frac14\zeta_2\, \eps -\frac76\zeta_3 \,\eps^2 
- \frac{47}{32} \zeta_4 \eps^3 
+ \bigg(\frac{7}{12} \zeta_z \zeta_3 - \frac{31}{10}\zeta_5\bigg) \eps^4 \\ 
&&+\,  \bigg(\frac{49}{36} \zeta_3^2 -\frac{949}{256} \zeta_6\bigg)  \eps^5 
+ \bigg(\frac{31}{20} \zeta_2 \zeta_5 + \frac{329}{96} \zeta_3 \zeta_4 
-\frac{127}{14} \zeta_7\bigg) \eps^6 + {\cal O}(\eps^7).
\eea
In term of this function, the leading-logarithmic 
amplitude in components reads:
\bea  \label{HnnComponents}\nn 
\Hhard^{(n,n),[1]} &=& \Hhard^{(n,n),[8_s]} = \Hhard^{(n,n),[10+\overline{10}]}
= \Hhard^{(n,n),[27]} = \Hhard^{(n,n),[0]} = 0, \\ 
\Hhard^{(n,n),[8_a]} &=& - \frac{2}{n!} \, N_c^{n+1} \, \sqrt{N_c^2 -1} \,(\hat\alpha_{g}^{(1)}(t))^n \, \frac{s}{t}.
\eea
Next, we consider $\Hhard^{(1,0)}$, 
whose result has been obtained in 
\eqn{H10b}. In components one obtains
\bea \label{H10Components}  \nn
\Hhard^{(1,0),[1]} &=&  i \pi \,  4 N_c^2 \, \hat \alpha_g^{(1)} \frac{s}{t} , \\ \nn
\Hhard^{(1,0),[8_s]} &=& i \pi \, N_c^2 \sqrt{N_c^2 - 1} \, \hat \alpha_g^{(1)} \frac{s}{t} , \\ \nn
\Hhard^{(1,0),[8_a]} &=& -2 N_c  \sqrt{N_c^2 - 1} (D_i^{(1)} + D_j^{(1)}) \frac{s}{t}  \\ 
\Hhard^{(1,0),[10+\overline{10}]}  &=&  0 , \\ \nn 
\Hhard^{(1,0),[27]} &=& i \pi \, 2 N_c \sqrt{(N_c + 3)(N_c - 1)} \, \hat \alpha_g^{(1)} \frac{s}{t}, \\  \nn
\Hhard^{(1,0),[0]} &=& i \pi \, 2 N_c \sqrt{(N_c - 3)(N_c + 1)} \, \hat \alpha_g^{(1)} \frac{s}{t}.
\eea
At two loops the NLL term reads
\bea  \label{H21Components} \nn
\Hhard^{(2,1),[1]} &=& -2 i \pi \, N_c^3 \, f_a^{(2,1)} \frac{s}{t}, \\ \nn
\Hhard^{(2,1),[8_s]} &=&  i \pi\, N_c^3 \sqrt{N_c^2 - 1} \,  (\hat\alpha_{g}^{(1)}(t))^2 \frac{s}{t} , \\ \nn
\Hhard^{(2,1),[8_a]} &=& -2 N_c^2  \sqrt{N_c^2 - 1} \Big[ \hat\alpha_g^{(2)} 
+ \hat\alpha_g^{(1)} (D_i^{(1)} + D_j^{(1)}) \Big] \,  \frac{s}{t}  \\ 
\Hhard^{(2,1),[10+\overline{10}]}  &=&  0, \\ \nn
\Hhard^{(2,1),[27]} &=&  i \pi \, N_c \sqrt{(N_c + 3)(N_c - 1)}\,   
\Big[(N_c+2) f_a^{(2,1)} +4(N_c+1)  (\hat\alpha_{g}^{(1)}(t))^2 \Big] \, \frac{s}{t} , \\ \nn
\Hhard^{(2,1),[0]} &=& i \pi \, N_c \sqrt{(N_c - 3)(N_c + 1)} \,
\Big[ (N_c-2) f_a^{(2,1)}  +4(N_c-1) (\hat\alpha_{g}^{(1)}(t))^2  \Big] \, \frac{s}{t},
\eea
where we have expressed the amplitude in terms 
of the functions 
\bea \label{f21} \nn
f_a^{(2,1)} &=&  K^{(1)} (2 \hat\alpha_g^{(1)} + K^{(1)}) + \mathbb{d}_2  \\ \nn
&=& -\frac{9}{2} \zeta_3 \eps - \frac{221}{32} \zeta_4 \eps^2 
+ \bigg(\frac{47}{12} \zeta_2 \zeta_3 - \frac{63}{2} \zeta_5 \bigg)\eps^3 \\ 
&&+\, \bigg(\frac{1193}{36} \zeta_3^2 - \frac{14585}{256}\zeta_6 \bigg) \eps^4
+ {\cal O}(\eps^5).
\eea
At NNLL accuracy we are able to make predictions 
for the real component of the hard function only. Given 
that this contribution corresponds to the odd amplitude, 
it implies that this correction affects only the $8_a$ and 
$10+\overline{10}$ representation:
\bea  \label{H20Components}\nn
\RE[\Hhard^{(2,0),[1]}] &=& 0, \\ \nn
\RE[\Hhard^{(2,0),[8_s]}] &=& 0 , \\ \nn
\RE[\Hhard^{(2,0),[8_a]}] &=& - 2 N_c \sqrt{(N_c^2-1)}  
\bigg\{ D_i^{(1)}D_j^{(1)} + D_i^{(2)}+ D_j^{(2)} \\ \nn
&&\hspace{3.0cm}-\, \pi^2 \left(\frac{N_c^2}{12} R^{(2)}  
- \frac{N_c^2+24}{4} \hat R^{(2)} \right)\bigg\} \, \frac{s}{t}, \\ \nn
\RE[\Hhard^{(2,0),[10+\overline{10}]}]  &=& -3\pi^2 N_c \sqrt{2(N_c^2-4)(N_c^2-1)} 
\hat R^{(2)}  \frac{s}{t}, \\ \nn
\RE[\Hhard^{(2,0),[27]}] &=& 0 , \\
\RE[\Hhard^{(2,0),[0]}] &=& 0.
\eea
where the function $\hat R^{(2)}$ 
has been defined in \eqn{R2hat}. Explicitly, 
up to ${\cal O}(\eps^5)$ one has 
\bea \label{R2hateps5}
\hat R^{(2)} &=& \frac{3}{4}\zeta_3 \eps + \frac{67}{64} \zeta_4 \eps^2 
+\bigg(\frac{21}{4} \zeta_5 -\frac{25}{24} \zeta_2 \zeta_3  \bigg) \eps^3
+\bigg( \frac{4423}{512} \zeta_6 - \frac{463}{72} \zeta_3^2 \bigg) \eps^4 +{\cal O}(\eps^5).
\eea

At three loops, the NLL component reads
\bea  \label{H32Components}\nn
\Hhard^{(3,2),[1]} &=& i \pi \, N_c^4 \, f_a^{(3,2)} \frac{s}{t}, \\ \nn
\Hhard^{(3,2),[8_s]} &=&   i \pi \, N_c^4 \sqrt{N_c^2-1} \, (\hat\alpha_{g}^{(1)}(t))^3 \, \frac{s}{t}, \\ \nn
\Hhard^{(3,2),[8_a]} &=&  - 2 N_c^3 \sqrt{N_c^2-1} \, \hat \alpha_g^{(1)} \bigg[ \hat \alpha_g^{(2)} 
+ \frac{\hat \alpha_g^{(1)}}{2} \Big(D_i^{(1)} + D_j^{(1)}\Big) \bigg] \frac{s}{t}, \\ \nn
\Hhard^{(3,2),[10+\overline{10}]}  &=&  0 ,  \\ \nn 
\Hhard^{(3,2),[27]} &=&   i \pi \, \frac{N_c}{2} \sqrt{ (N_c-1) (N_c+3)} \, 
\Big[ (N_c+ 2)^2 f_a^{(3,2)} + 8 (N_c+1)^2  (\hat\alpha_{g}^{(1)}(t))^3 \\ \nn
&&\hspace{4.5cm}+\, (N_c +1) (N_c+2) f_b^{(3,2)} \Big]  \frac{s}{t} , \\ \nn
\Hhard^{(3,2),[0]} &=&   i \pi \, \frac{N_c}{2} \sqrt{ (N_c+1) (N_c-3)}
\Big[ (N_c-2)^2 f_a^{(3,2)} + 8 (N_c-1)^2  (\hat\alpha_{g}^{(1)}(t))^3 \\
&&\hspace{4.5cm}+\, (N_c-1) (N_c-2) f_b^{(3,2)}\Big] \frac{s}{t}, 
\eea
where we have expressed the amplitude in terms 
of the functions 
\bea\label{f32} \nn
f_a^{(3,2)} &=&  -\frac{2}{3} \bigg[ K^{(1)} \Big(3 (\hat \alpha_g^{(1)})^2 
+ 3 \hat \alpha_g^{(1)} K^{(1)} + (K^{(1)})^2\Big) - \mathbb{d}_3 \bigg]  \\ \nn 
&=& -\frac{11}{6} \zeta_3 - \frac{11}{4} \zeta_4 \eps 
+ \bigg(\frac{11}{4} \zeta_2 \zeta_3 - \frac{119}{2} \zeta_5 \bigg) \eps^2 + {\cal O}(\eps^3), \\ \nn
f_b^{(3,2)} &=& 4 \hat \alpha_g^{(1)} \Big[K^{(1)} 
\Big(2 \hat \alpha_g^{(1)} + K^{(1)} \Big) + \mathbb{d}_2 \Big] 
= \frac{9}{2} \zeta_2 \zeta_3 \eps^2 + {\cal O}(\eps^3).
\eea
Notice also that $ (\hat\alpha_{g}^{(1)}(t))^3 = {\cal O}(\eps^3)$.
Last, the real part of the NNLL term at three loops 
reads
\bea \label{H31Components}\nn
\RE[\Hhard^{(3,1),[1]}] &=& 0, \\ \nn
\RE[\Hhard^{(3,1),[8_s]}] &=& 0 , \\ \nn
\RE[\Hhard^{(3,1),[8_a]}] &=& -2 N_c^2 \sqrt{N_c^2-1} 
\Big[\hat \alpha_g^{(3)}+\hat \alpha_g^{(2)} \Big(D_i^{(1)} + D_j^{(1)} \Big)
+\hat \alpha_g^{(1)} \Big(D_i^{(2)} + D_j^{(2)} \Big) \\ \nn
&&\hspace{-1.0cm}+\, \hat \alpha_g^{(1)} D_i^{(1)} D_j^{(1)} 
-\pi^2 \Big(\frac{N_c^2+24}{4} f_a^{(3,1)} + 2 (f_b^{(3,1)}+f_c^{(3,1)}) - N_c^2 f_d^{(3,1)} \Big) \Big]
\frac{s}{t}  , \\ \nn
\RE[\Hhard^{(3,1),[10+\overline{10}]}]  &=& N_c^2 \sqrt{\frac{(N_c^2-4)(N_c^2-1)}{2}} 
\Big[ 24 f_a^{(3,1)} - 2 f_b^{(3,1)} + f_c^{(3,1)} \Big] \frac{s}{t} ,\\ \nn 
\RE[\Hhard^{(3,1),[27]}] &=& 0 , \\
\RE[\Hhard^{(3,1),[0]}] &=& 0, 
\eea
where we have expressed the amplitude in terms 
of the functions 
\bea \label{f31} \nn
f_a^{(3,1)} &=&  \frac{1}{2} \hat \alpha_g^{(1)} 
\bigg[ K^{(1)} \Big(K^{(1)} + 2 \hat \alpha_g^{(1)} \Big) +2 R^{(2)} \bigg]  \\ \nn 
&=& -\frac{3}{16} \zeta_2 \zeta_3 \eps^2 + {\cal O}(\eps^3), \\ \nn
f_b^{(3,1)} &=& \frac{1}{6} \bigg[K^{(1)} \Big( K^{(1)} (3\hat \alpha_g^{(1)} +2 K^{(1)})
+ 3 \mathbb{d}_2 \Big) +6 R_A^{(3)} \bigg] \\ 
&=& \frac{5}{12} \zeta_3 + \frac{5}{8} \zeta_4 \eps  
+ \bigg(\frac{65}{4} \zeta_5 - \frac{19}{16} \zeta_2 \zeta_3 \bigg) \eps^2 + {\cal O}(\eps^3), \\ \nn
f_c^{(3,1)} &=& \frac{1}{3} \bigg[K^{(1)} \Big( K^{(1)} (3\hat \alpha_g^{(1)} +2 K^{(1)})
+ 3 \hat \alpha_g^{(1)} (2\hat \alpha_g^{(1)} + K^{(1)}) \Big) - 6 R_B^{(3)} \bigg] \\ \nn
&=&  -\frac{1}{6} \zeta_3 - \frac{1}{4} \zeta_4 \eps 
+ \bigg(\frac{11}{2} \zeta_5 + \frac{1}{4} \zeta_2 \zeta_3 \bigg) \eps^2 + {\cal O}(\eps^3), \\ \nn
f_d^{(3,1)} &=& \frac{1}{12} \bigg[ - \hat \alpha_g^{(1)} R^{(2)}+ 12 R_c^{(3)} \bigg] \\ \nn
&=& \frac{1}{864 \eps^3 }  - \frac{5}{1152 \eps} \zeta_2 - \frac{175}{1728} \zeta_3 
- \frac{425}{3072} \eps \zeta_4 + \bigg(\frac{23}{128} \zeta_2 \zeta_3 
- \frac{99}{64} \zeta_5 \bigg) \eps^2 + {\cal O}(\eps^3).
\eea

\section{Gluon-gluon hard function in a ``trace'' color basis}
\label{3LoopPredictTrace}

In SU($N_c$) gauge theory, the four-gluon 
amplitude can be written in a basis of single- 
and double-trace operators. We follow the 
definitions in \cite{Naculich:2011ep} 
(with traces normalized as 
$\TR\left[1\right]=N_c$ and 
$\TR\left[T^aT^b\right]=\tfrac12\delta^{ab}$):
\bea \label{cTrace} \nn
c^{[\TR1]} &=& \TR\left[ T^{a_1} T^{a_2} T^{a_3} T^{a_4}\right]+\TR\left[ T^{a_1} T^{a_4} T^{a_3} T^{a_2}\right], \\ \nn
c^{[\TR2]} &=& \TR\left[ T^{a_1} T^{a_2} T^{a_4} T^{a_3}\right]+\TR\left[ T^{a_1} T^{a_3} T^{a_4} T^{a_2}\right], \\ \nn
c^{[\TR3]} &=& \TR\left[ T^{a_1} T^{a_4} T^{a_2} T^{a_3}\right]+\TR\left[ T^{a_1} T^{a_3} T^{a_2} T^{a_4}\right], \\ \nn 
c^{[\TR4]} &=& \TR\left[ T^{a_1} T^{a_3}\right] \TR\left[ T^{a_2} T^{a_4}\right], \\ \nn  
c^{[\TR5]} &=& \TR\left[ T^{a_1} T^{a_4}\right] \TR\left[ T^{a_2} T^{a_3}\right], \\  
c^{[\TR6]} &=& \TR\left[ T^{a_1} T^{a_2}\right] \TR\left[ T^{a_3} T^{a_4}\right]. 
\eea
Within this color basis the tree level amplitude is easily 
obtained by noting that 
\be
f^{a_1a_4b}f^{a_2a_3b} = 2 \left(c^{[\TR1]} - c^{[\TR3]}\right),
\ee
and the amplitude reads (recall that 
$\MM^{(0)} = \Hhard^{(0)}$):
\bea  \label{H0ComponentsTrace}\nn 
\Hhard^{(0),[\TR1]} &=& -\frac{4s}{t}, \qquad  \Hhard^{(0),[\TR3]} = \frac{4s}{t}, \\ 
\Hhard^{(0),[\TR2]} &=& \Hhard^{(0),[\TR4]} = \Hhard^{(0),[\TR5]} = \Hhard^{(0),[\TR6]} = 0.
\eea
Explicit result for the color amplitude components
in the trace color basis can be obtained either by deriving 
a rotation matrix, which rotates from the orthonormal basis 
in \eqn{cgggg} to the trace basis in \eqn{cTrace}, or by 
obtaining an explicit matrix representation for the operators 
$\Tt$ and $\Tsu$ in the trace basis. We have performed 
the calculation in both ways, and here we report about the 
second method. 

To represent the color Casimirs as matrices 
acting on this basis, the first step is to express 
the generators on the external color-adjoint
gluons in terms of commutators inside the 
trace, which follow from the definition:
\be
 \T_1^b T^{a_1} \equiv -i f^{ba_1c} T^{c} = [T^{a_1},T^b].
\ee
Color contractions inside the traces can then be simplified using the 
SU($N_c$) identities
\be
\TR\big[T^a X T^a Y\big]=\frac12\TR[X]\TR[Y]-\frac{1}{2N_c}\TR[X Y].
\ee
Thus, for example,
\bea \nn
\Tt c^{[\TR1]} &=&\TR\left[ T^{a_1}\left( T^b T^b T^{a_2}T^{a_3}
 -2T^b T^{a_2}T^{a_3} T^b + T^{a_2}T^{a_3} T^bT^b\right)T^{a_4}\right]  \\
 &=& N_c \, c^{[\TR1]} -2 \, c^{[\TR5]}.
\eea
Proceeding similarly for the other basis elements, we obtain the matrix representation:
\be
 \Tt = \left[
 \def\sp{{\hspace{3mm}}}
 \begin{array}{c@\sp c@\sp c@\sp c@\sp c@\sp c}
 N_c & 0&0&0&0&-1 \\
 0 & 2N_c & 0 &1&0&1\\
 0&0&N_c&-1&0&0 \\
 0&2&0&2N_c&0&0\\
 -2&0&-2&0&0&0\\
 0&2&0&0&0&2N_c \end{array}\right],
 \quad
  \Tsu = \left[
 \def\sp{{\hspace{3mm}}}
 \begin{array}{c@\sp c@\sp c@\sp c@\sp c@\sp c}
 -\tfrac{N_c}{2} & 0&0&0&-1&-\tfrac12 \\
 0 & 0& 0 &-\tfrac12&0&\tfrac12\\
 0&0&\tfrac{N_c}{2}&\tfrac12&1&0 \\
 0&1&2&N_c&0&0\\
 -1&0&1&0&0&0\\
 -2&-1&0&0&0&-N_c \end{array}\right].
\ee
Similarly, in the trace basis the the color 
operator defined in \eqn{TConst}, and 
associated to the constant term of the 
quadrupole correction reads 
\beqa \label{TConstTrace} \nn
\frac{1}{2}\bigg\{ f^{abe}f^{cde} 
\bigg[ \{\T_t^a, \T_t^d \}  \Big(\{\T_{s-u}^b, \T_{s-u}^c \} 
+ \{\T_{s+u}^b, \T_{s+u}^c \} \Big)  \\ \nn
&&\hspace{-6.0cm}+\,   \{\T_{s-u}^a, \T_{s-u}^d \}
 \{\T_{s+u}^b, \T_{s+u}^c \} \bigg] -\frac{5}{8} C_A^2 \Tt \bigg\}_{gg} \\ 
&&\hspace{-8.0cm}= \, \left[
\begin{array}{cccccc}
 9 N_c & 
 -4 N_c & 
 -4 N_c & 
 -4 & 
 \frac12 (4+ N_c^2) & 
 \frac12 (4+ N_c^2) \\
 -4 N_c & 
 9 N_c & 
 -4 N_c & 
 \frac12 (4+ N_c^2) & 
 -4 & 
 \frac12 (4+ N_c^2) \\
 -4 N_c & 
 -4 N_c & 
 9 N_c & 
 \frac12 (4+ N_c^2) & 
 \frac12 (4+ N_c^2) & 
 -4 \\
 2 & 
 2+N_c^2 & 
 2+N_c^2 &
 6 N_c & 
 0 & 
 0 \\
 2+N_c^2 & 
 2 & 
 2+N_c^2 & 
 0 & 
 6 N_c & 
 0 \\
 2+N_c^2  & 
 2+N_c^2  &
 2 &
 0 & 
 0 & 
 6 N_c
\end{array}
\right] \, .
\eeqa

By using these result we obtain 
the following results: the LL hard function 
at all order reads 
\bea  \label{HnnComponentsTrace}\nn 
\Hhard^{(n,n),[\TR1]} &=& -\frac{N_c^n}{n!} \, (\hat \alpha_g^{(1)})^n\,  \frac{4s}{t}, \qquad  
\Hhard^{(n,n),[\TR3]} = \frac{N_c^n}{n!} \, (\hat \alpha_g^{(1)})^n \, \frac{4s}{t}, \\ 
\Hhard^{(n,n),[\TR2]} &=& \Hhard^{(n,n),[\TR4]} = \Hhard^{(n,n),[\TR5]} = \Hhard^{(n,n),[\TR6]} = 0.
\eea
Next, the we provide the NLL and the NNLL terms 
at each order of the perturbative expansion, starting 
at one loop. We obtain
\bea \label{H10ComponentsTrace}  \nn
\Hhard^{(1,0),[\TR1]} &=& \Big[ 2\, i \pi \, N_c \, \hat \alpha_g^{(1)}  - 4\Big(D_i^{(1)} + D_j^{(1)}\Big)\Big] \frac{s}{t} , \\ \nn
\Hhard^{(1,0),[\TR2]} &=& 0, \\ \nn
\Hhard^{(1,0),[\TR3]} &=& \Big[ 2\, i \pi \, N_c \,  \hat \alpha_g^{(1)}  + 4\Big(D_i^{(1)} + D_j^{(1)}\Big)\Big] \frac{s}{t}  \\ \nn
\Hhard^{(1,0),[\TR4]} &=& 8\, i \pi \, \hat \alpha_g^{(1)} \, \frac{s}{t} , \\ \nn 
\Hhard^{(1,0),[\TR5]} &=& 8\, i \pi \, \hat \alpha_g^{(1)} \, \frac{s}{t} , \\ 
\Hhard^{(1,0),[\TR6]} &=& 8\, i \pi \, \hat \alpha_g^{(1)} \, \frac{s}{t} .
\eea
At two loops the NLL term reads
\bea  \label{H21ComponentsTrace} \nn
\Hhard^{(2,1),[\TR1]} &=& \Big\{-4 N_c \Big[\hat \alpha_g^{(2)} 
+ \hat \alpha_g^{(1)} \Big(D_i^{(1)} + D_i^{(1)}\Big) \Big] \\ \nn
&&\hspace{1.5cm}+\, 2 i \pi \Big[ -2f^{(2,1)}_a+(N_c^2-4) (\hat \alpha_g^{(1)})^2 \Big] \Big\} \frac{s}{t}, \\ \nn
\Hhard^{(2,1),[\TR2]} &=&  8 i \pi \Big[ f^{(2,1)}_a+2 (\hat \alpha_g^{(1)})^2 \Big] \frac{s}{t} , \\ \nn
\Hhard^{(2,1),[\TR3]} &=& \Big\{ 4 N_c \Big[ \hat \alpha_g^{(2)} 
+ \hat \alpha_g^{(1)} \Big(D_i^{(1)} + D_j^{(1)}\Big) \Big] \\ \nn
&&\hspace{1.5cm}+\,  2 i \pi \Big[ -2f^{(2,1)}_a+(N_c^2-4) (\hat \alpha_g^{(1)})^2 \Big] \Big\}  \, \frac{s}{t}  \\ \nn
\Hhard^{(2,1),[\TR4]} &=&  4 N_c i \pi \Big[ f^{(2,1)}_a+4 (\hat \alpha_g^{(1)})^2 \Big]  \, \frac{s}{t}, \\ \nn
\Hhard^{(2,1),[\TR5]} &=& -8 N_c i \pi \Big[ f^{(2,1)}_a+ (\hat \alpha_g^{(1)})^2 \Big]   \, \frac{s}{t} , \\
\Hhard^{(2,1),[\TR6]} &=& 4 N_c i \pi \Big[ f^{(2,1)}_a+4 (\hat \alpha_g^{(1)})^2 \Big] \, \frac{s}{t},
\eea
where the function $f^{(2,1)}_a$ has been 
defined in \eqn{f21}. At NNLL the real 
component of the amplitude reads
\bea  \label{H20ComponentsTrace}\nn
\RE[\Hhard^{(2,0),[\TR1]}] &=& -\frac{1}{3} \Big[ 12 \Big(D_i^{(1)}D_j^{(1)} +D_i^{(2)} +D_j^{(2)} \Big) \\ \nn
&&\hspace{1.5cm}+\, \pi^2 \Big( 3 (N_c^2+ 12) \hat R^{(2)} -N_c^2 R^{(2)} \Big) \Big] \, \frac{s}{t}, \\ \nn
\RE[\Hhard^{(2,0),[\TR2]}] &=& 0, \\ \nn
\RE[\Hhard^{(2,0),[\TR3]}] &=& \frac{1}{3} \Big[ 12 \Big(D_i^{(1)}D_j^{(1)} +D_i^{(2)} +D_j^{(2)} \Big) \\ \nn
&&\hspace{1.5cm}+\, \pi^2 \Big( 3 (N_c^2+ 12) \hat R^{(2)} -N_c^2 R^{(2)} \Big) \Big] \, \frac{s}{t}, \\ \nn
\RE[\Hhard^{(2,0),[\TR4]}]  &=&  12 \pi^2 N_c \,  \hat R^{(2)} \, \frac{s}{t}, \\ \nn
\RE[\Hhard^{(2,0),[\TR5]}] &=&  0 , \\
\RE[\Hhard^{(2,0),[\TR6]}] &=&  -12 \pi^2 N_c \,  \hat R^{(2)} \, \frac{s}{t},
\eea
At three loops, the NLL component reads
\bea  \label{H32ComponentsTrace}\nn
\Hhard^{(3,2),[\TR1]} &=& 
\Big\{ -  i \pi N_c \Big[ 2 f^{(3,2)}_a + 2 f^{(3,2)}_b - (N_c^2 -12) (\hat \alpha_g^{(1)})^3 \Big] \\ \nn
&&\hspace{1.5cm}-\, 4 N_c^2 \hat \alpha_g^{(1)} \Big[ \hat \alpha_g^{(2)} 
+ \tfrac12 \hat \alpha_g^{(1)} \Big(D_i^{(1)} + D_j^{(1)} \Big) \Big] \Big\} \, \frac{s}{t}, \\ \nn
\Hhard^{(3,2),[\TR2]} &=&  2\, i \pi\,N_c (4 f^{(3,2)}_a + 16 (\hat \alpha_g^{(1)})^3 + 3 f^{(3,2)}_b)\,\frac{s}{t}, \\ \nn
\Hhard^{(3,2),[\TR3]} &=&  
\Big\{- i \pi N_c \Big[ 2 f^{(3,2)}_a + 2 f^{(3,2)}_b - (N_c^2 -12) (\hat \alpha_g^{(1)})^3 \Big] \\ \nn
&&\hspace{1.5cm}+\, 4 N_c^2 \hat \alpha_g^{(1)} \Big[ \hat \alpha_g^{(2)} 
+ \tfrac12 \hat \alpha_g^{(1)} \Big(D_i^{(1)} + D_j^{(1)} \Big) \Big] \Big\} \, \frac{s}{t}, \\ \nn
\Hhard^{(3,2),[\TR4]}  &=& 2 \,
i \pi \Big[(N_c^2+4) f^{(3,2)}_a + (N_c^2 + 2) f^{(3,2)}_b + 8 (N_c^2 + 1) (\hat \alpha_g^{(1)})^3  \Big] \, \frac{s}{t}, \\ \nn
\Hhard^{(3,2),[\TR5]} &=& 4  \,
i \pi \Big[(N_c^2+2) f^{(3,2)}_a + f^{(3,2)}_b - (N_c^2-4) (\hat \alpha_g^{(1)})^3  \Big] \, \frac{s}{t} , \\
\Hhard^{(3,2),[\TR6]} &=& 2  \,
i \pi \Big[ (N_c^2+4) f^{(3,2)}_a + (N_c^2 + 2) f^{(3,2)}_b + 8 (N_c^2+1) (\hat \alpha_g^{(1)})^3  \big] \, \frac{s}{t},
\eea
where the functions $f^{(3,2)}_a$, $f^{(3,2)}_b$ and  $f^{(3,2)}_c$ have been 
defined in \eqn{f32}. The real part of the NNLL term reads
\bea \label{H31ComponentsTrace}\nn
\RE[\Hhard^{(3,1),[\TR1]}] &=& - 4N_c \Big[ \hat \alpha_g^{(3)} + \hat \alpha_g^{(2)}\Big(D_i^{(1)} + D_j^{(1)}\Big) 
+ \hat \alpha_g^{(1)}\Big(D_i^{(2)} + D_j^{(2)}\Big)  + \hat \alpha_g^{(1)} D_i^{(1)} D_j^{(1)} \\ \nn
&&\hspace{2.5cm}+\, \pi^2 \Big( N_c^2 (4 f_d^{(3,1)} - f_a^{(3,1)}) - 4(4 f_b^{(3,1)} + f_c^{(3,1)}) \Big) \Big] \, \frac{s}{t}, \\ \nn
\RE[\Hhard^{(3,1),[\TR2]}] &=&  0, \\ \nn
\RE[\Hhard^{(3,1),[\TR3]}] &=& 4N_c \Big[ \hat \alpha_g^{(3)} + \hat \alpha_g^{(2)}\Big(D_i^{(1)} + D_j^{(1)}\Big) 
+ \hat \alpha_g^{(1)}\Big(D_i^{(2)} + D_j^{(2)}\Big)  + \hat \alpha_g^{(1)} D_i^{(1)} D_j^{(1)}  \\ \nn
&&\hspace{2.5cm}+\, \pi^2 \Big( N_c^2 (4 f_d^{(3,1)} - f_a^{(3,1)}) - 4(4 f_b^{(3,1)} + f_c^{(3,1)}) \Big) \Big] \, \frac{s}{t}, \\ \nn
\RE[\Hhard^{(3,1),[\TR4]}]  &=& -4 \pi^2 N_c^2 \Big[6 f_a^{(3,1)} - 2 f_b^{(3,1)} + f_c^{(3,1)} \Big] \, \frac{s}{t}, \\ \nn
\RE[\Hhard^{(3,1),[\TR5]}] &=&  0, \\
\RE[\Hhard^{(3,1),[\TR6]}] &=& 4 \pi^2 N_c^2 \Big[6 f_a^{(3,1)} - 2 f_b^{(3,1)} + f_c^{(3,1)} \Big] \, \frac{s}{t},
\eea
and the functions $f^{(3,1)}_a$, $f^{(3,1)}_b$, $f^{(3,1)}_c$ 
and $f^{(3,1)}_d$ have been defined in \eqn{f31}.

\section{Gluon Regge trajectory and impact factor
in ${\cal N}=4$ SYM}
\label{TrajectoryImpact}

In section \ref{dip_comparison} we have shown how 
to extract the impact factors and Regge trajectory
from a given amplitude. These ingredients are 
necessary to obtain a complete description of the 
$1\to 1$ transition up to NNLL in the high-energy 
logarithm. As discussed in section \ref{dip_comparison}, 
the recent calculation of the gluon-gluon amplitude 
up to three loops in ${\cal N}=4$ SYM \cite{Henn:2016jdu} 
allows us to obtain the Gluon Regge trajectory at NNLO 
in this theory, which was previously unknown. According to 
\eqns{impactfactorOneLoop}{ImpactFactorTwoLoops}, 
we are able to extract also the gluon impact factor in 
this theory. This information represents the last ingredient 
which is necessary in order to obtain a complete description 
of the $1\to 1$ transition up to NNLL in the high-energy 
logarithm, and we collect it in this appendix. 
We express the gluon Regge trajectory
in terms of the coefficients $\hat \alpha_g^{(i)}$, which 
enters directly the fixed-order amplitude coefficients 
provided in appendix \ref{3LoopPredictOrtho} and 
\ref{3LoopPredictTrace}.

At one loop the gluon Regge trajectory in 
${\cal N}=4$ SYM is of course identical 
to the QCD case, i.e. 
\be
\hat\alpha_g^{(1)}|_{{\cal N}=4 \, \rm SYM} = \frac{1}{2\eps}(\rGamma - 1) = 
-\frac14\zeta_2\, \eps -\frac76\zeta_3 \,\eps^2 + {\cal O}(\eps^3).
\ee
The gluon impact factor at one loop reads 
\bea \label{D1gSYM} \nn
D_g^{(1)}|_{{\cal N}=4 \, \rm SYM} &=& N_c \bigg[ \zeta_2 
+ \eps \frac{17}{12} \zeta_3  + \eps^2 \frac{41}{32}\zeta_4 
+ \eps^3 \bigg(-\frac{59}{24}\zeta_2\zeta_3 +\frac{67}{20} \zeta_5 \bigg) \\ 
&&\hspace{1.0cm}+\, \eps^4\bigg(-\frac{35}{18}\zeta_3^2 - \frac{7}{6}\zeta_6 \bigg) 
+ {\cal O}(\eps^5)\bigg] \, .
\eea
It is easy to check that the impact factor in ${\cal N}=4$ SYM
correspond to the highest trascendental weight of the $N_c$ term
of the correponsing QCD impact factor, see \eqn{D1g}.
At two loops the Regge trajectory reads
\be
\hat\alpha_g^{(2)}|_{{\cal N}=4 \, \rm SYM} = N_c \bigg[ - \frac{\zeta_3}{8} 
- \eps \frac{3}{16}\zeta_4 +\eps^2 \bigg(\frac{71}{24}\zeta_2 \zeta_3 
+ \frac{41}{8}\zeta_5 \bigg)+{\cal O}(\eps^3) \bigg] \,,
\ee
which corresponds to the ${\cal O}(\as/\pi)^2$ coefficient of 
$-H_{1\to 1}/C_A$ in \eqn{3loop_traj}. Once again, 
it corresponds to the term with highest trascendental weight 
of the $N_c$ term of the QCD result, see \eqn{alphahat}.
The impact factor at two loops reads
\be \label{D2gSYM}
D_g^{(2)} = N_c^2 \bigg[ - \frac{\zeta_2}{32\eps^2} - \frac{\zeta_4}{64}
+\eps \bigg(\frac{17}{24} \zeta_2\zeta_3 -\frac{39}{16} \zeta_5 \bigg)
+\eps^2 \bigg(-\frac{659}{288} \zeta_3^2 - \frac{5531}{512} \zeta_6 \bigg) 
+ {\cal O}(\eps^3) \bigg] \,. 
\ee
Last, the Regge trajectory at three loops (with meaning explained below \eqn{3loop_traj}) reads
\be
\hat\alpha_g^{(3)}|_{{\cal N}=4 \, \rm SYM} = 
N_c^2 \bigg[-\frac{\zeta_2}{144}\frac{1}{\eps^3}
+\frac{5\zeta_4}{192}\frac{1}{\eps}
+\frac{107}{144}\zeta_2\zeta_3
+\frac{\zeta_5}{4} + {\cal O}(\eps)\bigg] .
\ee
Notice that the difference in the single pole 
compared to the ${\cal O}(\as/\pi)^3$ coefficient of 
$-H_{1\to 1}/C_A$ in \eqn{3loop_traj} is due to the 
subtraction of $K^{(3)}$, see \eqn{alphaghat}.

\bibliography{Draft-v98}
\bibliographystyle{JHEP}

\end{document}